**First Edition**

# Practice Problems for Hardware Engineers

**SHAHIN NAZARIAN**

First Edition

# Practice Problems for Hardware Engineers

SHAHIN NAZARIAN

# Practice Problems for Hardware Engineers

PRACTICE PROBLEMS FOR HARDWARE ENGINEERS

Copyright @ 2021 by Shahin Nazarian

# About the Author

Shahin Nazarian is a faculty member of the Viterbi School of Engineering (Electrical and Computer Engineering Department) at the University of Southern California.

His teaching and research activities span over a broad range of topics related to design, optimization and verification of digital hardware and software systems. He has designed several USC undergraduate and graduate courses on logic design, SoC design, system verification, software, and algorithm design.

He has worked as a consultant and designer for numerous projects involving semiconductor and software companies such as Apple, Broadcom, Fitbit, Fujitsu, Google, GoPro, Huawei, Hulu, Intel, LG, Nintendo, Nokia, Qualcomm, Samsung, Semcon, Synopsys, TI, TiVo, Ubiquiti Networks, Xilinx, and ZTE.

He is the founder and President of Vervecode, Inc. which is active on entrepreneurship, investment, and research in various areas of computer science, engineering, and mathematics.

# About this Book

This book is to help undergraduate and graduate students of electrical and computer engineering disciplines with their job interviews. It may also be used as a practice resource while taking courses in VLSI, logic and computer architecture design. The first edition consists of more than 150 problems and their solutions which the author has used in his VLSI, logic, and architectures courses while teaching at USC. The author wishes this book to be available free of charge, subject to the copyright policy on page 3.

# Contents

# Chapter 1 – CMOS Implementation

1. Given $F = \overline{A \cdot B + C \cdot D \cdot E}$

   a. Design a CMOS gate to realize the function F.
   b. Size transistors in step A, such that the rise and fall delays are equal to the rise and fall delays of the following inverter, respectively.

   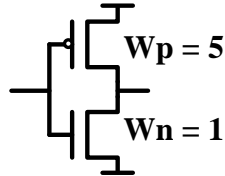

   c. Calculate the ratios between worst-case delay and best-case delay, i.e., $T_r^{worst}/T_r^{best}$ and $T_f^{worst}/T_f^{best}$.

2. Implement the following logic using compound CMOS gates:
$$\bar{X} = AB(C+D) + E(F + G(P+Q))$$

   a. Draw the CMOS transistor level schematic of pull-down and pull-up network.
   b. Size the pull-down and pull-up networks. (assume $\mu_n/\mu_p = 4$)
   c. Based on your sizing results, what is the ratio of best case and worst case fall/rise delays?

3. Give the CMOS implementation for following equation and give the sizing. (assume $\mu_n/\mu_p =$
$$\overline{OUT} = XY(Z + W + V + U)$$

4. Given the following circuit:

   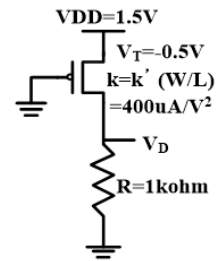

   a. Assume the pmos transistor were in linear region. Write an equation for the output voltage.
   b. Repeat part a, for saturation.
   c. Calculate which region, linear or saturation is the correct one.

5. A PMOS transistor is fabricated with the following specifications:
   - $Zero\ substrate\ bias\ threshold\ voltage = -0.5v$
   - $Substrate - bias\ coefficient = -0.25v^{0.5}$
   - $Channel\ length\ modulation\ coefficient = -0.05v^{-1}$
   - $Substrate\ Fermi\ Potential = 0.2v$
   - $k' = 20\ uA/V^2$
   - $Suppose\ \sqrt[2]{0.41} = 0.6$

   a. Given ISD=8.48uA, VS=1.2V, VB=1.8V, VG=0.2V, VD=0V. What is the transistor region? Calculate W/L.
   b. Given VS=1.2V, VB=1.8V, VG=0V, VD=1V. What is the transistor region in this case? Calculate ISD.

6. Given $\bar{F} = A(\bar{B} + C) + D + \bar{E}$:

a. Design a CMOS compound gate to realize this function.
b. Find the Euler paths for both PDN and PUN. Can you find them identical?
c. Draw the layout in stick diagram for $\bar{F}$ compound gate, according to the Euler paths you derived.

7. Given $\bar{F} = A(B + \bar{C} + D) + (\bar{E} + F)G\bar{H}$:
a. Draw the CMOS Gates to realize the function
b. Size all transistors for BEST case
c. Find a Common Euler Path for both PUN and PDN. (Hint: Try to arrange transistor order properly to find a common Euler Path)
d. Use the Common Euler Path to draw Stick Diagram.

8. Based on the schematic below, answer the questions:
a. Is there a Euler path in the following pull-down network? If there is, please identify it. If not, please redesign the pull-down network so that the functionality is the same and a Euler path exists.

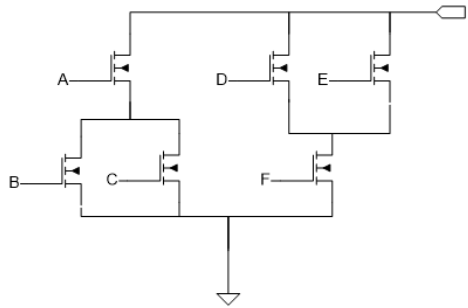

b. Draw the corresponding transistor level schematic of pull-up network.
c. Find the common Euler path in pull-up and pull-down network.

9. Let us denote the scaling factor as S. How would the sheet resistance change if we use
a. Constant-field scaling?
b. Constant-voltage scaling?

10. Voltage is scaled down by S, and all dimensions are scaled down by M. What is the scaling factor for I, Delay, P, Energy, Power Density?

11. What is function F?

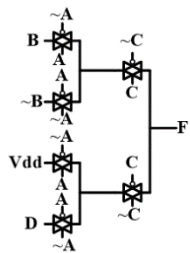

12. What are the voltage values at internal nodes 1-6? Consider $V_{dd} \gg V_{tn}$

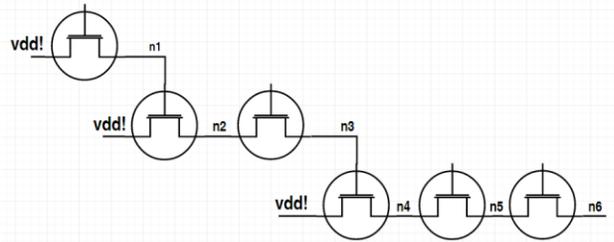

**13.** Assume the nmos threshold voltage VT is roughly 1/5 of VDD, what is the maximum voltage that one can find at A, B, C and D, respectively?

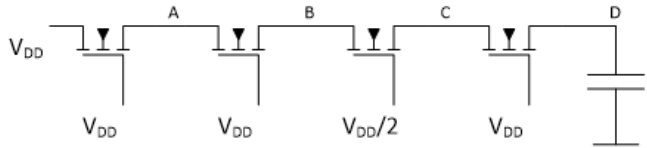

**14.** Design three masks for diffusion, n-well, contact (window), and metal layers to realize a 1kΩ resistor using p-type diffusion with sheet resistance 200Ω per square. The window and metal layers are for the contacts on both ends of the resistor. Also, the design should follow these spacing rules:
  a. The minimum feature size is 1um for line width and window opening.
  b. 0.5um extension of metal feature and diffusion feature over the contact is required.
  c. 1um extension of n-well around diffusion is required.

**15.** For the given circuit please calculate the delay of the top three critical paths.
  - XOR=XNOR = 1ns
  - NAND = 500ps
  - NOR = 750ps

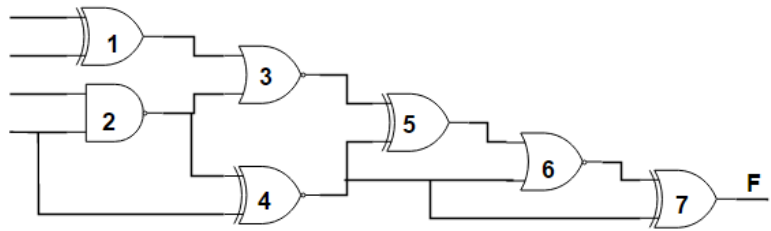

**16.** A combinational design has the worst-case delay of 5ns. It is divided into 5 subcircuits, (a.k.a. pipeline stages) with the delay values of 0.2, 1.8, 0.8, 1.1, 1.1ns, respectively. Sequential logic elements (registers) are used between the sub-circuits to build a 5-stage pipeline circuit.
  a. Draw the 5-stage pipeline of the circuit described above.
  b. What is the maximum clock frequency this pipeline can operate?
  c. The total delay of processing 1000 sets of serial inputs using the original (non-pipelined) design is 5us (the propagation delay of the original circuit is 5ns, therefore the total delay is $1000 \times 5ns = 5\mu s$. What is the total delay of the pipelined circuit to process the 1000 sets of inputs? Ignore the delay of the registers between each two stage

**17.** Consider an 8-bit RBS (Ripple Borrow Subtractor) which uses 8 single-bit full subtractors back to back. Let us number the blocks from 0 to 7 and label the 8-bit subtractor's inputs Bin[0], X[0:7] and Y[0:7] as well as its outputs S[0:7] and Bout[0:7]. Assume for each single bit full adder block, the propagation delay from X,Y inputs to its S output is 1.0ns but from Bin to S is 1.5ns . The propagation delay from X,Y to its Bout output is 2.0ns but delay from Bin to Bout is 0.5ns. Assume Bin, X, and Y change at time 0 and that results in a change of value in both S and Bout outputs.
   a. When is the earliest time S[0] becomes stable and correct? Answer the same question for S[3], Bout[3], S[6], and Bout[7]?
   b. What is the propagation delay from input X[7] to S[7]? How about Bin[5] to Bout[5]?
   c. Which path is the critical path and what is the propagation delay of the critical path?
   d. We analyzed the timing of a 4-bit RCA and discussed why two types of single-bit full adder blocks are required to minimize the delay of the critical path. Use that discussion and argue how many types of single-bit full subtractor blocks are required for an 8-bit RBS and for each of the 8 blocks, which type should be used to minimize the overall delay.
   e. If we assume the data are given from flip-flop and written into flip-flop, what is the maximum clock frequency at which this subtractor can operate? Can you make the operation faster by approx. four times.(Hint: prev. question)? If yes, how? If no, explain why?

**18.** Below is a circuit for carrier generation for a 4-bit adder. Assume that $\mu_n/\mu_p = 2$.
   a. Consider the pmos pull-up sub-network and nmos pull-down sub-network for input signal P1, G1 and G0. Each of the sub-networks has three transistors only. Prove that they are logically equal.
   b. Please size the transistors such that the circuit has equal worst-case rise and fall delay and they are equal to those of a minimum-sized inverter with equal worst-case rise and fall delay.
      i. What about Pn and Gn?
   c. If all the transistors are minimum sized (including the ones in the inverter), what's the worst-case rise and fall time of the node at the input of the inverter? Suppose the gate capacitance of a minimum-sized nmos transistor is C and its resistance is R. We ignore all the diffusion capacitances.

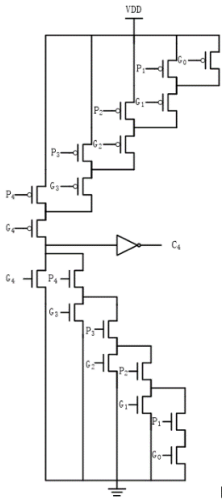

**19.** How does the current density-delay product scale with the CMOS process technology for full scaling or constant field scaling? Ignore the effects of interconnect capacitances.

20. Consider the NMOS pass transistor below. It is fabricated with the following physical parameters:
    - $N_D = 10^{21} cm^{-3}$
    - $N_A(substrate) = 10^{17} cm^{-3}$
    - $N_A(channel\ stop) = 10^{19} cm^{-3}$
    - $W = 10 \mu m$
    - $Y = 5 \mu m$
    - $L_M = 1.5 \mu m$
    - $L_D = 0.25 \mu m$
    - $X_j = 0.3 \mu m$
    - $tox = 100 Å$

    Also, $V_T = 0.5v$. The body terminal of the NMOS is connected to ground. Assume all the p-n junctions are abrupt. Ignore body effect and all other second order effects.

    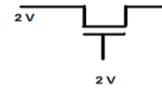

    a. Calculate the total oxide capacitance of the NMOS.
    b. Calculate the total drain diffusion capacitance of the NMOS.

    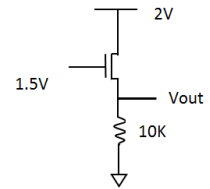

21. Create 4- input NAND and NOR gates using pass transistors.

22. Given the following circuit:
    a. Assume there is no body effect and no channel length modulation. The parameters for the NMOS is as follows. W=10um, L=1um, Vt=0.5V, $Cox = 14fF/um^2$, $K'_n = 220uA/V^2$, $L_D = 0.1um$, calculate Vout and gate capacitance.

23. An NMOS transistor is fabricated with the following specifications:
    - Overlap length: $L_D = 3 \mu m$
    - Channel length $L = 6 \mu m$
    - Gate oxide capacitance $C_{ox} = 7e-8\ F/cm^2$
    - Transistor width $W = \mu m$
    - Junction length $Y = 10 \mu m$
    - Junction depth $X_j = 2 \mu m$
    - Gate oxide thickness $t_{ox} = 50nm$
    - Capacitance per unit area $C_{j0} = 1.5e-8\ F/cm^2$
    - Capacitance per unit length for sidewalls $C_{jsw} = 6.5\ pF/cm$

a. Calculate the Oxide-related capacitance $(C_{gb}, C_{gd}, C_{gs})$ in saturation region.
b. Calculate the junction capacitance based on the above parameters. (Note: take voltage equivalence factor as 1)

24. An NMOS transistor is fabricated with the following physical parameters:
- $N_D = 10^{20} \text{cm}^{-3}$
- $N_A(\text{substrate}) = 10^{16} \text{cm}^{-3}$
- $N_A(\text{channel stop}) = 10^{19} \text{cm}^{-3}$
- $W = 8\mu m$
- $Y = 5\mu m$
- $L = 2\mu m$
- $L_D = 0.35\mu m$
- $X_j = 0.2\mu m$
- $M_j = M_{jsw} = 0.5$

a. Calculate the drain diffusion capacitance for $V_{DB} = 2.5V$? Repeat this for 4V.
b. Calculate the overlap capacitance between the gate and drain for an oxide thickness of $t_{ox} = 50\text{Å}$.

25. Implement the following logic using compound CMOS gates:
$$F' = ABC + DE + GH$$
a. Draw the transistor level schematic of pull-down and pull-up network.
b. Size the pull-down and pull-up networks. (assume $\mu_n/\mu_p = 4$)
c. Based on your sizing results, what is the ratio of best case and worst-case fall/rise delays? (use resistive model)

26. The delay as a function of Vdd: Let's look at the fall delay for a CMOS inverter, with $V_{TN} = 0.5v$ and $k_n = 200\,\mu A/v^2$. Suppose the input of the inverter is 0, and it changes to 1 with zero transition time. The output voltage will fall from Vdd to zero.
a. At the very moment the input changing to 1 (the output voltage is still Vdd), what region (OFF, linear, or saturation) is the NMOS in for (a) Vdd = 2V and (b) Vdd = 0.8V? Explain why.
b. When the output voltage level drops to Vdd/2, calculate the output current of the NMOS for (i) Vdd = 2V and (ii) Vdd = 0.8V.
c. Compare the current values with different Vdd values in (b). Given that the output current is approximately inverse proportional to the propagation delay, can you find the delay increases superlinearly, linearly, or sublinearly as the Vdd decreases?

27. Given the CMOS implementation below:
a. What is the logic of S?
b. Show S is a mirror logic

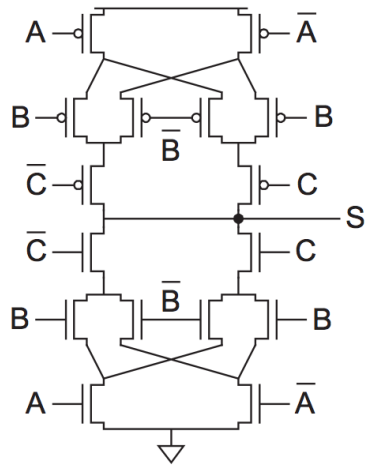

28. For the depletion-load NMOS inverter, we have:
    - k'driver=k'load=25uA/V2
    - (W/L)driver=4, (W/L)load=1
    - Vdd=1.8V
    - VT,load=-0.3V, VT,driver=0.4V.
    Ignore the body effect of depletion NMOS.

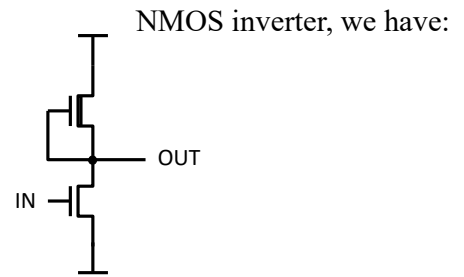

i) Fill out the following table: the regions for the two NMOS.
ii) Calculate the noise margin.

| Vin | Vout | driver | Load |
|---|---|---|---|
|  | VOH |  |  |
|  | VOL |  |  |
| VIL | high |  |  |
| VIH | Low |  |  |

29.
  i) Complete the following form for calculation of the noise margin and VM of the following gate.

| Vin | Vout | PMOS | NMOS |
|---|---|---|---|
|  | VOH |  |  |
|  | VOL |  |  |
| VIL | high |  |  |
| VIH | low |  |  |
| VM | VM |  |  |

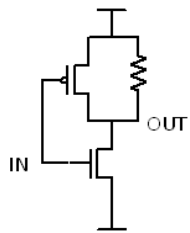

  ii) Compare VOH, VOL and VM between this gate and the standard CMOS inverter (you don't need to calculate the exact value).

**30.** Given the following Inverter:

  i)   What is the value for VOL?
  ii)  Write the necessary equations to solve VOH.
  iii) Write the necessary equations to solve VIL.
  iv)  Write the necessary equations to solve VIH.

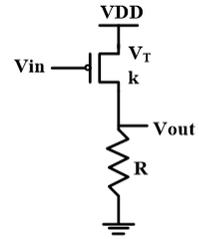

**31.** Calculate noise margin at nodes a and b.

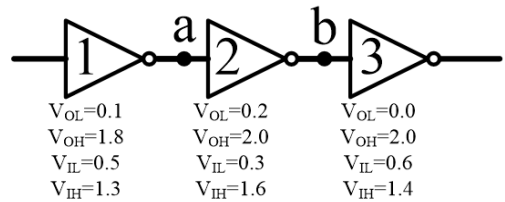

$V_{OL}=0.1$   $V_{OL}=0.2$   $V_{OL}=0.0$
$V_{OH}=1.8$   $V_{OH}=2.0$   $V_{OH}=2.0$
$V_{IL}=0.5$   $V_{IL}=0.3$   $V_{IL}=0.6$
$V_{IH}=1.3$   $V_{IH}=1.6$   $V_{IH}=1.4$

**32.** What is the maximum voltage value at node i? Repeat for OUT.

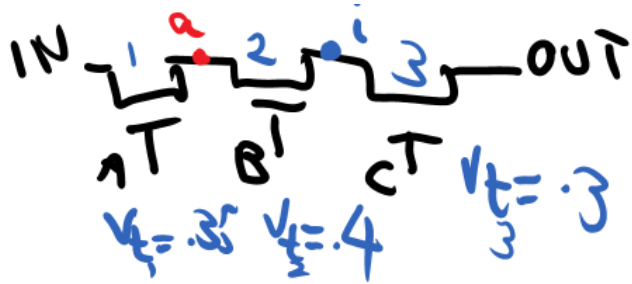

# Chapter 2 – Boolean Implementation

33. For the given Boolean equation answer the following questions:

$$\bar{F} = AB + CE + D$$

a. Draw the CMOS implementation of the given compound gate (assume $\frac{\mu_n}{\mu_p} = 4$).
b. Assuming a copy of this cell is connected to its output, as load. Which input of the second gate should you connect the output of first gate, to realize the worst case? Repeat for the best case. Please answer with brief reasoning.
c. For the worst case given above, what is the Fanout4 / Fanout1 delay. Ignore all the internal capacitances and wire capacitances. For a transistor of width W, input cap is Cg.
d. What is the best case and worst-case delay for a load of CL (For pull up and pull down)?

34. You may assume all inputs (A, B, C, D, E, F, G) and their complements are available in this problem.
a. Design a compound CMOS gate that realizes the following Boolean equation:
$$\bar{Z} = ABC + HDE + FG$$

b. Size the transistors such that the worst case rising and falling delays of the compound gate matches those of a minimum size equal rise-fall Inverter. Assume $\mu_{nmos} = 4\mu_{pmos}$ and $C_L$ be the load capacitance.
c. Compare its area to that of the minimum size inverter. The area of a gate can be approximated by the summation of the WL of the transistors in that gate.
d. Find the best case rise and fall delay for this logic, what is ratio between worst-case to best-case delays.

35. Calculate the Boolean expression for the following transmission gate implementation.

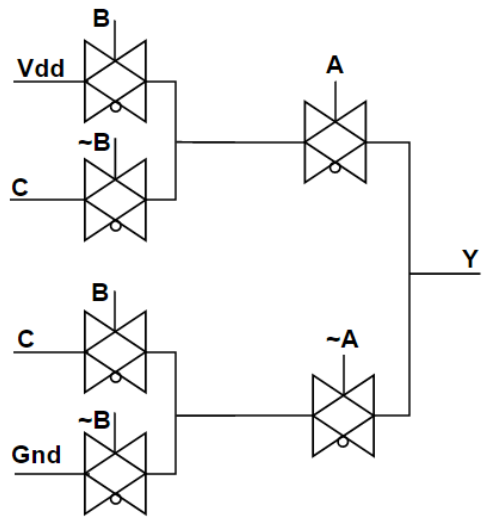

36. Below is a cell taken out of a Manchester carry

chain adder. Only one of Pi and Gi will be logic high at the same time. However, they can be both logic low at the same time.

a. Find out the Boolean equation for Ci. Use Ci-1, Pi and Gi.
b. Why do we use a transmission gate instead of a pmos or nmos pass transistor?
c. is also a circuit with Si as output below. Please derive the Boolean equation for Si. Can we use it to generate sum bit in a full adder? Why or why not?

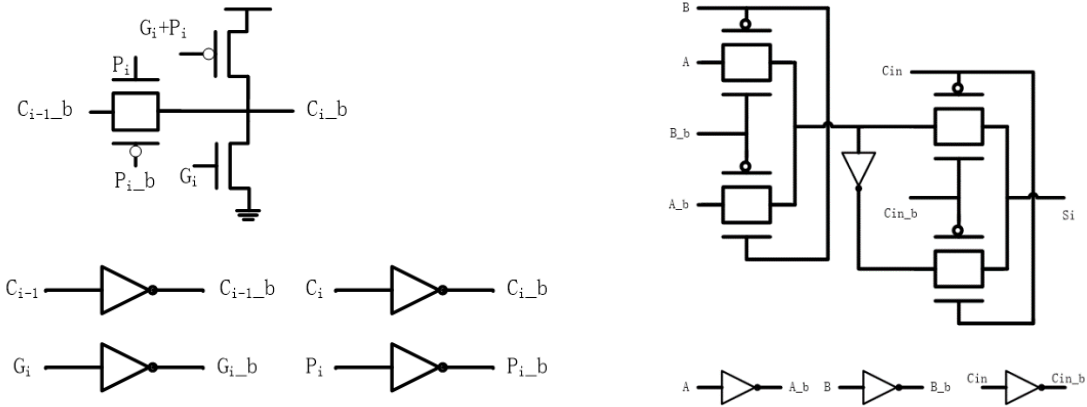

37. Suppose you are asked to design a converter to translate a 4-bit code In3In2In1In0 to a 2-bit code Out1Out0. The schematic diagrams are shown below.
a. Find out the Boolean equations for Out0 and Out1 based on In0, In1, In2 and In3.
b. For each circuit, find out the same Euler path for nmos pull down and pmos pull up networks (One for each circuit is enough, you don't need to enumerate all the common Euler paths).
c. Based on the Euler paths you have found, draw a stick diagram for Out0 and Out1 respectively.

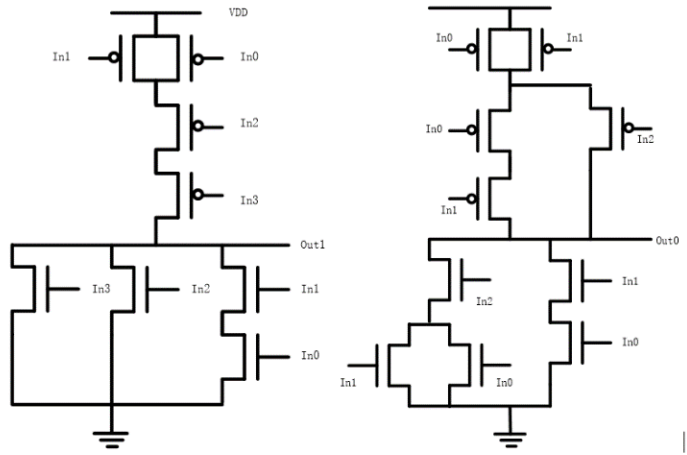

**38.** Based on the stick diagram provided, draw the transistor-level schematic. Write the corresponding Boolean equation.

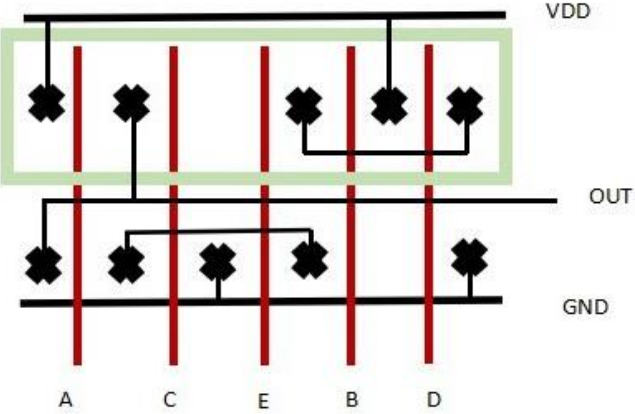

**39.** Draw the pass transistor-based (or TG-based) structure for the flowing Boolean equation.

$$F = AB + \bar{C} + D$$

# Chapter 3 – Sequential Logic

**40.** Check whether there is any Setup or Hold violation. Note: Numbers are chosen to be simple, but they are not realistic.

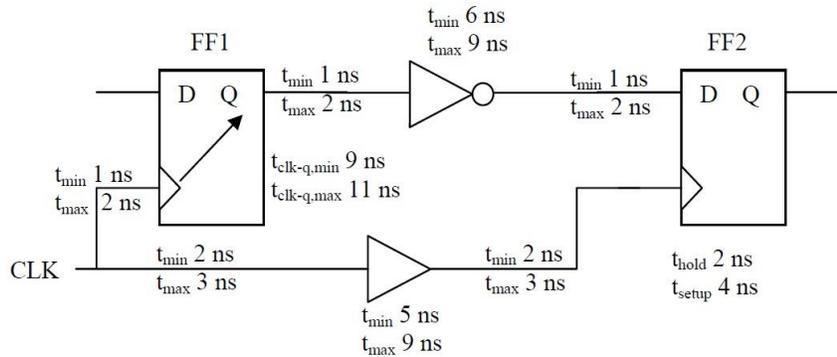

**41.** Calculate the maximum operating frequency.

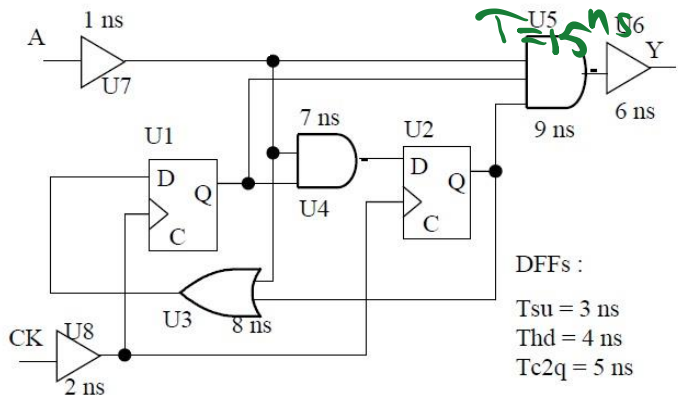

DFFs :

Tsu = 3 ns
Thd = 4 ns
Tc2q = 5 ns

**42.** Given the following information for a pipeline stage, verify whether there is setup and hold time violation:

Tcq min =0.5ns, Tcq max 1.3ns, Tskew = ±1.15ns, Tcomb-min = 1ns, Tcomb-max = 5ns, Tclk = 7.5ns, Tsetup = 1.1ns, Thold = 0.9ns.

**43.** Design a D-FF with a-hi synchronous set and reset.

**44.** Consider a long wire with overall resistance of 50k ohm and overall capacitance of 20fF. For simplicity approximate the propagation delay by time constant.
  a. What is the delay of the total wire?
  b. Now cut the wire in half and put a buffer with delay of 0.25ns and now calculate the delay of the wire. Did buffer insertion help?
  c. Next use a buffer with delay 1ns and follow the above procedure. Did buffer insertion help?
  d. Back in part (a) try adding two buffers to see whether it works better.
  e. Try adding 3 buffers.

**45.** We have a CMOS inverter with NMOS parameters as follows:

- $\mu_n C_{ox}$=50 µA/V2

- $(W/L)_n$ =12

- $V_{tn}$=0.7V

- $C_{load}$=10pF

- $VDD$=3V

Calculate output falling slew using ACC and DIFF method, between $V_{out}$=90%$VDD$ and $V_{out}$=10%$VDD$.

**46.** The worst case average output current of a 3 input NAND logic in the presence of a 50fF load is given as $10\mu A$ for falling output. That for the rising output is given as 8microAmps. What are the worst case falling and rising propagation delay of the NAND logic? Assume VDD is 1V.

**47.** Assume the effective internal cap seen at any terminal of an nMOS (pMOS) with size W is approximated as C (C) and the effective channel resistance of that nMOS (pMOS) as R (2R). Assume the resistance and capacitance of each internconnect piece between the NAND and NOR is 2R and 4C, respectively. The external load seen at Z is 16C.

a) Size each gate such that its output current strength is the same as a reference inverter (W, 2W).
b) Calculate the worst Elmore delay from primary inputs to primary output Z, for both rising and falling Z.
c) Find the input vectors that result in worst case delay for the NAND and NOR gates in this circuit.

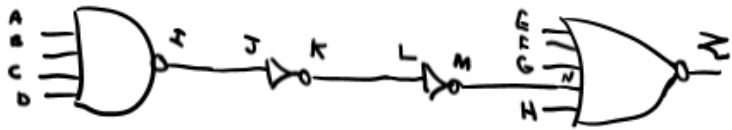

**48.** Use both capacitance-oriented and resistance-oriented formula for the Elmore delay (i.e., time constant) from node B to A as a function of R and C and verify that results are identical. In this case, a step input is connected to B and the delay to output A is asked. Repeat this for C to A, for B to C and C to B. In calculating all the delays, you may use the time constant instead.

**49.** Find the (propagation) delay from point:
  **a)** P to A
  **b)** P to B
  **c)** B to D
  **d)** C to P

You may approximate the propagation delay as time constant multiplied by 0.7. Alternatively, you may use the time constant itself to approximate the delay.

# Chapter 5 – Ring Oscillator

**50.** Calculate the duty cycle and the period of the output waveform of the marked node.

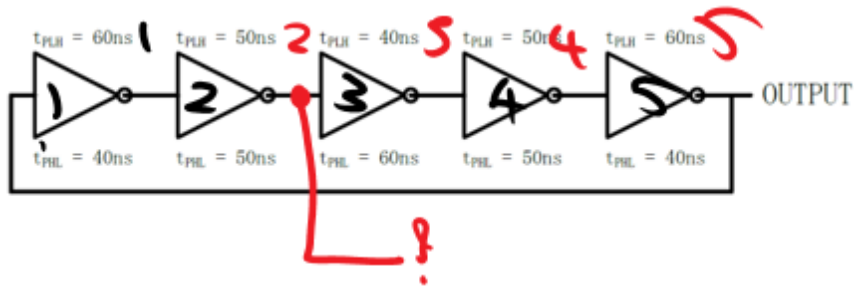

**51.** Consider a ring oscillator using 2N+1 CMOS inverters, each having a falling propagation delay TPHL = 3ns and a rising propagation delay TPLH = 5ns. Calculate the oscillation frequency. Pick a certain node and specify the cycle time, the time the signal for that node is high and the time the signal is low. If duty cycle is defined as the clock high time/ clock period; then find the duty cycle of the clock.

**52.** Assume a ring oscillator using 5 CMOS inverters, provides a signal with 45% duty cycle and a period of 2ns. Design the TPHL and TPLH for the inverters such that your assumption remains valid.

**53.** For a 5-inverter ring oscillator, assume that $Tdf = 30ns, Tdr = 50ns$. Calculate the time period T. If at time 0, the input of the first inverter changes from 0 to 1, when will we see the first time output changes from 1 to 0?

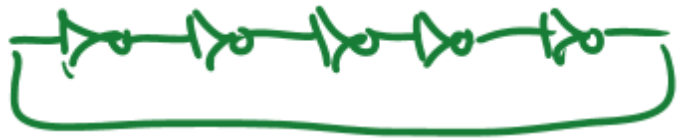

# Chapter 6 – Dynamic Logic

54. Given F=a+b.c.z

   a. Design a circuit that realizes F using a 1-stage domino logic (i.e., a dynamic + a static stage).
   b. Size the dynamic stage of the circuit. You may ignore the internal caps. Assume the reference inverter is sized as W, 2W (for its nMOS and pMOS, respectively.)
   c. To have a more realistic sizing, consider the internal caps and redo the sizing (i.e., progressive sizing) Assume the summation of transistor widths for the circuit should add up to less than 50W.
   d. Find the input pattern that results in worst-case charge sharing. Does this happen during precharge, evaluation or both?
   e. Find the input pattern that gives the best-case scenario concerning charge sharing issue.
   f. Describe the three solutions to resolve the charge sharing problem.
   g. Assume an equivalent capacitance of C to model the internal parasitics of a transistor of size W. Calculate the capacitance values on every internal and output node of the circuit using your progressive sizing of part (c).
   h. Using the cap values of previous parts and the worst-case charge sharing in previous parts, calculate the degraded voltage at the output of the dynamic stage.
   i. Would the degraded output (of the dynamic stage) calculated in part (i) result in Boolean error?
   j. Based on the progressive sizing you calculated in part (c) calculate the Elmore delay of F' for rising output during precharge and falling output during evaluation. To simplify the calculation, you are allowed to round up the W fractions.
   k. Based on the results of j, find the worst case scenario (i.e., the one with worst-case delay) for the falling F'. Also write the input vector that results in that worst case.

55. Design a circuit that implements F=a+b.c.z or complement of F using domino CMOS logic. Use two domino stages (i.e., Dynamic + Static + Dynamic + Static stages.) You have both complementary as well as the original inputs a, b, c, and z if needed for each of the designs.

56. Design a dual rail 2-input XNOR gate using domino gates.

57. Design an XNOR gate using domino logic. Do sizing and avoid charge sharing problem. Aspect ratio = 2.

   58. Implement the logic using domino CMOS logic F = (A+B+C)(D+E)+FGH

What are the problems with this domino? And how can you solve the problems?

**59.** Design a NP-domino logic circuit that implements the following circuit. (Your design should use three gates, one corresponding to each gate in the circuit below.)

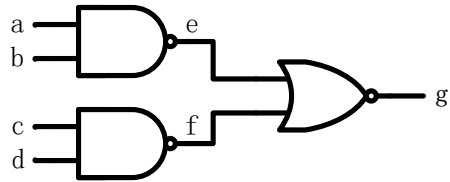

**60.** Design an NP domino logic that implements OUT = AB + CD(W+P)

**61.** Design an NP domino logic that implements OUT = !{ABC+(DWE)'} in 3 stages.

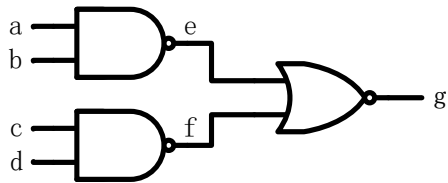

**62.** How would you design a fast S477 given C223?

**63.** Find the maximum frequency. Find the hold.

Delta clock to Q of all FFs is 0.

Assume FFs are designed to have negligible delta D to Clock (i.e., setup time is zero).

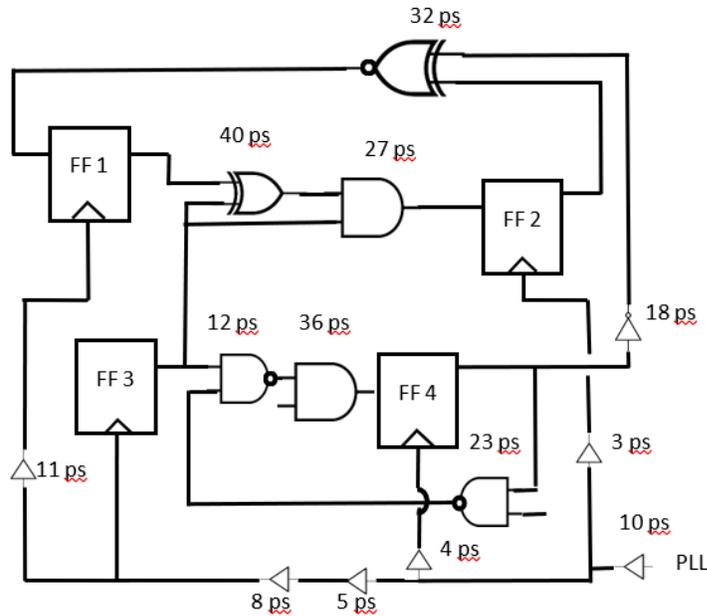

**64.** Given P = ~(A+B'C+W(D+EFG)), find the input setting for the worst case delay. Assume an NMOS transistor with a size of W, has an effective drain or source cap of C, and an equivalent channel resistance of R. Those for the PMOS are reported as C and 4R. Assume right before falling output starts, all the circuits nodes are charged. Similarly assume that all the nodes are discharged right before output rising starts.

**65.** In logic design, complex logic cells (gates) such as AOI and OAI are often used to combine the functions of several cells into a single cell, thus reducing the chip area and parasitic components of the circuit. Consider an OAI452 whose logic function is W = {(A+B+C+D)(E+F+G+H+I)(J+K)}'
   a) Draw a full CMOS circuit diagram for W.
   b) Draw a domino CMOS logic circuit that realizes W or W' whichever more convenient for you to design.
   c) Draw an equivalent circuit for the circuit in part (b) by using equivalent transistor sizes. W/L for each NMOS and PMOS is 10/2 and 40/2 respectively. Assume A, E, H and J are high and the rest of the data inputs are low.
   d) Using the equivalent circuit in part (c) we would like to calculate the propagation delay of the final output during precharge. To simplify the delay calculation, you may add the individual propagation delay values of the each stage in the circuit (i.e., the dynamic part and the inverter.) Assume an internal capacitive load of 100fF at the output of the dynamic stage and output of the inverter both. For all NMOS transistors k'n = 40µA/V2 and γ=0. For the PMOS transistor k'p = 20µA/V2. VDD = 2V and for all transistors |VT0|= 0.4V. Hint: First calculate the propagation delay of the dynamic stage w.r.t. the clock signal (for rising dynamic stage output.) Then calculate the falling propagation delay of the inverter using the step pulse at the input (for falling static stage output.) Finally adjust delay of the static stage considering the non-ideal transition time of the input of the static stage (i.e., the output of the dynamic stage.) Assume transition time is defined based on 20%VDD and 80%VDD crossing points.
   e) Design a dual rail Domino logic for W.
      Note: You have the complementary of the primary inputs available if needed for each design.

**66.** For a given pipeline stage find out if there is setup and hold time violation:
Tcq min =0.5ns, Tcq max 1.3ns, Tskew = ±1.15ns, Tcomb-min = 1ns, Tcomb-max = 5ns, Tclk = 7.5ns, Tsetup = 1.1ns, Thold = 0.9ns, Tdq = 2.1ns

# Chapter 7 – Super Buffer

**67.** For one inverter, Cd=Cg=1fF, Cload=1pF, how many inverters do you need to minimize the delay? (Hint: you can get an approximate value of α)

**68.** An inverter drives a long wire of 9mm long on metal one with width 3λ (λ = 0.125μm) to the input of another inverter. Both these inverters have identical unit size transistors. In order to speed up the circuit, two inverters are inserted in the wire, so each wire segment is 3 mm long. Verify how much speed up this method can achieve based on Elmore delay. Assume delay is equal to RC and omit the series 0.69 here.
Assume the fringing field capacitance of metal one is 50fF/1mm and there is no other capacitance related to metal one. Assume the sheet resistance is 0.025Ω/sq. Assume Cg (n or p) = 100fF. Assume the pull up/pull down resistance of each inverter is 1000Ω. Assume the diffusion capacitance Cd at source or drain is 40fF.

**69.** Assume a wire fringing capacitance to substrate of 23.0aF/um, and wire to substrate capacitance of 8.0aF/um2, an ohms/sq value of 0.08, and a repeater delay of 0.05ps. For a wire with length = 10,000 um and width = 0.4um, how many repeater stations should be added to the wire to reduce the wire delay to a minimum? Use the expression 0.9×Rtot×$C_{tot}$ for estimated wire delay, in this problem. When computing the fringing capacitance contribution, assume the given capacitance value includes both sides of the wire.

# Chapter 8 – SRAM

### 70. Transistors W/L Ratio

Consider the CMOS SRAM cell shown in the following figure. Transistor M1 and M2 have (W/L) values of 4/4. Transistor M3 and M4 have (W/L) values of 2/4. M5 and M6 are to be sized such that the state of the cell can be changed for the bit-line voltage $V_c$=0.5V. Assuming that M5 and M6 are the same size, calculate the required (W/L).
Use the following parameters:
   $V_{DD} = 5V$
   $V_{T0,n} = 0.7V$
   $V_{T0,p} = -0.7V$
   $k'_n = 20\mu A/V^2$
   $k'_p = 10\mu A/V^2$
   $\gamma = 0.4V^{1/2}$
   $|2\emptyset_F| = 0.6V$

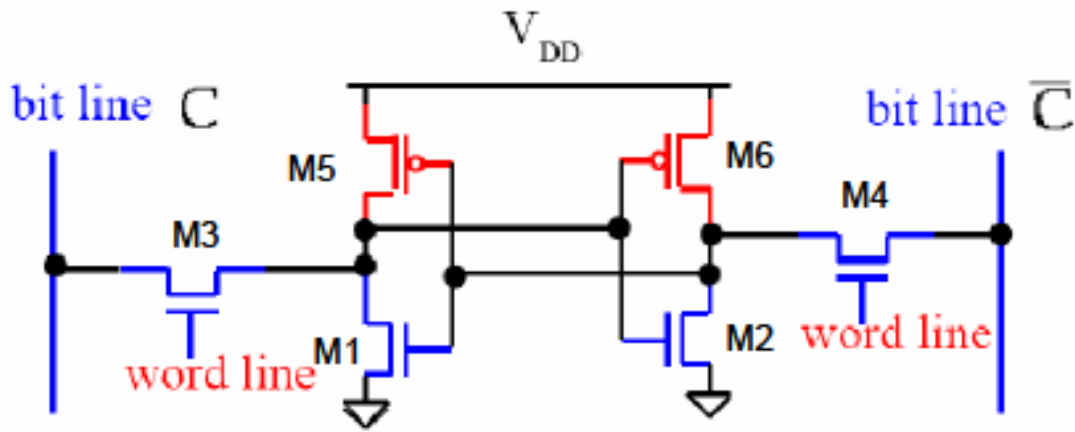

### 71. Decoder
Consider a 10-to-1024 Decoder with inputs named A9 – A0. Assume both original and complementary values for inputs are given, and up to 3-input NAND or NOR are allowed to use.
   (a) Draw the basic Decoding Circuitry with the smallest number of gates. Also calculate total amount of transistors.
   (b) Based on idea of Pre-Decoding, draw the Decoding Circuitry that pre-decodes 5 sets of input pairs [A9, A8], [A7, A6], [A5, A4], [A3, A2], [A1, A0]. It then uses 2 input and 3 inputs gates for the second stage of decoding. Finally it uses 2 input gates. You are also allowed to use inverters between any two stages if necessary. Also calculate total amount of transistors. How many transistors have you saved with Pre-Decoding idea compared to Part a)?

## 72. Voltage Calculation

Considering the SRAM cell circuit in figure below. Assume $V_{TN} = |V_{TP}| = 0.5V$, $V_{dd}=2V$. $k'_n = 60\mu A/V^2$, $k'_p = 30\mu A/V^2$. Ignore body effect and channel length modulation. The sizing of the SRAM cell is as follows: $(W/L)_{M3, M4} = 2$, $(W/L)_{M1, M2} = 4$. $(W/L)_{M5, M6} = 3/2$.
  (1) Calculate the maximum value of $V_1$ during Read 0.
  (2) Calculate the minimum value of $V_1$ during Write 0 when 1 is stored in the cell previously.

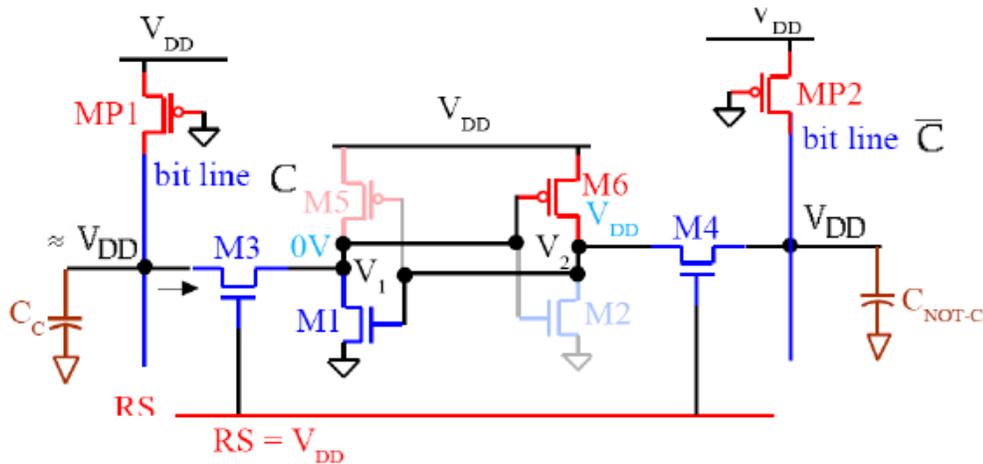

## 73. Capacitance Calculation

Considering a 256 × 256 SRAM, where each 6T cell is 2 μm wide and 1.5 μm tall. All transistors in the SRAM cell has L = 0.1 μm. Also the NMOS access and pull-down transistors' width is 0.25 μm, and the PMOS pull-up transistors' width is 0.12 μm. The M2-layer bit-line has width of 0.2 μm. Other circuit parameters are: Vdd = 1.2V, Cg = Cd = 1fF/ μm. The interconnect has capacitance of Cpp = 0.1fF/ μm2, Cfr = 0.05fF/ μm/edge, and resistance of 0.1 Ω/square.
a) What is the total capacitance load on each bit-line in this memory?
b) What is the worst-case Elmore delay on bit-line during read or write operation using Π-model?

## 74. Transistors Sizing

In Mr Bruin's SRAM design, he sizes 4w, 5w, 6w to M6, M3, M4 respectively. And he finds the read and write operations don't work properly. Assuming the width of the template invertor is 1w for Nmos and 4w for Pmos, what changes would you make to the transistor sizes assuming you are allowed to use 4w, 5w, 6w sizings only.

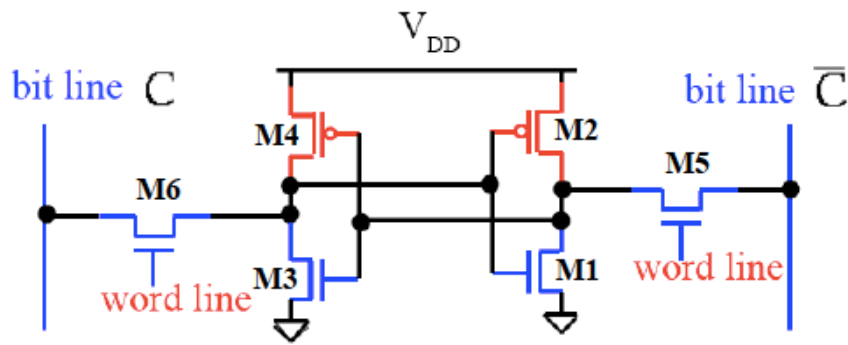

## 75. Sizing Principle
Please help Mr Bruin understand the sizing principle by writing the current relation during a read operation. Vt,n Vt,p are known to you.

## 76. Sense Amplifier
What kind of transistor (PMOS or NMOS) is used as the isolation transistor inside Sense Amp? How about the column mux in a multi-bank SRAM design, assuming the mux is placed before the sense amp or write circuit to save some area?

## 77. Read Operation Delay Calulation
Consider a 1k-bit SRAM storing 16-bit data, meaning basic structure will be 64×16. Assume word line resistance is Rword per memory cell, word line capacitance is Cword per memory cell, bit line resistance is Rbit per memory cell, bit line capacitance is Cbit per memory cell. Use Dgate for single gate delay, and Dmux for delay of one stage of mux, and use only 2-input gate for decoding. Ignoring precharging delay and sense amplifier time, write approximately worst case delay expression for read operation for the basic 1k-bit memory, a 2kit memory consisting of two (32×32) blocks, and a 4k-bit consisting of four (16×64) blocks.

## 78. Voltage of Writing
Assume VTN0 = 0.4V, VTP0 = ‐0.4V, and VDD=1.8V, k'n =mnCox=30mA/V2, k'p=mpCox=15mA/V2. Ignore second order effects such as the body effect and channel length modulation if necessary. Assume the transistors are already sized as follows: (W/L)M4, M3 =4/2, (W/L)M1, M2 =8/2, (W/L)M5, M6 =4/3. Note: The following figure shows the SRAM cell and transistor regions at the beginning of Write 0.
a) Calculate the minimum value of V1 that the above sizing is done for a safe     Write 0.
b) Repeat this for V2 for a safe Write 1, if the initial value of V1 is 0.

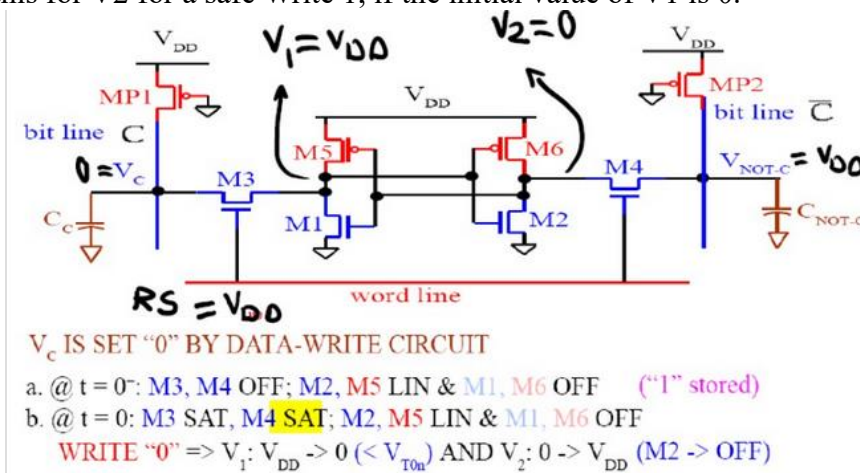

## 79. Voltage of Reading Operation
Consider the following 6-T SRAM cell and the corresponding Word, Bit, and Bit_bar lines. Assume that all the six transistors in the SRAM cell are minimum-sized and assume that it's a 2:1 template, Vdd = 1V, and Vtn = |Vtp| = 0.3V. Also assume that SRAM cell is storing a logic-0, i.e. before it is activated to perform the read operation described next, the voltages at nodes x and y are 0V and Vdd, respectively. Finally, assuming Bit and Bit_bar are precharged to Vdd - Vtn before a rising transition is applied to Word line and ignore body effect, compute the maximum voltage that will be seem at node x during the above mentioned read operation.

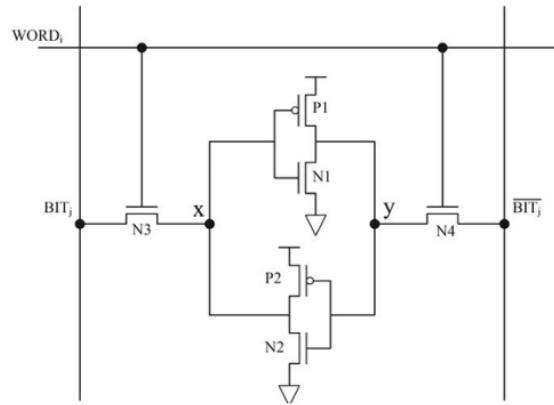

## 80. Resistance Calculation
Consider SRAM cell in your HW1 question 5) with parameter as follow: k'n = 50 μA/V2, VDD = 2.5V, (W/L)2 = (W/L)4 = 6/2, (W/L)1 = (W/L)3 = 3/2, VTN = 0.5V. To make sure VQ lower than VTN during reading 0, calculate smallest Resistance of RL.

## 81. Row Decoder
You are contemplating modifying the classical 6T SRAM cell where you want to use two pMOS transistors in place of the two nMOS transistors whose gates are driven by the word line. To try this idea out, you need another design group in your company to design a suitable row decoder.
Please write a complete set of requirements (specifications) for the row decoder that will ensure that this other design group will be able to design a decoder that works correctly and fast for your modified cell design. Explain the reason for each requirement.

## 82. Memory Architecture
Assume a 512Mbyte Memory system is made of 8 64 Mbyte and is connected to a 64-bit (8 byte) wide data bus. Each chip consists of 4 banks * 16K rows * 1K cols * 8bits (4 x 16k x 1k x8 bits). Note that each column address would refer to a group of 8 bits (i.e. the memory is byte-oriented.) However they are 8 chips each providing a byte, and we should be able to address any certain byte among the 8byte coming from the 8 chips.
Given 32 address bits, show the address breakdown and block level view of the memory.
Show the address breakdown (i.e. which bank, row and column) 0x004f1ad8 refers to.

33
Shahin Nazarian USC All Rights Reserved.

83. Determine the logical effort of all the gate inputs (A, B and C) and the gate parasitic delay. Assume that the drain, gate and source capacitances are the same. When computing the parasitic delay you may ignore the capacitance between stacked NMOS transistors. μn/μp=2.

Hint: the effective resistance can be roughly calculated as $\frac{R_1 R_2}{R_2 - R_1}$, if $R_1$ is connected to GND and $R_2$ is connected to Vdd.

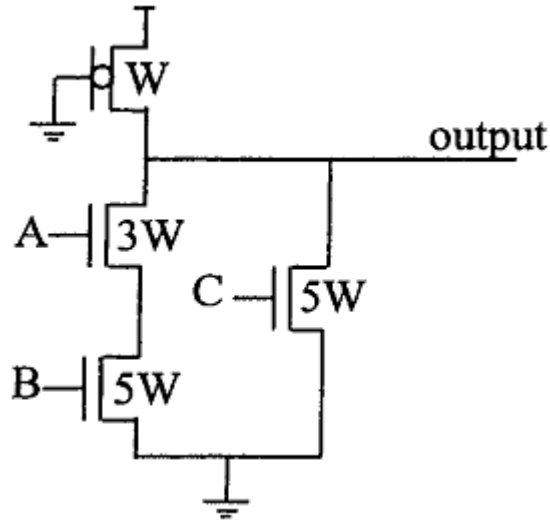

84. Estimate the minimum delay of a 10:1024 decoder driving an electrical effort of H = 20 using static CMOS. Assume the decoder is implemented with 8 stages (NAND3-INV-NAND2-INV-NAND2-INV-INV-INV)
Assume we have a chain of infinite number of inverters with the same size. If un/up=2, Cg/Cd=1, and W is the minimal width for both PMOS and NMOS. Compare the following 2 designs: 1) each inverter is sized that Wp=2W and Wn=W 2) Each inverter is sized that Wp=Wn=W. Which design has a better performance on delay and why?
A chain of 3 inverters is driving a load of 64C, where C is equal to total input capacitance for the unit sized inverter. If the first inverter is unit sized, determine the sizes for the remaining 2 inverters so that the total delay can be minimized.
Exploring the following design alternatives for an eight-input AND gate.
    (a)

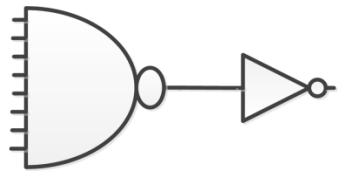

    (b)

(c)

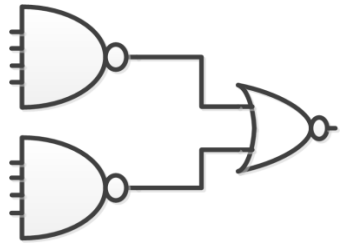

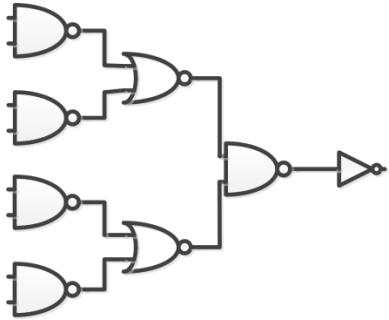

85. Calculate the logical effort of each gate type in (a) ~(c). Assume inverter template with $W_n = 1$ and $W_p = 2$.

    2) Write the path delay expression of each designs namely (a), (b) and (c). Denote the path electrical effort by $H$.

    3) The 8-AND gate is to be used in two different system designs, one with $H = 1$ and the other with $H = 12$. Find the optimal design choice among (a) ~(c) for the two systems.

    The ratio of PMOS to NMOS widths is denoted by µ for equal rise and fall time. Derive logical effort formulas for the following (suppose the template is a balanced inverter, and $\frac{\mu_n}{\mu_p} = \mu$)

    (a) Total logical effort for an n-input NAND gate
    (b) Per input logical effort for an n-input NAND gate
    (c) Total logical effort for an n-input NOR gate
    (d) Per input logical effort for an n-input NOR gate

86. Considering a CMOS compound gate implementing function Y=~((AB+C)D). All the NMOS transistors are minimum sized and $W_p = 2W_n$.
Please specify the logical effort delay of a two-stage circuit where only input A of the first stage changes, and other inputs have constant non-controlling values. Assuming the reference template is a 2/1 sized inverter. $C_{in} = 3C_g$, and $C_L = 20C_g$. Note that there are rising and falling two cases.

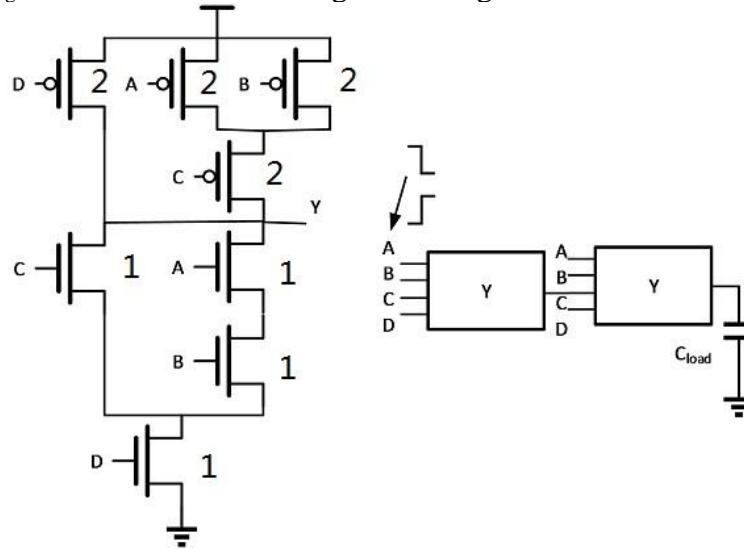

87. Optimize the circuit in Figure 1.9 to obtain the least delay along the path from A to B when the electrical effort of the path is 4.5.

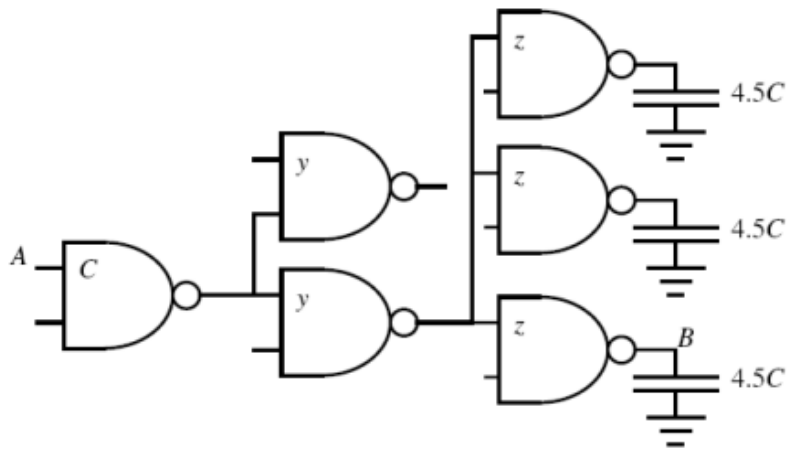

88. Design a CMOS circuit to implement function $f = \overline{(a + b) \cdot c + d}$, assuming mobility ratio as $\mu_n / \mu_p = 2$. Size the transistors so that it has approximately equal rising and falling delays. Then based on the sizing, calculate relevant **g** and **p** for every input. Use the following notation: $C_g$ for gate capacitance with minimum-size transistor, $C_d$ for diffusion capacitance with minimum-size transistor, R for resistance with minimum-size nMOS transistor. Also use Inverter with minimum-size nMOS and twice of minimum-size pMOS as template. Keep all transistors minimum length. (Use worst case resistance for this calculation, ignore internal capacitance?)

89. Assume the reference template is a 4/1 sized inverter. The goal is to design a circuit that implements a 6-input AND gate. Each circuit input should see an input cap Cin of 20Cg. The external load C¬out at the circuit output should be about 1650Cg.

a) (15pts) Consider a 6 input NAND gate followed by an Inverter and calculate its optimal logical effort delay. Note: it is not necessary to calculate the size scaling factors of the two gates.
b) (15pts) Compute the optimal number of stages. If higher than two, add proper number of inverters to the output node of the design of part (a) and recalculate the optimal delay and argue whether it's lower than the optimal delay in part (a).
c) (15pts + 5 extra credit pts) Explore architectures and report the optimal delay of the best circuit you can come up with.
Note: 10 points for any design better than the one in part (b). Maximum of 5 extra credit points for the topmost design(s). Partial extra credit may be assigned depending on the design and calculations.
d) (15pts) In design of part (a) size all the pMOS and nMOS transistors to minimum size (of 1) and calculate the worst case delay assuming each interconnect (between two gates) has a wire cap of 0.5Cg and negligible resistance.

90. In this problem assume that $\mu\_p/\mu\_n=2$ and all gates have equal rise and fall delays. Consider the following circuit: The input capacitance of each input of the NAND4 gate is 10C.
a) You are free to scale each gate, except the 4-input NAND gate. Scale them in a proper way to get the minimum worst-case delay. How much is the delay in this case?
b) Now you are free to add as many as buffers that you want. How many buffers do you need to make this circuit work faster? How much is the new delay after adding buffers to this circuit?

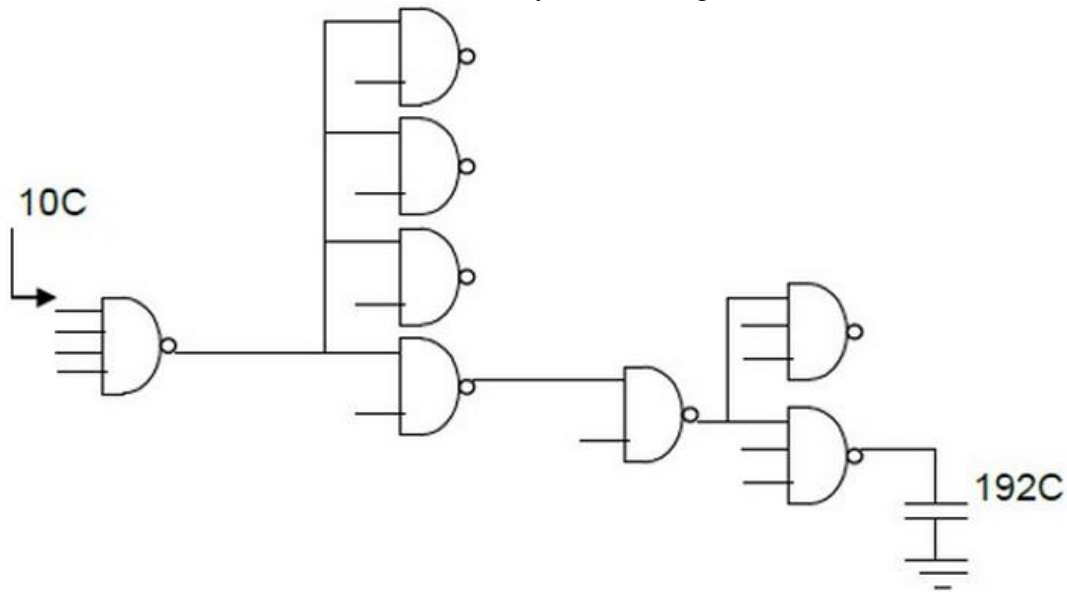

91. Consider two different implementations of an 8-bit decoder with H=9.
Design 1: Pre-decoder: NAND4 + INV, row decoder: NAND2 + INV
Design 2: Pre-decoder: NAND2 + INV, row decoder: NAND4 + INV
Assume the reference 2/1 sized inverter's FO4 delay is 100ps.
   a. Calculate the path effort of both designs, with all gates symmetrically sized.
   b. Can you estimate the propagation delay of each of the design when the gates are sized to achieve minimum delay?

c. Now assume that you would like to minimize the propagation delay only for rising transition at the output, i.e., rising delay. Each gate can be sized to have the rising delay and falling delay differ by a factor of 2. Size the Design 2 to minimize rising delay and get the approximation delay result.

92. A clock buffer use a 1-2 fork that has maximum input capacitance of 100fF. Both true and complementary outputs must drive loads of 300fF. Compute the input capacitance of each inverter to minimize the worse-case delay from input to either output. Estimate the worse-case delay using logical effort. Assume the reference is 2/1 sized inverter.

Assume we have a chain of infinite number of inverters with the same size. If un/up=2, Cg/Cd=1, and W is the minimal width for both PMOS and NMOS. Compare the following 2 designs:

1) Each inverter is sized that Wp=2W and Wn=W
2) Each inverter is sized that Wp=Wn=W

Which design has a better performance on delay and why?

93. Consider an (m + 1, m) fork with input A and outputs B and C. Given that the total input capacitance at input A must be no more than $20C_g$ and that each of the two outputs of the fork must drive an external load capacitance of $1000C_g$, design a fork that uses optimal number of optimally-scaled versions of the inverter template (nMOS and pMOS transistors have widths of 1 and 4) to minimize the worst-case delay for the fork. Clearly show the size of every transistor in your design.

# Chapter 10 – Power Optimization

### 94. Dynamic Power

Consider a 3-input static logic gate implementing Boolean function, F = AB+C. The gate is driving a *100 fF* total load at *500 MHz* and *VDD = 1.2 V*.

a) First, assume that these inputs are uncorrelated. The input signal probabilities are given as: *p(A) =p(B) = 0.2; p(C) = 0.667*. What is the average power dissipation at the output of this gate?
b) Next, assume that A is dependent on B and C: 25% of the total time A = B and the rest of the time A = C'. *p(B) = 0.2; p(C) = 0.667*. Calculate the average power dissipation.

### 95. Power Supply Scaling

In recent years, as the designers of the state-of-the-art circuits put more effort in minimizing the energy consumption, the ultra-low voltage operation, in particular, operating VLSI circuits in the near-threshold (NT) regime, has gained increasing attention due to the significant energy efficiency enhancement that comes by operating this regime.

a) If the threshold voltage is 0.2V and we reduce the supply voltage from 1.0V to 0.5V, calculate the power consumption reduction for both switching power and short circuit power. Assume all other parameters are the same.
b) What is the main disadvantage of operating in near-threshold regime?

### 96. Leakage Power

When the supply voltage is very low, we need to pay more attention to the leakage power consumption. Explain the reason.

### 97. Dynamic Power

Consider a 4-input static logic gate implementing Boolean function, z = A(B+C)+BCD. The input signal probabilities are given as: $p(A) =p(D) = 0.25; p(B) = 0.33; p(C) = 0.5$.

a) Assuming that the inputs are uncorrelated and that all signals are temporally independent, what is the average power dissipation at the output of this gate if it is driving a 100 fF total load at 500 MHz and VDD = 1.2V?
b) Next, assuming kn=kp=200uA/V2, Vtn=|Vtp|=0.3V, $\tau$ in=100ps and $\tau$ out=250ps, calculate short circuit power dissipation.

### 98. Adiabatic Logic

Consider a dual-rail adiabatic logic gate realizing outputs AB+C and !(AB+C). Assume that the load capacitance at each output of this gate is 200fF, Vc,max=2V, the effective ON resistance of each conducting transmission gate is 5k$\Omega$, and the rising/falling transition time for the clocked power line is 100ns.

a) Show the schematic circuit diagram of this adiabatic gate.
b) Calculate the energy consumption for the following scenario. When the clocked power line is low, the input changes from A=1 and B=C=0 to A=B=C=1; As the clocked power line ramps up, the output changes. Only account for energy dissipation due to transition at the output nodes and ignore all internal capacitances.

*Hint: You must first determine if one or both outputs change in order to do the correct calculation.*

### 99. Bus Splitting

Consider splitting a bus that connects $N$ modules A1, A2, ⋯ AN into $m$ buses, each connecting a group of $N/m$ modules. Each bus, Bi, is also connected to its two neighbor buses, i.e., Bi-1 and Bi+1 through a transfer bus with a controllable switch. The capacitance of the original single bus is *CBUS* while that of each of the smaller buses is *CBUS/m* and the capacitance of a transfer bus is *mCBUS/N*.

Suppose that we can assign modules to smaller buses such that 80% of the data communication occurs inside the group and does not need to use transfer buses. The remaining 20% of the data communication originated within a group goes to adjacent groups i.e., no communication goes across multiple transfer busses.

a) Calculate the percentage of power saving of the split bus architecture in terms of $N$ and $m$.
b) Given N, find the optimal value of $m$ that maximizes the power saving using the aforesaid bus splitting architecture.

#### 100. Switching Power
Considering a circuit implementing F = A'BD + AC'D + BC'D, with probabilities of each input as p(A) = 0.3, p(B) = 0.4, p(C) = 0.5, and p(D) = 0.6. Assume all inputs are uncorrelated, and the circuit is fed by Vdd of 2V, running at 1GHz clock, and driving load of 200fF, calculate the switching power consumption.

#### 101. Glitch Reduction
Draw the circuit above (F = A'BD + AC'D + BC'D) with exactly the same logic as expression (three 3-input NAND followed by one 3-input NAND). Study the solution to this problem that lists a glitch situation that could happen. Also study the modification that is done to avoid that glitch.

#### 102. Bus Encoding
Based on what you have learned from Bus Encoding, for a driver that outputs a 3-bit counting result (000, 001, 010, etc.), transitions on bit lines may be too many during some situation. One good and common encoding algorithm to reduce power consumption due to transitions is using Gray Code. Provide a simple encoding algorithm to convert the 3-bit binary code to gray code, and calculate how many transitions can you save after applying the algorithm during a complete counting-up sequence (000→001→010→…→111). Give a summary about why Gray Code is good for reducing transitions in representing continuous binary values.

#### 103. Switching Power
Consider driving a fixed load of 8Cg where Cg is the input capacitance of a minimum-sized equal rise/fall time inverter with either a 3-input CMOS NOR gate or a 3-input CMOS NAND gate. Assume that three input signals to the NAND or NOR gate are spatially and temporally independent and have a signal probability of 0.5. Furthermore, assume that transistors in NAND and NOR gates have been sized so that each gate has a worst-case rise/fall propagation delay equal to that of an equal rise/fall time, minimum-sized inverter. Ignore all diffusion and wire capacitances and the body effect. Report the ratio of the power dissipation of the 3-input NOR gate to that of the 3-input NAND gate. Hint: A minimum-sized CMOS inverter denotes a CMOS inverter with (W/L)p=(W/L)n=1:1. We assume μn/μp=2, so an equal rise/fall time, minimum-sized CMOS inverter is one with (W/L)p=2:1 and (W/L)n=1:1. Make sure to calculate the power dissipation on both input pins and the output pin.

#### 104. Switching Power
A 180 nm standard cell process can have an average switching capacitance of 150 pF/mm2. You are synthesizing a chip composed of random logic with 0.2 as the average switching activity factor, where switching activity refers to the number of transitions per cycle. Estimate the power consumption of your chip if it has an area of 70 mm2 and runs at 450 MHz at VDD = 0.9 V.

#### 105. Switching Power

We know that the inverter becomes a buffer when switching the position of Nmos and Pmos. Assume Vt,n=|Vt,p|=Vt, please give the average switching power dissipation of this buffer in one clock period. Hint: refer to page 5 in Power unit slide.

**106. Leakage Power**

What will happen to the three leakage current when we decrease the supply voltage and the transistor width?

**107. Dynamic Power**

In order to implement OR4 logic to drive a 10fF on-chip capacitance, we end up with two options,
   (A) NOR4(1X)+ INV(5X)
   (B) NOR2(2X) + NAND2(?X)

Assume the reference template is a 2/1 sized inverter, and the gate and diffusion capacitance per unit width is C = 1.25fF/µm.
   a) Calculate the size ratio of NAND2 in option (B) to make the two options achieve the same delay. Assume the minimum CMOS transistor width $W_{min}$=80nm.
   b) Which option saves more dynamic power? and why? Internal capacitances inside the gates are ignored.

108. **Dynamic Power**

Given a function expression f(A,B,C) = (A + BC + B'C')', we assume that all inputs toggle with a probability $p_1 = 0.5$.
   a) Derive the activity factor of the output $\alpha_{0->1}$
   b) Sketch one implementation of this logic using basic gates (INV, NOR, NAND, XOR) and estimate the dynamic power consumption of this circuit EXCLUDING the power consumed at primary inputs. Assume $V_{dd}$=1V, f=1GHz, the capacitance at output of each gate is 50fF. Internal capacitances inside the gates are ignored.

109. **Leakage Power**

Suppose that the leakage current of a single transistor is:
$I_{leak} = I_0 10^{(V_{GS}-V_{th}+\lambda_d V_{DS})/S}$ for NMOS transistor,
and $I_{leak} = I_0 10^{(V_{SG}-|V_{th}|+\lambda_d V_{SD})/S}$ for PMOS transistor.
Where $\lambda_d = 0.1$ and the sub-threshold swing S = 100mV/decade. The supply voltage $V_{dd}$ = 1V. The implementations of NOR2 and INV gates are shown on the right.

   a) Calculate the voltage value $V_x$ at node x, when $V_A = V_B = 1$ in the NOR2 gate.
   b) Compare the leakage current of NOR2 gate and INV gate, when $V_A = 1$ in the INV gate.

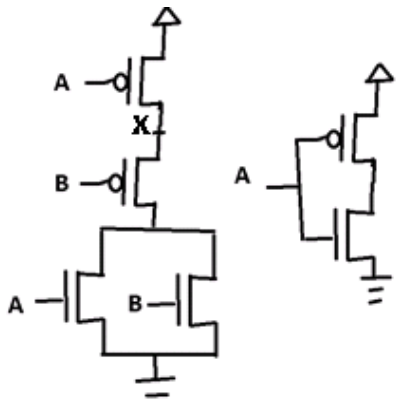

# Chapter 11 – Multi-Gate

**110.** Explain why FinFET devices have attracted a lot of attention as an alternative to the bulk CMOS when technology scales beyond the 32nm technology node.

**111.** FinFET structure allows for fabrication of separate front and back gates. Explain the two different connection modes for an N-type or P-type fin.

**112.** Based on the two connection modes, draw the structure of three kinds of FinFET-based 2-input NAND gate. Note that for multiple-input logic cells (e.g., NAND, NOR), there is another IG mode connection where the front gate and back gate are driven by different input signals.

**113.** The figure below is a FinFET layout. Specify which logic gate it is. Specify the number of fins for this gate.

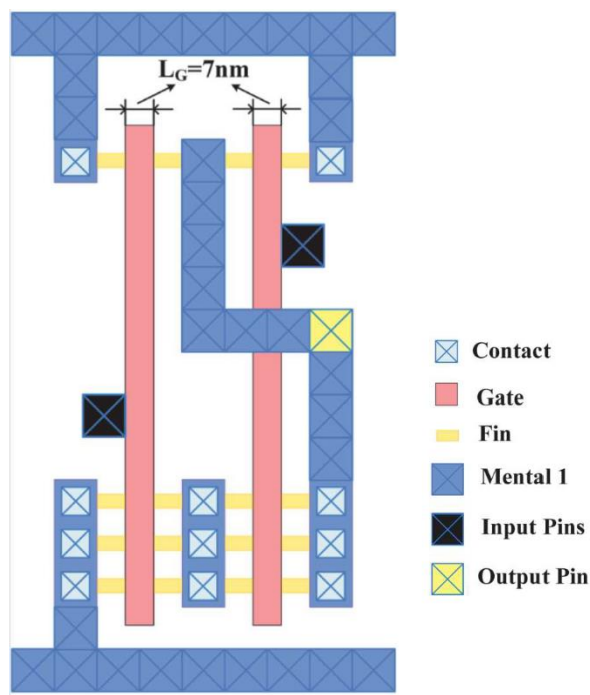

# Chapter 12 – Logic Effort and Data Path

**114.** Design a CMOS compound gate to implement F=[AB(C+D)]' such that the worst case delay is equal to a 3/1 reference inverter. Then use the 3/1 inverter as the reference template and find all the logical effort (g) and parasitic delay (p) values for the compound gate. You should consider all the inputs and both rising and falling condition.

**115.** Assume N 1-bit full-adder blocks of Unit 8 page 5 are used to create an N-bit RCA. The goal is to minimize the delay of the critical path. However we should try to reduce the area as well. Therefore specify all the transistors that should be minimal sized.

**116.** The 1-bit Full-adder schematic is shown below. If you want to design an N-bit carry ripple adder where N is a very large number, size each transistor so that the critical path delay can be minimized. Assume that we use the same a-bit Full-adder in each stage. The minimal size of transistor is 1 and the maximal size of transistor is 16. For simplicity, you don't need to use progressive sizing. Use logical effort to calculate the delay.

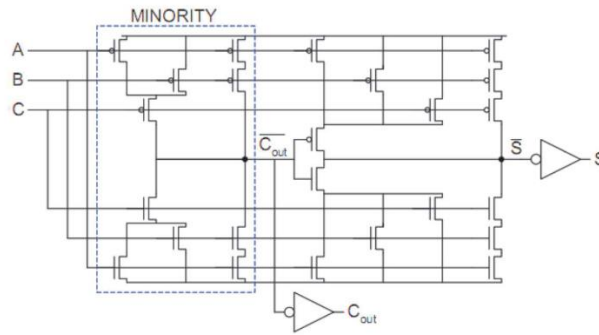

**117.** Now we use inverting Full-adder to eliminate the inverters in the critical path. The new design is as follows. Prove that logically we can still use identical inverted Full-adder for each stage.

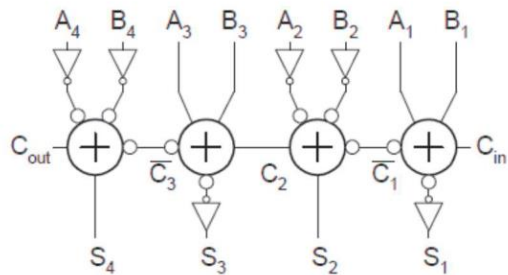

**118.** Consider the path from A to B involving three two-input NAND gates shown in Figure 1.7. The input capacitance of the first gate is C, and the load capacitance is also C. What is the least delay of this path, and how should the transistors be sized to achieve least delay? Use un/up = 2. The NAND2 has the same effective resistance as a reference size inverter.

**119.** Using the same network as in the previous example, find the least delay achievable along the path from A to B when the output capacitance is 8C.

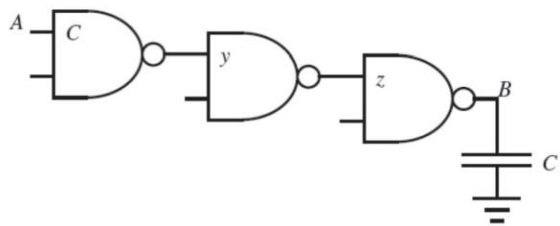

(8 points) In this problem, assume that the ratio of mobility of electrons to the mobility of holes has a value equal to 2, i.e., assume that $\mu = \frac{\mu_n}{\mu_p} = 2$.

a. Design a template for a single complementary CMOS gate to implement the function $f = \overline{ab + bc + ac}$. Your template must use the minimum number of transistors.
b. Size the transistors such that the above template's worst-case rise delay is (approximately) equal to its worst-case fall delay.
c. Compute all relevant $g$ and $p$ values for the above template using the inverter described on Page-1 of this exam as the reference template.

**120.**

**121.** Ladner-Fischer. Identify critical path of the adder. Using a 2/1 sized inverter as the reference template. Assume that each gate is 1X, diffusion and gate capacitances are equal. Cin = 3Cg, Cload = 10Cg. Calculate logical effort for critical path of the adder.

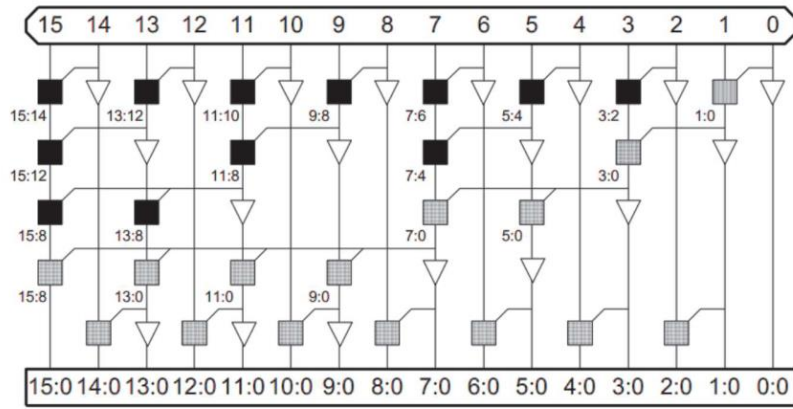

**122.** Assume that diffusion and gate capacitances are equal, using a 3/1 sized inverter as the reference template. The goal is to implement an 10-input NAND gate. Each input of the gate should see an input capacitance Cin of 10 Cg. The output load for this gate is 5000Cg. Explore architectures and report the optimal delay of the best circuit you can come up with including the schematic of your design and the size of each gate.

**123.** In this problem, assume that $\mu_n/\mu_p=2$. Compute the logical and parasitic efforts of the following gates using 2 inputs NAND as the reference template. Assume diffusion and gate capacitances are equal.

**124.** Fork Design a fork generating A and A' from primary input A, assuming capacitance in primary is 5Cg, and each branch is driving load of 800Cg. Find the number of stages of 2:1 Inverters that lead to minimum delay, calculate worst case delay using Logical Effort, and find input capacitance of each gate. Use 2:1 Inverter as template, and $\mu = 2$.

**125.** In the following three problems, assume that the ratio of mobility of electrons to the mobility of holes has a value equal to 4. Design a CMOS inverter and assign size to each transistor such that the gate's worst-case rise delay is (approximately) equal to its worst-case fall delay.

**126.**

i. Build a 16-input AND logic with Structure of NAND4-NOR4 structure, assuming primary input cap is Cg, $\mu = 2$, and load cap is 3Cg. Calculate worst case delay using logical effort.

ii. If are allowed to add Inverters following the NOR4, how many Inverters you will add to reach minimum delay? Calculate worst case delay using Logical Effort and find input cap for each gate.

iii. Repeat part 2) with load cap being 300Cg.

**127.** Design a minimum-delay fork where the total capacitance at the fork input is 10Cg and each branch drives an external load of 2000Cg. Compute the optimal scaling for each gate and compute the worst-case delay for your final design.

---

**Solution.**

A fork produces both the true and the complemented versions of its input driving the same load. The two branches must therefore have different parities (one has $N_1$ inverters, the other has $N_2 = N_1 \pm 1$), but they share the input capacitance:

$$C_{in,1} + C_{in,2} = 10\,C_g, \qquad C_L = 2000\,C_g \text{ (each branch)}.$$

For an inverter chain of $N$ stages with first-stage size $a$ (in units of $C_g$), the stage effort is

$$f = \left(\frac{C_L}{a}\right)^{1/N}, \qquad D = N f \;\;[\text{+ } Np].$$

**Choice of $N_1, N_2$.** The effort per branch is $H = 2000/a \approx 2000/5 = 400$, so the optimum number of stages is $\ln H \approx 5.3$. Since the two branches must have opposite parity, use

$$N_1 = 5 \;\;(\text{odd}), \qquad N_2 = 6 \;\;(\text{even}).$$

**Balancing the two branch delays.** Equate stage-by-stage delay: $N_1 f_1 = N_2 f_2 \;\Rightarrow\; 5 f_1 = 6 f_2 \;\Rightarrow\; f_1 = 1.2\,f_2$.

With $a_1 = 2000/f_1^{5}$ and $a_2 = 2000/f_2^{6}$ and $a_1 + a_2 = 10$, solving gives

$$f_2 \approx 2.74, \qquad f_1 \approx 3.29,$$
$$a_1 \approx 5.27\,C_g, \qquad a_2 \approx 4.73\,C_g.$$

**Gate sizes.**

Branch 1 (5 inverters, drives 2000 $C_g$):
$$5.27,\; 17.3,\; 57.0,\; 187.4,\; 616 \;\;\to\; 2000\,C_g$$

Branch 2 (6 inverters, drives 2000 $C_g$):
$$4.73,\; 12.96,\; 35.5,\; 97.3,\; 266.7,\; 730.5 \;\;\to\; 2000\,C_g$$

**Worst-case delay.**

$$D_1 = 5 f_1 = 5(3.29) \approx 16.44\,\tau$$
$$D_2 = 6 f_2 = 6(2.74) \approx 16.44\,\tau$$

Including the intrinsic (parasitic) delay $p_{inv}=1$ per inverter,

$$D_{worst} = N f + N p \approx 16.44 + 6 \approx 22.4\,\tau.$$

Both branches are balanced, so the fork delay $\approx 16.4\,\tau$ (excluding parasitics) / $\approx 22.4\,\tau$ (including parasitics).

# Chapter 13 – Flip-Flop

a) **(10 Pts)** In the NAND Bistable DFF design discussed in class, D changes from the old value 1 to 0. Show waveforms of important signals (CP, D, S, R, Q, Q' and the output of gate 4) during this change, assuming no setup/hold time violation. You need to show the delay of gates in your waveform.
b) **(10 Pts)** In the NAND Bistable DFF design, NAND gates 1 to 6 have delay $t_1$ to $t_6$. Please give the setup/hold time expression using $t_1$ to $t_6$, assuming no wire delay. Based on your analysis, what should we do to decrease the setup time/hold time of the DFF?
c) **(10 Pts)** Design a NOR Bistable DFF. You may use the idea of NAND Bistable DFF and modify it to a NOR-based.

**128.**

**129.**

a) Highlight any differences:
    I. Between implicit P-FF and explicit P-FF.
    II. Between Explicit-Pu ls ed FF (EPFF) and Static Explicit-Pulsed FF (SEPFF) .
b) Explain in detail the functionality of Hybrid Latch FF (HLFF).

**130.**

a) Explain why the Conditional Capture FF (CCFF) uses less power than the HLFF.
b) How does Conditional Discharge FF (DET-CDFF) realize conditional discharge?

**131.**

a) Think of any power optimization techniques for FFs and explain them in detail. Also add any example FF that you know of that uses that idea. List any two FF structures featured with low power

b) After the chip is fabricated and handed over to your hand , explain what we could do if it has setup time violations. What about hold time violations?

c) Considering process variation, in a chain of DFFs, explain under what conditions the soft-edge FF will help to make the circuit robust.

d) Why people are more conservative about hold time in corner analysis ?

e) Claim: For an N-stage soft-edge-FF based design, an N-stage setup check is all we need (to detect any setup violation ). Argue whether or not the claim is valid.

f) Claim: For an N-stage soft-edge-FF based design, I-stage setup checks are all we need. Argue whether or not the claim is valid.

# Chapter 14 – Time Borrowing

You are given the following latch-based pipeline with four combinational logic blocks $CLB_1$, $CLB_2$, $CLB_3$, and $CLB_4$, with worst-case delays $\Delta_1$, $\Delta_2$, $\Delta_3$, and $\Delta_4$, respectively.

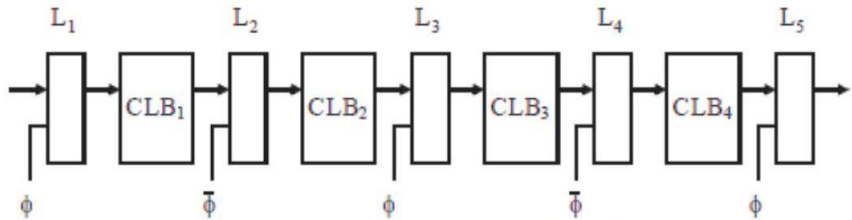

Each latch has the following parameters: $\Delta_{CQ}^L$, $\Delta_{DQ}^L$, $\Delta_{DC}^L$, and $\Delta_{CD}^L$, where the symbols have their usual meanings (with the super-script 'L' added to denote a latch). You are given a clock with period $T$, 40% duty cycle, and zero skew. You are also given the above clock's complementary version.

Write a complete set of inequalities that must be satisfied to guarantee that the above pipeline will not suffer due to maximum delay problems (assuming that external time borrowing is allowed).

**132.**

**133.** You are given the following latch-based sequential design with two combinational logic blocks CLBl, CLB2. Each with a worst-case delay D. Assume each latch has zero setup time, zero hold time and zero clock-to-Q delay. You are given a clock with 50% duty cycle. What is the minimal possible clock period of this design?

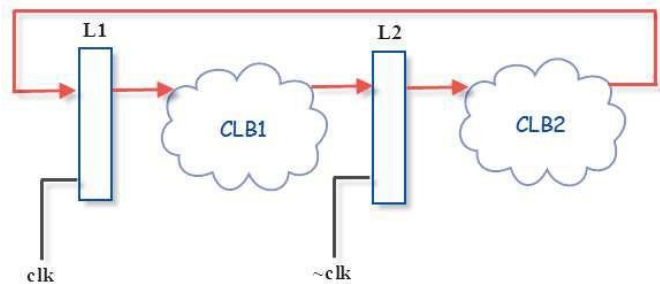

# Chapter 15 – Domino and dual rail dynamic logic

**134.** How many domino stages are required to implement the logic using domino CMOS logic F = (A+B+C)(D+E)+FGH? What are the problems of using domino? And how can you solve the problems?

**135.**

a. Given F=a+b.c.z:
i. sign a circuit that realizes F using a 1-stage domino logic (i.e., a dynamic + a static stage).
ii. Design a circuit that implements F=a+b.c.z or complement of F using domino CMOS logic. Use two domino stages (i.e., Dynamic + Static + Dynamic + Static stages.) You have both complementary and un-complementary inputs a, b, c, and z if needed for each of the designs.
b. Design a domino circuit that implements F=ab+cd.
c. Draw a two-stage domino logic (i.e., D + S + D + S stages) for F= D*(AB+C). Use (AB+C) as first stage and D as the second stage. Which transistor(s) will cause charge sharing for stage 1? Write down the possible input combinations that result in charge sharing, however if the transistor with input A would cause the problem. For example, you can write ABC=000->111. Draw a circuit to prevent charge sharing problem for stage 1.
d. We would like to implement F= ~(AB+C). Draw the domino logic, assuming the complements of A,B and C are available. (Hint: use De Morgan's law)
e. Draw a one-stage dual rail structure to implement F= (AB+C).
f. Design an XNOR gate using domino logic that does not have a charge sharing problem.
g. Design a dual rail 2-input XNOR gate using domino gates.
h. Design a NP-domino logic circuit that implements the following circuit. Your design should use three gates, one corresponding to each gate in the circuit below

**136.** In logic design, complex logic cells (gates) such as AOI and OAI are often used to combine the functions of several cells into a single cell, thus reducing the chip area and parasitic components of the circuit. Consider an OAI452 whose logic function is W = {(A+B+C+D)(E+F+G+H+I)(J+K)}'

a. Draw a full CMOS circuit diagram for W.

b. Draw a domino CMOS logic circuit that realizes W or W' whichever more convenient for you to design.

c. Design a dual rail Domino logic for W.

NOTE: You have both complementary and un-complementary inputs if needed for each design

**137.** Consider the following two domino buffer designs

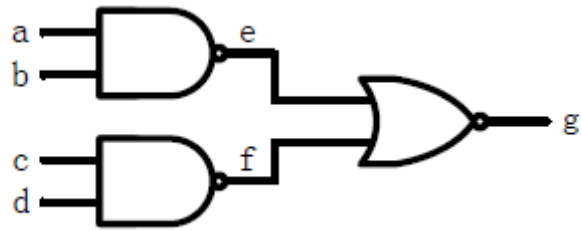

a) In the first design, do the two dynamic gates pre-charge at the same time (i.e., in parallel)?
b) Will the two dynamic gates in the second design pre-charge properly? If so, will the two dynamic gates pre-charge in parallel or sequential?
c) Which of the two buffer designs, when size optimally lead to lower evaluate delay for the same value of H? Explain the reasons behind your answer.

**138.** For the circuit below, specify functions of x and y in terms of A, B, and C.

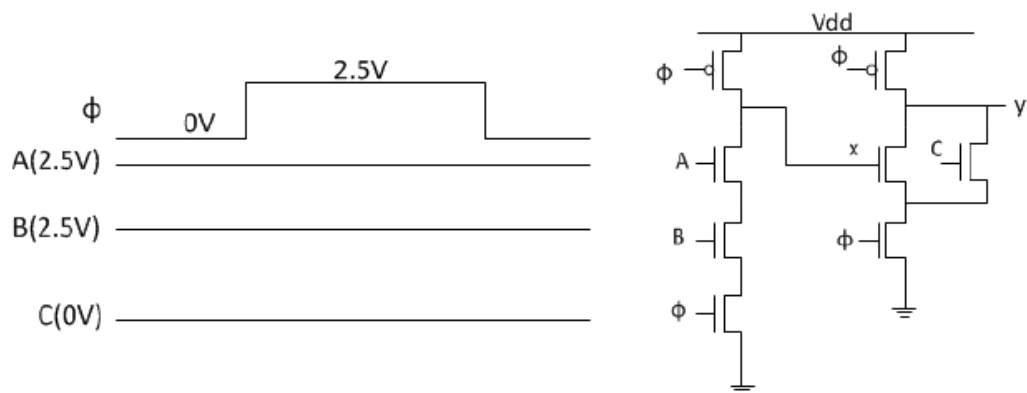

a) Sketch the waveforms at x and y for the given input's waveforms. Do x and y evaluate to the values that you expected from their Boolean functions?
b) Redesign the gates to eliminate any race conditions. Sketch the waveforms at x and y for the new gates.

Important: you may NOT and any additional gates (including a static CMOS inverter); You do however have access to both ∅ and ~∅ and can change the type or design of either gate.

**139.** Design a foot-less domino circuit to implement. F=AB+CD. Each input has capacitance corresponding to 30 $\lambda$ of transistor width. The output must drive a load equivalent to 500 $\lambda$ of transistor width. The domino inverter is HI-skewed such that Wp = 4Wn, Size the transistors to achieve the MINIMUM worst-case evaluation delay, and estimate this delay using logical effort. Assume the reference is a 2/1 sized inverter and, in this technology, the minimum width of a transistor is 10 $\lambda$.

**140.** Draw circuit for dual-rail 4-input NAND using Dynamic Logic, assuming complementary values are also available.

**141.** Draw circuit for dual-rail implementing F= ABC+DE using Dynamic Logic, assuming complementary values are also available. Show cases that lead to charging sharing that result in voltage drop at the output node of Dynamic Part of the circuit. Use Cp to represent any parasitic capacitance for one transistor.

# Chapter 16 – Scripting

**142.** Print an array in reversed case-insensitive order.

**143.** Concatenate 5 strings and put the result into a new string.

**144.** Write a program named "1.pl" or "1.py". After typing the command "perl 1.pl number1 number2 number3" or "python 1.py number1 number2 number3", the output should print the largest one of the 3 numbers to the screen.

**145.**

a) What is the importance of Perl/Python warnings? How do you turn them on?

b) Explain at least 2 different types of data Perl/Python can handle.

c) Given a file named input.txt, count the word "memory" occurrence (case insensitive).

d) Write a subroutine that takes three arguments, adds them together, and returns the result.

# Chapter 17 – Miscellaneous

146. Soukup's Algorithm
   a) Based on your experience of Soukup's implementation, assume three files for a design: SRAM.v, SRouter.v, and tb.v.
   b) How would you specify the address of source and target?
   c) How would you enhance the Soukup's code to handle multi-layer (3D) IC routing problem?
   d) After synthesizing the design using a 2ns clock period, there is a -0.5ns slack. Please propose several ways to improve the timing performance.
   e) After the design is fixed to satisfy the timing constraint, assume you'll find out that the area is too large to fit in the available frame that is provided by the foundry. Please propose several ways to address this issue.

**147) DCNN design**
   a) What is the major benefit to implement DCNN in hardware?
   b) Now we find that for a 1024-input neuron, the adder becomes very slow. One idea is to divide the large adder into 2 stages and make it a pipelined design. How to achieve this in Verilog?
   c) If we want to change the activation function to tanh: f(x)=tanh(x), how to write a Verilog code for this function? What would the hardware be after synthesis?

**148) APR**
   a) What are the libraries that you need to import to the encounter? Please briefly describe their functionalities.
   b) What is placement? And what is routing? Which one is more difficult?
   c) Placement and routing for ASIC and FPGA are different. Please briefly explain the difference. Can encounter be used for APR in an FPGA board, e.g., Xilinx?
   d) Why clock cycle sometimes need to be increased to have correct function after APR?

**149) DC, Synthesis**
   a) What is synthesis? What is the basic synthesis flow using design compiler?
   b) What's the difference between target and link library?
   c) What design constraints should be applied when synthesis a design using DC? And what's the purpose of these constraints?
   d) How to solve the timing issue in the design?

**150) STA**
   a) How to solve the hold violation in the design after ARP?

b) What is STA? What's the advantage and disadvantage of STA?
c) What is false path? What is Multi-cycle path? What is virtual clock? What is timing arc?

### 151) CAM and FIFO
a) What is CAM? What is the difference between TCAM and Binary CAM? Describe how TCAM works.
b) Could you implement a synthesizable TCAM using Verilog? How?
c) What is metastability? How to handle crossing clock domain issue?

### 152) Misc.
a) (5 pts) What logic (hardware) you think is intended as the outcome of the synthesis tool given the following Verilog module? Write any syntax/synthesis issue(s) you see in the following design.
```
module check(in, c1, c2, out);
  input in,1, c2;
  output out;
  reg out;

  always@(in, c1, c2)
  begin
    out<=0;
    if(c1 & c2) out<=in;
  end
endmodule
```

b) (5 pts) Among the following FSM design styles using always blocks which ones will be incorrect (the design will not work)? Explain why? Note, we do not worry about design efficiency here. All we worry about is to have a design with correct functionality.
　　i. One block for S.M., another block for the combined O.F.L. and N.S.L.
　　ii. One block for combined S.M. and N.S.L. and another for O.F.L.
　　iii. One block for combined S.M. and O.F.L., and another for the N.S.L.
　　iv. One block for combined S.M., N.S.L. and O.F.L.
　　v. One separate block for each of S.M., N.S.L. and O.F.L. (3 total)

c) (3 pts) Discuss one of the solutions to collision problem in Hash.
d) (3 pts) What is the complexity of finding a linked-list item and then erasing it.
e) (3 pts) What is the difference between Espresso vs Latte?
f) (5 pts) Explain why many of the VLSI design slow flow steps are implemented using heuristics rather than algorithms.

### 153) Timing

a) (3 pts) When exactly should the clock enable be high during initialization of the project design? At the beginning, in the middle or at the end?

b) (4 pts) Timing analysis of a synthesized netlist shows that the critical path has an overall delay of 3.74ns, and the setup time plus clock-to-Q delay of a FF in the library is 100ps. We need to pipeline the circuit so it can operate with a clock frequency of 2GHz. How many pipeline stages are required.

c) (10 pts) Set up an equation that can be used to find the balancing point for two clock subnetworks that are connected by a 0.04mm long 0.001mm wide interconnect wire cap per unit area of $0.5fF/(\mu m)^2$. Assume the two subnetworks have almost identical delays and they will load the clock network by as much as 20fF and 15fF, respectively.

d) (5 pts) Clock __________ is very likely to have negligible effect on __________. Choose from the following and explain the reason.
Skew/setup        skew/hold     jitter/setup     jitter/hold

e) (15 pts) Differentiate between clock jitter and clock skew. What is timing uncertainty? How would use measure the timing uncertainty caused by skew vs jitter.

f) ( 10 pts) Explain Clock De-skewing with an example.

## 154) DDR

a) (3 pts) What would you check to see whether the Denali DDR2 memory is initialized successfully?

b) (6 pts) What is the meaning of each of the DDR2 controller I/O signals in the following list?
- a. SZ:
- b. NOTFULL:
- c. VALIDOUT:
- d. READY:
- e. FETCHING:
- f. DQS:

c) (1 pt) If we set {csbar, rasbar, casbar, webar} = 4'b0111, which command will be operated?

d) (1 pt) In the project, 'SSTL18DDR2DIFF.v' and 'SSTL18DDR2.v' files are given. Choose the folder name(s) where you put in these file. ('design', 'tb', 'work', 'sim', 'script')

e) (2 pts) What part of the memory architecture the DDR2 family has targeted the most to be able to increase the throughput? Why?

f) (4 pts) The number of CAS latency cycles in a DDR3 SDRAM can be lower than that in a DDR SDRAM. Argue whether or not this is true.

g) (3 pts) Briefly explain the relation between posted CAS and AL (Additive Latency)

h) (7 pts) What is the main functionality of a DDR2 controller? Explain its operation at system level.

i) (5 pts) What is the primary advantage of DDR2 SDRAMs over DDR?

j) Project related
(Note: Some of these are not sample exam problems and are only meant for general practice of DDRx technologies)
    a. (3 pts) What are the different precharge modes in DDR2?
    b. (3 pts) How many rows from the same bank can be activated simultaneously?
    c. (3 pts) If you plan to set BL=8, AL=2, and CL = 5, what parameters do you need to set up in each Mode Register?
    d. (2 pts) What is the purpose of on-die termination in DDR2, given that it also has OCD?
    e. (2 pts) Which SSTL standard is used in DDR2 and what is its reference voltage level?
    f. (3 pts) What is the logic value RAS, CAS, AL, BA for self-refresh entry mode?
    g. (4 pts) Explain the interleaved addressing mode in DDR2.
    h. (3 pts) Find the timing diagram for CK, CKE, address and DM for power up mode and list the commands.
    i. (2 pts) If we want the data out of two consecutive Read bursts to be concatenated, how many clock cycles do we need to consider between the two READ commands?
    j. (2 pts) If we want to write only a single word in the entire burst, how do we indicated that?
    k. Why do we need FIFOs in DDRx controllers?
    l. What is the difference between Scalar Read/Write and Burst Read/Write?
    m. What bits should we use to select different mode Registers?
    n. Why can we combine return address FIFO and return data FIFO? Why can't we combine input data FIFO and input command/address FIFO?
    o. Design a pure combinational logic to general the "put" signal of input DATA FIFO (IN_put) and input CMD FIFO (CMD_put) based on the command issued by testbench (CMD[2:0]=3'b000 for nop, 3'b001 for scalar read and 3'b010 for scalar write). Fill out the sensitivity list and then complete the logic.
        always@(CMD, CMD_NOTFULL )
            begin

    ….
    end

  p. Assume a ddr2 project work, similar to your ddr3 work in phase 2. What will happen if we remove the lines "set_dont_touch [ find cell process_logic_ddr2/ring_buffer4/DELAY*]" and "get_attribute [ find cell process_logic_ddr2/ring_buffer4/DELAY*] dont_touch" in the ddr2_controller.tcl file?

  q. In our project phase2, is it possible to remove the "full" signal of the return FIFO? Why or why not?

## 155) Testing

  a) (5 pts) Generate a test for the faulty circuit:

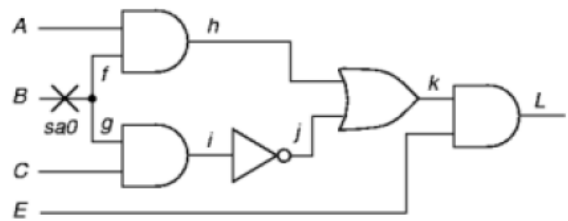

b) (5 pts) Does the test you found for the previous part, also tests SA0 at E?

c) (20 pts) Provide an example to show the difference between how test pattern generation (e.g., ATPG) and fault simulation work.

d) (10 pts) Explain whether the short could create a voltage value at the two output nodes which is different from VDD/2? If so, argue whether IDDQ testing would be able to detect the fault? Draw a circuit diagram that can be used to approximate the shorted voltage and also the IDDQ current.

e) How to to detect that short current.

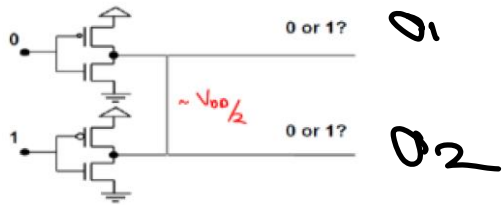

Shahin Nazarian   USC   

f) (6 pts) Perform fault simulation for each of the following test vectors: and indicate which of the marked faults can be detected by each test vector:

   I ) 10110      II) 01000

   a. SA0 at G and SA1 at M are detected.

   b. Note of the 3 faults can be detected.

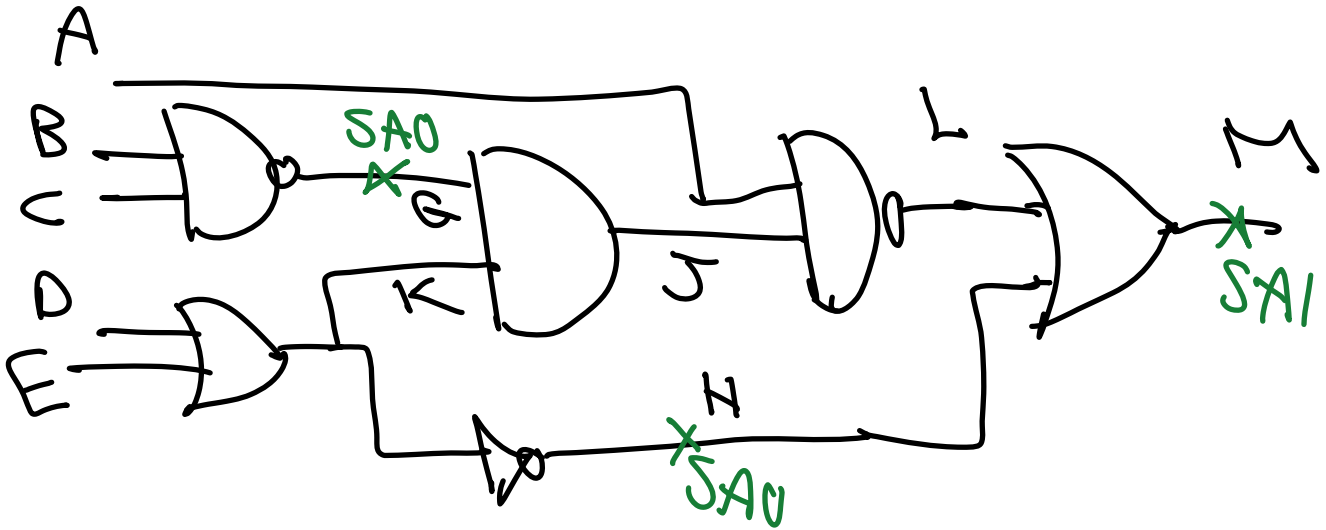

g) (12 pts) Find at least one test for each of the following faults in the previous circuit:

   I ) SA0 at J    II ) SA1 at K   III) SA0 at K   IV) Short between L and H

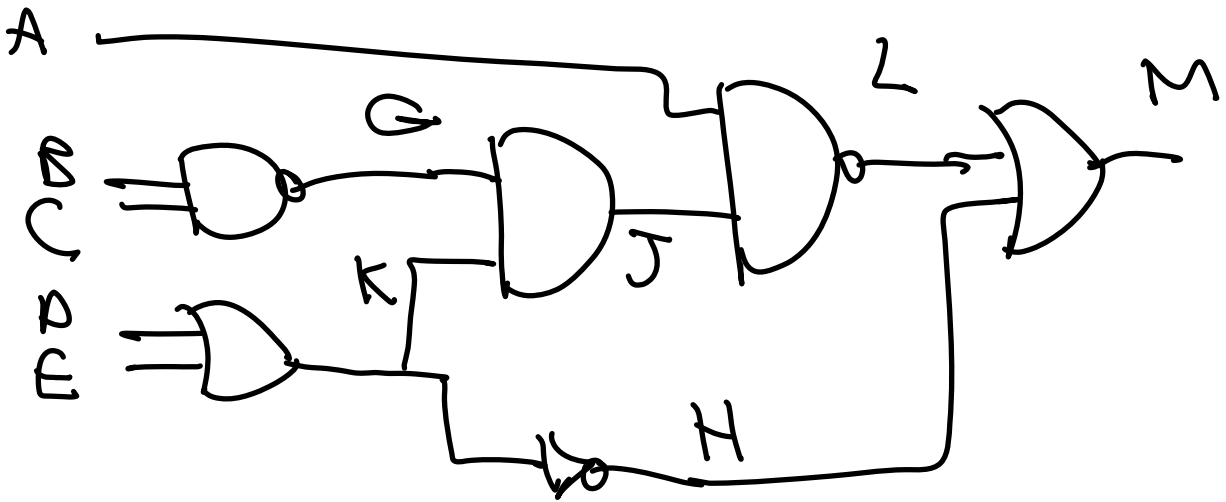

### 156) DLL & PLL

a) (15 pts + 10 extra) How does a DLL work? How would you differentiate it from PLL? Which one, DLL or PLL uses Phase?
b) ( 5 extra pts ) Draw the block diagram of a DLL.
c) (5 extra pts) Draw the high level block diagram of a digital PLL.

### 157) MISR & BIST

MISR the BIST out of Z!Xussia

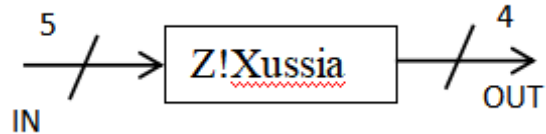

### 158) FIFO
Claim: In a special case, Grey Counter would create an unsafe (EMPTY or FULL) situation. False/True.

### 159) Twisted Bitline
The analogy used was related to:
    I.     Your roommate and music
    II.    Your manager and IQ tests
    III.   Bill Clinton and apple stock price
    IV.   None of the above
    V.    Two of the above

### 160) Synchronous System Interface
Briefly explain the application of SSI (Synchronous System Interface) in SoC (System on Chip).

### 161) File Structure of Verilog
In ddr2_controller.v, write the Verilog code to generate ck based on clk.

### 162) LFSR
Draw a modular LFSR with the characteristic polynomial as: $f(x) = 1 + x^2 + x^7 + x^8$
Also specify the matrix representation of the LFSR.

### 163) Memory
Sources and drains are at zero, Gates are at 0, and wells are at a relatively high voltage. What is going on?!
Name one DRAM scaling challenge.

### 164) Finite State Machine
Generate a sequence of "1 8 17 10 2 9 18 11" using FSM
Generate such a sequence using NOR-based ROM.
What's the limit of implementing it using rom? Think of a sequence of 10,000 numbers

# Chapter 18 – Verilog Basics

## 165)

### a) Verilog Syntax/compile errors

There are 4 syntax/compile errors in the following Verilog code. Identify them and very briefly explain your reason. Ignore functional mistakes, i.e., we are not worried how the code would behave; all we want is to compile with no error. Also fill in the blank part (the value of out in the commented statement)

```
module FSM1(clk, in, reset, out);
input clk, in, reset;
output [3:0] out;
reg out;
parameter zero=0, one=1, two=2, three=3;
always @(state)
   begin
      case (state)
         zero:
            out = 0000;
         one:
            out = 0001;
         two:
            out = 0010;
         three:
            out = 0100;      // out = __________ in decimal
         default:
            out = 0000;
      endcase
   end
always @(posedge clk or posedge reset)
   begin
      if (reset)
         state = zero;
      else
         case (state)
            zero:
               state = 1;
            one:
               if (in)
                  state = 0;
               else
                  state = 2;
            two:
               state = 3;
            three:
               state = 0;
            four:
               state = 0;
   end
endmodule
```

### b) Case Equality

Fill in the k values.
```
initial
  begin
    k = 5;
    a = 6'b0011XX;
```

```
        b = 6'b0011ZZ;

        If(a !== b)
            begin
                    k = k + 1;
                    k = k*k;
            end
        else k = 10;
        // k = ______________
        if(a != b)
            begin
                    k = k + 1;
                    k = k^2;
            end
        else k = 10;
        //k = ________________
   end
```

c) **Operators**
   Write the result of each operation.
   4'b0011 & 4'b1100 =
   ^4'b0011 =
   4'b0011 << 1 =
   {4'b1100, 4'b00XX} =
   {3'b 111, 2{3'b100}} =

d) **Syntax**
   What does #2 in the following code do?
   ```
   begin a = 0;
   c = 1;
   b = #2 a + c;
   end
   ```

e) **Syntax**
   What is the difference between ( = = , ! = ) and ( = = = , ! = = )?

f) **Latches in synthesis**
   Give an example inferring latches in a case statement. Explain how to avoid the latches.

# Chapter 19 – Verilog Coding

## 166)

a) **Behavioral RTL Verilog – Transparent Latch**

Design of a positive-level (i.e, transparent when clk=1) transparent latch, with an asynchronous active high reset. Use a procedural statement.

**module** TR_LATCH( in, clk, out, reset)
    input in, clk;
    output out
    //your code here
**endmodule**

b) **Behavioral Dataflow Verilog – 8-to-1 Multiplexer**

Design a module that implements a 8-to-1 Mux. Dataflow style should be used. The goal is to have a module for this Mux that uses only one assign statement. Note: Designs with more than one assign statement will receive partial credit.

**module** MUX_8TO1( d, select, q );
input[7:0] d;
input[2:0] select;
output   q;
//your code here
**endmodule**

c) **Behavioral RTL Verilog – 4-bit Shift Register**

Design a 4-bit shift register with synchronous with the block diagram of Figure1.) It performs logical shift to right or left depending on the value of *dir* signal. The *rst* signal loads the input data (*D*) into the output (*Q*). If *rst* is high, at every positive edge of *clk*, the content of shift register (Q) shifts to right by one bit if *dir* is high, and to left by one bit if *dir* is low. SER_IN overwrites LSB or MSB in left shift or right shift respectively. To better visualize how the shift register works, an example is illustrated in Figure 1: for D=0111 and SER_IN=1 both right and left shifts are presented.

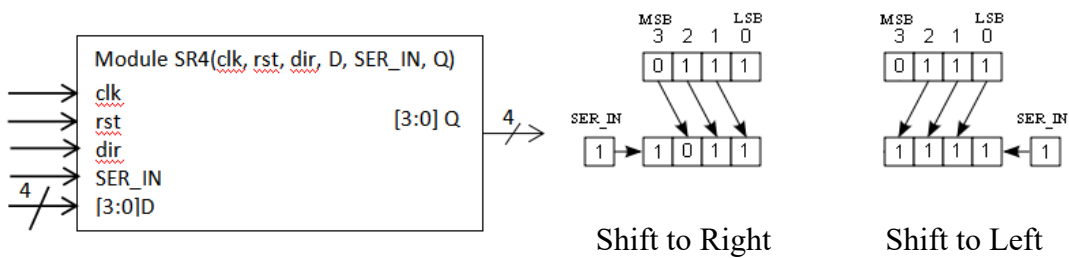

                                                Shift to Right      Shift to Left
                  **Figure 1**                                  **Figure 2**

**module** SR4(clk, rst, dir, D, SER_IN, Q);

```
    input clk, rst, dir, SER_IN;
    input [3:0] D;
    output [3:0] Q;
//your code
endmodule
```

d) **Behavioral and Structural RTL Verilog Combined – 16-bit Shift Register**

Design a 16-bit shift register using the 4-bit shift register of the previous problem. The 16-bit shift register performs logical shift to right by one bit if *dir* is high, and to left by one bit if *dir* is low. You need to instantiate only four SR4 modules and connect them appropriately. Note that the single bit *SER_IN* input would overwrite LSB or MSB of 16-bit shift register.

```
module SR16(clk, rst, dir, D, SER_IN, Q);
    input        clk, rst, dir, SER_IN;
    input [15:0] D;
    output [15:0] Q;
//your code here
endmodule
```

e) **Behavioral RTL Verilog– FSM Design**

Figure 3 shows the state diagram of a Moore FSM machine. Complete the Velilog code of this FSM. Don't create new variables but use only existed ones.

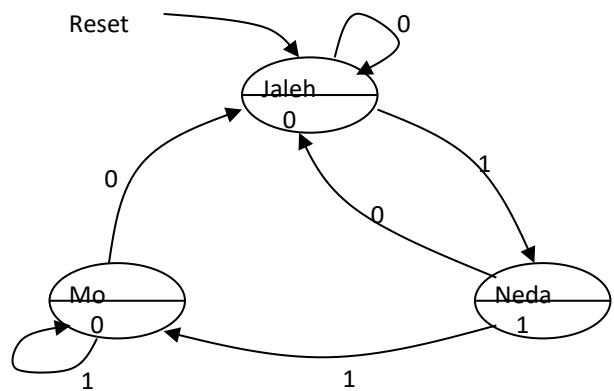

**Figure 3**

```
module FSM2 (input wire Reset, input wire In, output wire Out);
    localparam Jaleh = 2'b00, Neda = 2'b01, Mo = 2'b10;
    reg [1:0] CurrentState, NextState;
```

```
always @ (posedge Clock) begin
    if (Reset) begin
                                            // missing
    End
    else begin
                                            // missing
        End
    End
always @ ( * ) begin
    NextState = CurrentState;
    case (CurrentState)     // case statements are missing

    Endcase
End
assign Out =                                // missing
Endmodule
```

### f) FSM – Verilog Design of a Crosswalk Controller

A crosstalk controller circulates among the following states in order (Figure I).
- State Solid: A solid "Hand" sign is displayed for 32 cycles by asserting the output signal "HAND"
- State Walk: The "Walk" sign is displayed for 10 cycles by asserting the output signal "WALK"
- State Blink: The blinking "Hand" sign is displayed for 8 cycles while counting down from 8 to 1 on a display, i.e., the NUM[3:0] should count down from 1000 to 0001 in binary and meanwhile the output "HAND" signal should toggle between 1 and 0 every clock cycle.

The controller changes the state from Blink back to Solid and repeats the process. The controller would display a solid "Hand" sign for 32 cycles every time the "reset" input is asserted (Note: reset is active low and asynchronous.) NUM_ON should be set to 1 during the Blink state and 0 in the other two states. Design a state machine (Moore or Mealy) in Verilog to control the movement through each phase of display and generate the countdown number of display (NUM[3:0]).

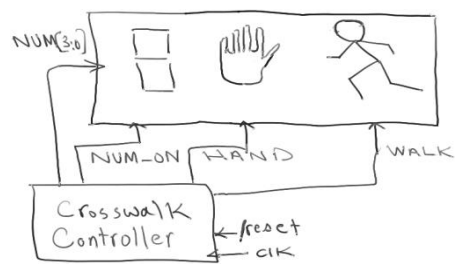

g) **Hierarchical Parametric FIFO Design and Verification**

The following design and verification task should be done based and your experience with the assignments.

    a. Design a 64×Width-bit wide 2-clk FIFO of depth 320, using a smaller 64-bit wide FIFO of depth 32.

        module FIFO (clkr, clkw, reset, data_in, put, get, data_out, fillcount, empty, full, … );

    b. Write a testbench for your design.

h) **33-Dimensional Maze Router**

Assume it is the year 2577, and we have finally discovered all the 33 dimensions, including my favorite, the Zaydaf, along which the cost of moving would be negative (-1) if you have black tea and face no blockage. Our goal is to design a 33-dimensional maze router to find our way from a vacation resort in Black Eye through galaxies back to Sombrero, where our main base is. The whole cosmos is divided into grids of equal size. It is possible to proceed in each dimension with the given cost. Assume the whole space is provided as a 33-dimensional grid with all the blockage, e.g., the black holes and dark matters marked in it with the cost of a practical ∞. For simplicity ( :D ) assume the size of each dimension is limited and provided in the testbench you would need to write, as a parameter. You would need to design the RTL. Do not try to synthesize this at home:D

Note: The top design earns extra points.

i) **Write a Verilog code of the following design.**

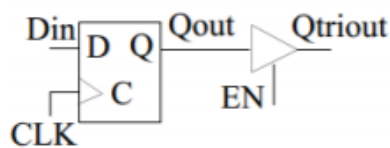

j) **Verilog Design – Gray Counter:**

Design a 6bit gray counter. This counter has a synchronous reset signal.

# Chapter 20 – SOLUTIONS

# 20.1: Chapter 1 - CMOS Implementation

1.
   a.
   b.

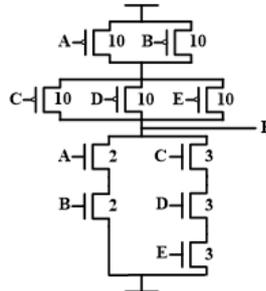

Answer not unique. Fractional numbers accepted.

   c. I recommend using the resistive model:

Assume $w_p=10$ results in resistance $R_p$
the worst case $R_{PUN}=2R_p$
best case is $5R_p/6 \Rightarrow$

$T_r^{worst}/T_r^{best} = 12/5$

Similarly we can calculated the ratio for the PDN as:

$T_f^{worst}/T_f^{best} = 2$

2.
   a.

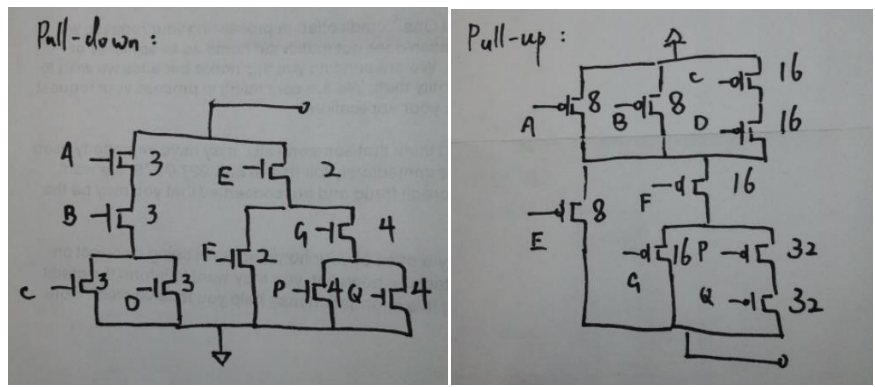

   b. See the above.

c. Fall delay: 5/13
Rise delay: 8/21

3.

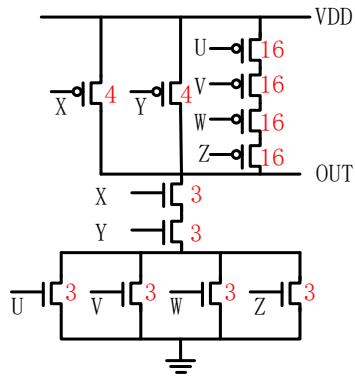

4.

a. $V_D^2 + 4V_D - 0.75 = 0$

b. $\frac{V_D}{0.2} = 1$

c. Saturation

5.

a. We first calculate the threshold voltage:

$V_T = V_{T0} + \gamma\left(\sqrt{|-2\phi_F + V_{SB}|} - \sqrt{|-2\phi_F|}\right) = -0.5 - 0.25\left(\sqrt{|-2\times 0.2 - 0.6|} - \sqrt{|-2\times 0.2|}\right)V = -0.6V$

Since $V_{SG} - V_T < V_{SD}$, the transistor is in saturation region

$I_{SD} = \frac{1}{2}k'\frac{W}{L}(V_{GS} - V_T)^2(1 + \lambda V_{DS}) \Rightarrow \frac{W}{L} = \frac{2I_{SD}}{k'(V_{GS} - V_T)^2(1 + \lambda V_{DS})} = 5.$

b. Since $V_{SG} - V_T > V_{SD}$, the transistor is in linear region.

$I_{SD} = \frac{k'}{2}\frac{W}{L}\left[2(V_{GS} - V_T)V_{DS} - V_{DS}^2\right] = 10\mu A$

6.

a.

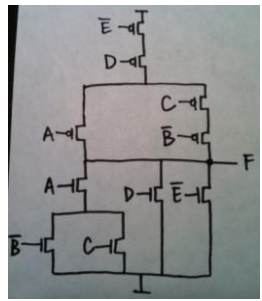

b. $\bar{E} > D > A > \bar{B} > C$
c.

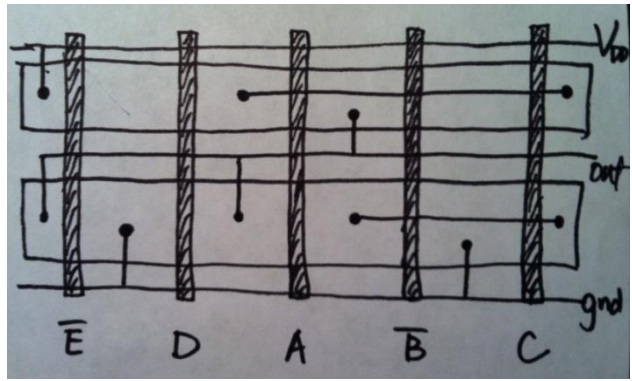

7.

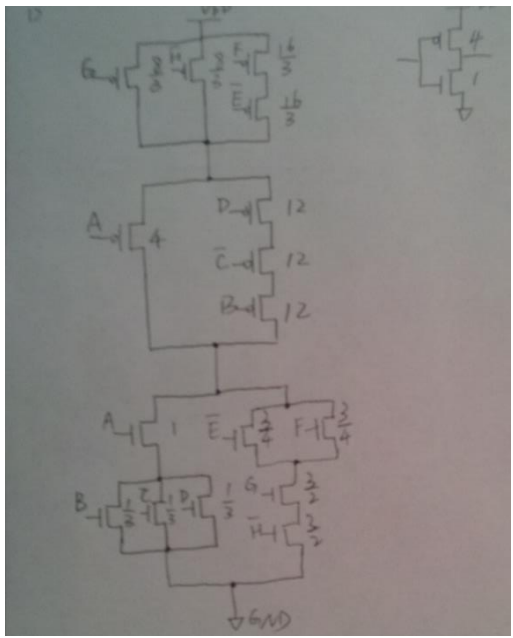

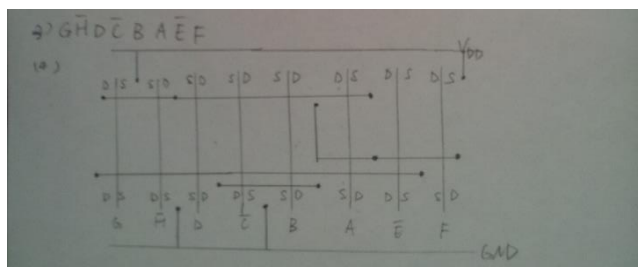

8.
a. No, there isn't. Rearranging the transistors works:

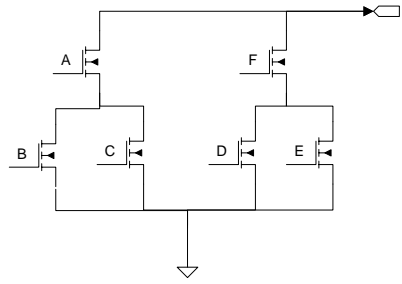

b.

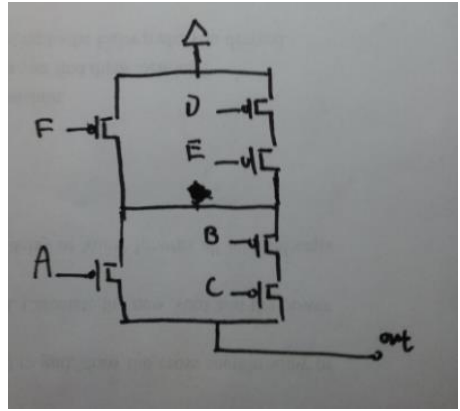

c. Common Euler path: B C A F D E

**9.**
R=V/I
Therefore, in constant field scaling R is unchanged; in constant voltage scaling R is changed to 1/S.
R=Rsheet*L/W and Rsheet = R*W/L
Therefore, in constant field scaling Rsheet is unchanged; in constant voltage scaling Rsheet is changed to 1/S.

**10.**
$$V_{new} = \frac{1}{S} \times V_{old}$$

$$I_{new} = \frac{M}{S^2} \times I_{old}$$

$$Cload_{new} = \frac{1}{M} \times Cload_{old}$$

$$Delay_{new} = \frac{1}{S} \times \frac{M^2}{S} \times \frac{1}{M} \times Delay_{old} = \frac{S}{M^2} \times Delay_{old}$$

$$P_{new} = \frac{M}{S^3} \times P_{old}$$

$$Energy_{new} = \frac{1}{M \times S^2} \times Energy_{old}$$

$$Power\_Density_{new} = \frac{M^3}{S^3} \times Power\_Density_{old}$$

**11.** $F = C \cdot (A \cdot B + \bar{A} \cdot \bar{B}) + \bar{C} \cdot (A + \bar{A} \cdot D)$

**12.**

$V_{n1} = V_{dd} - V_{tn}$
$V_{n2} = V_{n3} = V_{dd} - 2V_{tn}$
$V_{n4} = V_{n5} = V_{n6} = V_{dd} - 3V_{tn}$

**13.**
A, B: $V_{DD} - V_T$
C, D: $V_{DD}/2 - V_T$

**14.**

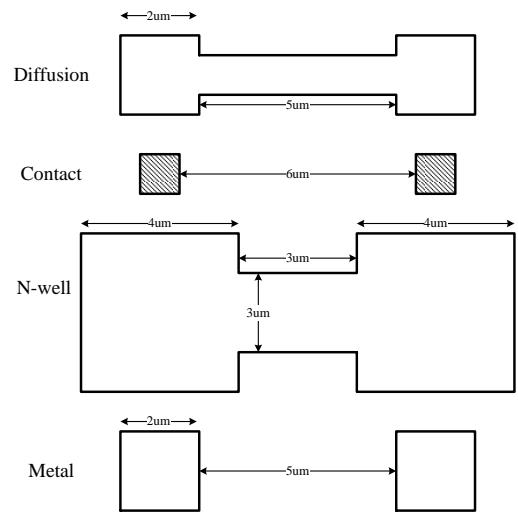

**15.**
First – 1-3-5-6-7 delay: 4.5ns
Second – 2-4-5-6-7 delay: 4.25ns
Third – 2-3-5-6-7 delay: 4ns

**16.**
a. Flop – comb(0.2) – flop – comb(1.8) – flop – comb(0.8) – flop – comb(1.1) – flop – comb(1.1) –flop  (DRAW this as schematics view)
b. Max frequency = (1/1.8ns) = 555.5 MHZ
c. Total time = (1000 +4) x 1.8ns = 1.807usec

**17.**

a. S[0] → 1.5ns; S[3] → 4.5ns, Bout[3] → 3.5, S[6] → 6.0, and Bout[7] → 5.5?
b. X[7] to S[7] → 1.0ns; Bin[5] to Bout[5] → 0.5ns
c. X,Y [0] to Bout[0], then Bout[0] to Bin[7], then Bin[7] to S[7]; total delay = 6.5ns
d. Three types of subtractor needed, first one to reduce X[0], Y[0] to Bout delay, second one for Bin to Bout minimum, third one for Bin to S delay minimum.
e. Max frequency of operation = (1/6.5ns) = 153.8 MHZ, to improve speed by four times, we have to pipeline by four stages and each stage has a combinational logic delay smaller than (6.5/4) = 1.625 ns

**18.**

a. From PMOS part, for that branch to conduct a VDD to the remaining part of the PMOS network: $\left(\overline{P_1} + \overline{G_0}\right)\overline{G_1}$

From NMOS part, for that branch to conduct a GND to the remaining part of the NMOS network: $\overline{G_1 + P_1 G_0}$

Using DeMorgan's Law, we have: $\overline{G_1 + P_1 G_0} = \overline{G_1}\left(\overline{P_1 G_0}\right) = \overline{G_1}\left(\overline{P_1} + \overline{G_0}\right)$

b. The sizing is shown in figure below.
c. Using a lumped model, the worst-case fall time is: 5R×(2C) = 10RC and the worst-case rise time is: 10R×(2C) = 20RC. 2C is the capacitance of the inverter with minimum sized NMOS and PMOS. Notice that this inverter has worst-case rise time about 2 times worse than worst-case fall time.

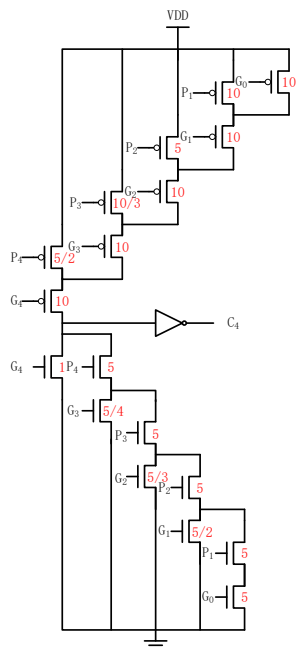

**19.**

| Parameter | Constant Electric Field (Full Scaling) | Constant Voltage |
| --- | --- | --- |

| W, L, t_ox | 1/S | 1/S |
|---|---|---|
| V_DD | 1/S | 1 |
| Current Density×Delay | 1 | S |

$$Current\_Density \times Delay = \frac{I}{WL} \times \frac{C_{gate}V_{DD}}{I} = \frac{C_{gate}V_{DD}}{WL} = C_{ox}V_{DD} = \frac{\varepsilon_{ox}}{t_{ox}}V_{DD}$$

**20.** $\varepsilon_{si} = 11.7 \times \varepsilon_o$

$\varepsilon_{ox} = 3.9 \times \varepsilon_o$  $n_i = 1.45 \times 10^{10} cm^{-3}$
$L = L_M - 2L_D = 1.5 - 2(0.25) = 1\mu m$

Total oxide cap = Cgd + Cgs + Cgb
Oxide cap: Overlap cap + Gate- channel cap
The NMOS above is in saturation region.
Hence:
- Cgb (total) = 0
- Cgd (total) = CoxWL_D
- Cgs (total) = CoxWL_D + 2/3 CoxWL
- $C_{ox} = \frac{\varepsilon_{ox}}{t_{ox}} = \frac{3.9 \times 8.85 \times 10^{-14}}{100 \times 10^{-8}} = 3.453 \times \frac{10^{-7}F}{cm^2}$
- $C_{gd} = C_{ox} \cdot W \cdot L_D = 3.453 \times 10^{-7} \times 10 \times 10^{-4} \times 0.25 \times 10^{-4} = 8.63 fF$
- $C_{gs} = C_{ox} \cdot W \cdot L_D + \frac{2}{3}C_{ox} \cdot W \cdot L = 8.63\ fF + \frac{2}{3} * 3.453 \times 10^{-7} \times 10 \times 10^{-4} \times 1 \times 10^{-4}$
- 8.63 fF + 23.02 fF = 31.65 fF

Total oxide cap = 0 + 8.63 + 31.65 = **40.28 fF**

a) Junction cap = Sidewalls cap + Channel side cap

For Drain Junction, Vdb = Vd - Vb = 2V – 0V = 2V.

Drain: V = -2 V  → reverse bias 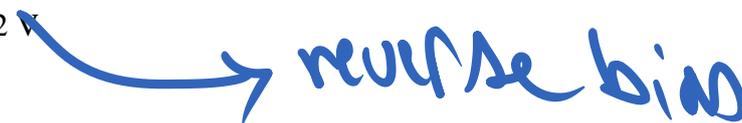

$$\phi_o = \frac{kT}{q}\ln\frac{N_A \cdot N_D}{n_i^2} = 0.026\ln\frac{10^{17} \cdot 10^{21}}{(1.45 \times 10^{10})^2} = 1.058V$$

$$C_{jo} = \sqrt{\frac{\varepsilon_{si} \cdot q}{2}\left(\frac{N_A \cdot N_D}{N_A + N_D}\right) \cdot \frac{1}{\phi_o}}$$

$$= \sqrt{\frac{11.7 \times 8.85 \times 10^{-14} \times 1.6 \times 10^{-19}}{2} \cdot \frac{10^{17} \cdot 10^{21}}{10^{17} + 10^{21}} \cdot \frac{1}{1.058}}$$

$$= 8.849 \times 10^{-8} F/cm^2$$

$$C_j(V) = \frac{W(Y + X_j) \cdot C_{jo}}{\sqrt{1 - \frac{V}{\phi_o}}} = \frac{53 \times 10^{-8} \times 8.849 \times 10^{-8}}{\sqrt{1 + \frac{2}{1.058}}} = 27.5\ fF$$

For sidewall capacitance calculation:

$$\phi_{osw} = \frac{kT}{q} \ln \frac{N_A(sw) \cdot N_D}{n_i^2} = 0.026 \ln \frac{10^{19} \cdot 10^{21}}{(1.45 \times 10^{10})^2} = 1.178 V$$

$$C_{josw} = \sqrt{\frac{\varepsilon_{si} \cdot q}{2} \left(\frac{N_A(sw) \cdot N_D}{N_A(sw) + N_D}\right) \cdot \frac{1}{\phi_{osw}}}$$

$$= \sqrt{\frac{11.7 \times 8.85 \times 10^{-14} \times 1.6 \times 10^{-19}}{2} \cdot \frac{10^{19} \cdot 10^{21}}{10^{19} + 10^{21}} \cdot \frac{1}{1.178}}$$

$$= 8.345 \times 10^{-7} F/cm^2$$

$$C_{jsw}(V) = \frac{P \cdot X_j \cdot C_{josw}}{\sqrt{1 - \frac{V}{\phi_{osw}}}} = \frac{(2 \times 5 + 10) \times 0.3 \times 10^{-8} \times 8.345 \times 10^{-7}}{\sqrt{1 - \frac{V}{\phi_{osw}}}}$$

$$= \frac{5.007 \times 10^{-14}}{\sqrt{1 - \frac{V}{\phi_{osw}}}} F$$

$$C_{jsw}(-2V) = \frac{5.007 \times 10^{-14}}{\sqrt{1 + \frac{2}{1.178}}} = 30.48 \times 10^{-15} F$$

$$C_{db} = C_j(-2V) + C_{jsw}(-2V) = 27.5 + 30.48 = \mathbf{57.98 fF}$$

### 21.

One possible Solution:

NAND:
Y = (ABCD)' = A' + B' + C' + D' = A' + A (B' + C' + D')
= A' + A (B' + B(C'+D')) = A' + A(B' + B(C'+ CD'))

NOR:
Y = (A+B+C + D)' = A'B'C'D' = A'B'C'D' + A(0) = A' (B'C'D'+ B.0) + A(0)
= A' (B'(C'D' + C(0))+ B.0) + A(0)

These equations are ready to be easily implemented by pass transistors.

**22.**

Assume NMOS is in saturation region.

$$I_D = \frac{k_n}{2}(V_{GS} - V_t)^2$$

$$= \frac{1}{2} K_n' \frac{W}{L}(V_{GS} - V_t)^2$$

$$= \frac{1}{2} \times 220\,\mu A/V^2 \times \frac{10u}{1u} \times (V_{GS} - 0.5V)^2$$

$V_{GS} = 1.5V - I_D \cdot 10k\Omega$

$\Rightarrow I_D = 135\,\mu A$ or $I_D = 74.1\,\mu A$

if $I_D = 135\,\mu A$, $V_{GS} = 0.15V < V_t$   so, it is not the case.

if $I_D = 74.1\,\mu A$, $V_{GS} = 0.759V > V_t$   so, $I_D = 74.1\,\mu A$

$V_{out} = I_D \cdot 10k\Omega = 0.741V$

$V_{DS} = 2V - V_{out} = 1.259V > V_{GS} - V_t$   so, it is in the saturation region.

In saturation

$C_{gb} = 0$

$C_{gd} = C_{ox} W L_D = 14\,fF/\mu m^2 \times 10\,\mu m \times 0.1\,\mu m = 14\,fF$

$C_{gs} = \frac{2}{3} C_{ox} WL + C_{ox} W L_D = \frac{2}{3} \times 14\,fF/\mu m^2 \times 10\,\mu m \times 1\,\mu m + 14\,fF/\mu m^2 \times 10\,\mu m \times 0.1\,\mu m$

$= 107.3\,fF$

$C_g = 14\,fF + 107.3\,fF + 0 = 121.3\,fF$

**23.**

a. Cgb=0, Cgd = 105e(-16)F, Cgs=245e(-16)F

b. Cjunction = Csidewall + Cbottom = (2Y+W)*Cjsw + (Xj+Y)W*Cj0 = 106.25e(-16)F

**24.**

a) $\Psi_0 = \dfrac{kT}{q} \ln \dfrac{N_A N_D}{n_i^2} = 0.026 \times \ln \dfrac{10^{16} \times 10^{20}}{(1.45 \times 10^{10})^2} V = 0.94V$

$C_{j0} = \sqrt{\dfrac{\varepsilon_{Si} q}{2} \left( \dfrac{N_A N_D}{N_A + N_D} \right) \dfrac{1}{\Psi_0}}$

$= \sqrt{\dfrac{11.7 \times 8.85 \times 10^{-14} \times 1.6 \times 10^{-19}}{2} \dfrac{10^{16} 10^{20}}{10^{16} + 10^{20}} \dfrac{1}{0.94}} [F/cm^2] = 2.98 \times 10^{-8} [F/cm^2]$

$C_j(-2.5) = \dfrac{(Y + x_j) W \times C_{j0}}{\left(1 + \dfrac{V_{DB}}{\Psi_0}\right)^{M_j}} = \dfrac{(5.2) \times 8 \times 10^{-8} \cdot 2.98 \times 10^{-8}}{\sqrt{1 + \dfrac{2.5}{0.94}}} = 6.48\,fF$

$C_j(-4) = \dfrac{(Y + x_j) W \times C_{j0}}{\left(1 + \dfrac{V_{DB}}{\Psi_0}\right)^{M_j}} = \dfrac{(5.2) \times 8 \times 10^{-8} \cdot 2.98 \times 10^{-8}}{\sqrt{1 + \dfrac{4}{0.94}}} = 5.41\,fF$

$\Psi_{sw0} = \dfrac{kT}{q} \ln \dfrac{N_A N_D}{n_i^2} = 0.026 \times \ln \dfrac{10^{19} \times 10^{20}}{(1.45 \times 10^{10})^2} V = 1.118V$

$C_{jsw0} = \sqrt{\dfrac{\varepsilon_{Si} q}{2} \left( \dfrac{N_{A(sw)} N_D}{N_{A(sw)} + N_D} \right) \dfrac{1}{\Psi_0}}$

$= \sqrt{\dfrac{11.7 \times 8.85 \times 10^{-14} \times 1.6 \times 10^{-19}}{2} \dfrac{10^{19} 10^{20}}{10^{19} + 10^{20}} \dfrac{1}{1.118}} [F/cm^2] = 8.207 \times 10^{-7} [F/cm^2]$

$C_{jsw}(-2.5) = \dfrac{(2Y + W) x_j \times C_{jsw0}}{\left(1 + \dfrac{V_{DB}}{\Psi_0}\right)^{M_{jsw}}} = \dfrac{(2 \times 5 + 8) \times 0.4 \times 10^{-8} \cdot 8.207 \times 10^{-7}}{\sqrt{1 + \dfrac{2.5}{1.117}}} = 32.84\,fF$

$C_{jsw}(-4) = \dfrac{(2Y + W) x_j \times C_{jsw0}}{\left(1 + \dfrac{V_{DB}}{\Psi_0}\right)^{M_{jsw}}} = \dfrac{(2 \times 5 + 8) \times 0.4 \times 10^{-8} \cdot 8.207 \times 10^{-7}}{\sqrt{1 + \dfrac{4}{1.117}}} = 27.61\,fF$

$C_{DB}(-2.5) = C_j(-2.5) + C_{jsw}(-2.5) = 6.48\,fF + 32.84\,fF = 39.32\,fF$

$C_{DB}(-4) = C_j(-4) + C_{jsw}(-4) = 5.41\,fF + 27.61\,fF = 33.02\,fF$

b) $C_{ox} = \dfrac{\varepsilon_{ox}}{t_{ox}} = \dfrac{3.9 \times 8.85 \times 10^{-14}}{50 \times 10^{-7}} = 6.903 \times 10^{-8} [F/cm^2]$

$C_{ov} = WL_D \dfrac{\varepsilon_{ox}}{t_{ox}} = 0.35 \times 8 \times 10^{-8} \times \dfrac{3.9 \times 8.85 \times 10^{-14}}{50 \times 10^{-7}} = 1.93\,fF$

$C_{intrinsic} = WL_{eff} \dfrac{\varepsilon_{ox}}{t_{ox}} = 8 \times (5 - 0.35 \times 2) \times 10^{-8} \times 6.903 \times 10^{-8} = 23.75\,fF$

$C_G = C_{GS} + C_{GB} + C_{GD} = \begin{cases} \text{Linear}: C_{intrinsic} + 2C_{ov} = 23.75\,fF + 2 \times 1.93\,fF = 27.61\,fF \\ \text{Saturate}: \dfrac{2}{3} C_{intrinsic} + 2C_{ov} = 19.69\,fF \end{cases}$

### 25.

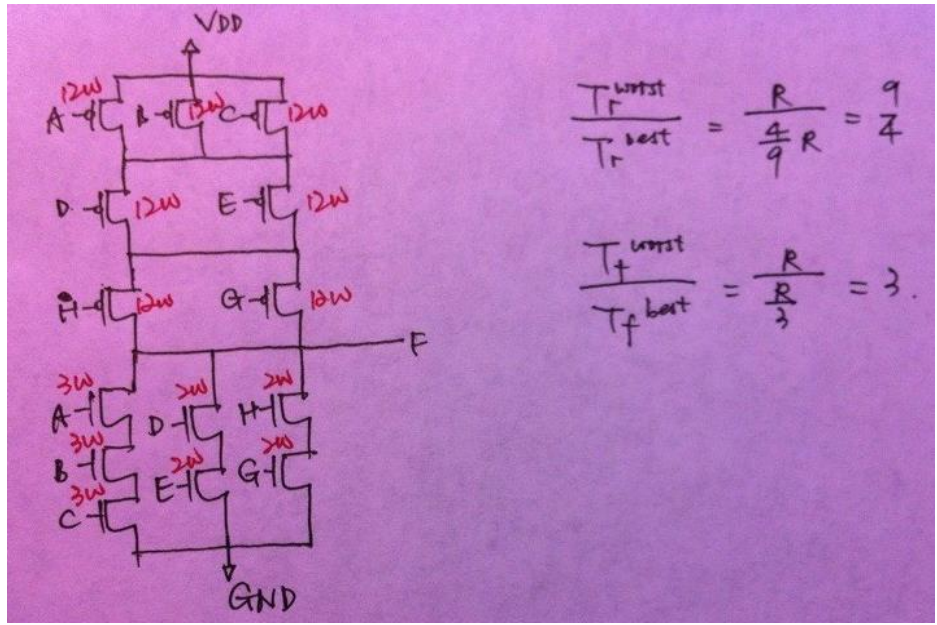

### 26.
a. Saturation, because Vgs - Vt = 1.5V and Vds = 2V
   Saturation, because Vgs - Vt = 0.3V and Vds = 0.8V
b. (a) 200uA; (b) 9uA.
c. Vdd,a/Vdd,b = 2/0.8 = 2.5 whereas Ia/Ib = 25. So the delay increases superlinearly as the Vdd value decreases.

### 27.

34. a) $AB\bar{C} + A\bar{B}C + \bar{A}B\bar{C} + \bar{A}\bar{B}C$

b) $\bar{A}\bar{B}C + \bar{A}B\bar{C} + ABC + AB\bar{C}$
can be shown to be
$\bar{S}$

**28.**

| Vin | Vout | driver | load |
|---|---|---|---|
| 0 | $V_{OH}$ | cut-off | linear |
| Vdd | $V_{OL}$ | linear | saturation |
| $V_{IL}$ | high | saturation | linear |
| $V_{IH}$ | low | linear | saturation |

$V_{OH}$=Vdd, $V_{OL}$=0.008;
$V_{IH}$=0.573V, $V_{IL}$=0.467V ($20V_{IL}^2$-$16V_{IL}$+3.11=0).

**29.**

| Vin | Vout | PMOS | NMOS |
|---|---|---|---|
| 0 | $V_{OH}$ | linear | cut-off |
| Vdd | $V_{OL}$ | cut-off | linear |
| $V_{IL}$ | high | linear | saturation |
| $V_{IH}$ | low | saturation | linear |
| $V_M$ | $V_M$ | saturation | saturation |

$V_{OH}$ is the same (Vdd), $V_{OL}$ is larger than zero, $V_M$ is larger than that of CMOS inverter.

**30.**   i) Vin=VDD, Vout=$V_{OL}$=0V.

ii)

Vin=0V, Vout=$V_{OH}$.

$$0.5 \times k \times [2(V_{SG} - |V_t|)V_{SD} - V_{SD}^2] = V_{out}/R$$

$$\rightarrow 0.5 \times k \times [2(VDD - |V_t|)(VDD - V_{out}) - (VDD - V_{out})^2] = \frac{V_{out}}{R}$$

iii)

Vin= $V_{IL}$

$$0.5 \times k \times [2(V_{SG} - |V_t|)V_{SD} - V_{SD}^2] = V_{out}/R$$

$$\rightarrow 0.5 \times k \times [2(VDD - V_{in} - |V_t|)(VDD - V_{out}) - (VDD - V_{out})^2] = V_{out}/R \quad (1)$$

$$\frac{d}{dV_{in}} 0.5 \times k \times [2(VDD - V_{in} - |V_t|)(VDD - V_{out}) - (VDD - V_{out})^2] = \frac{\frac{d}{dV_{in}} V_{out}}{R}$$

$$\rightarrow 0.5 \times k \times \left[2(-1)(VDD - V_{out}) + 2(VDD - V_{in} - |V_t|)\left(-\frac{dV_{out}}{dV_{in}}\right)\right.$$

$$\left. - 2(VDD - V_{out})\left(-\frac{dV_{out}}{dV_{in}}\right)\right] = \frac{\frac{dV_{out}}{dV_{in}}}{R}$$

$$\rightarrow k \times [-(VDD - V_{out}) + (VDD - V_{in} - |V_t|) - (VDD - V_{out})] = -\frac{1}{R}$$

$$\rightarrow k \times [2V_{out} - V_{in} - |V_t| - VDD] = -1/R \quad (2)$$

iv)

$V_{in} = V_{IH}$

$$0.5 \times k \times (V_{SG} - |V_t|)^2 = \frac{V_{out}}{R}$$

$$\rightarrow 0.5 \times k \times (VDD - V_{in} - |V_t|)^2 = V_{out}/R \quad (1)$$

$$\frac{d}{dV_{in}} 0.5 \times k \times (VDD - V_{in} - |V_t|)^2 = \frac{\frac{d}{dV_{in}} V_{out}}{R}$$

$$\rightarrow k \times (VDD - V_{in} - |V_t|) \times (-1) = \frac{\frac{dV_{out}}{dV_{in}}}{R}$$

$$\rightarrow k \times (VDD - V_{in} - |V_t|) = 1/R \quad (2)$$

**31.** For node a:

$NM_L = V_{IL2} - V_{OL1} = 0.3 - 0.1 = 0.2$

$NM_H = V_{OH1} - V_{IH2} = 1.8 - 1.6 = 0.2$

For node b:

$NM_L = V_{IL3} - V_{OL2} = 0.6 - 0.2 = 0.4$

$NM_H = V_{OH2} - V_{IH3} = 2.0 - 1.4 = 0.6$

# 20.2: Chapter 2 – Boolean Implementation

**32.**

$$V_a^{max} = V_{DD} - 0.35$$

$$V_i^{max} = V_{DD} - 0.4$$

$$V_{OUT}^{max} = V_{DD} - 0.4$$

**33.**

a. 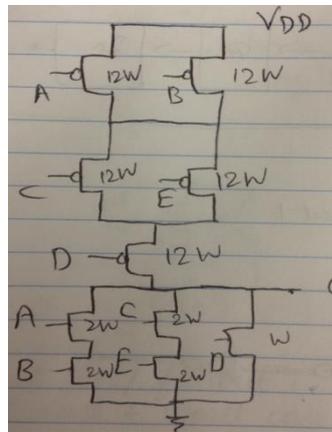

b. Output connected to input D of second gate will give best case, as that input overall has the least amount of gate capacitance acting as load. However, if the output is connected to any other inputs, then their nmos have higher sizes hence bigger load to drive. Thus worst case is realized.

c. The ratio is 4.

d. Pull up & pull down worst case: $RC_L$
Best case pull-up: $2/3\ RC_L$
Best case pull-down: $1/3\ RC_L$

**34.**

a. Draw schematics and let the instructor know if you don't get to the following results.

b. For pull-down network, nMOS, A, B, C, D, E = 3 W; nMOS F,G = 2 W
For pull-up network, pMOS A, B, C, D, H, E, F, G = 3x4 W;

c. Total Area = (22 + 96) WL, Inverter area = 5WL, ratio = 118/5

d. Best case fall delay = worst case delay/3; best case rise delay = worst case delay/ (18/7)

**35.**

AB + BC + AC

**36.**

a. Notice that the PMOS transistor will only conduct when both NMOS and transmission gate is cut off. When $P_i = 1$, $G_i$ will be 0, and $C_{i\_b}$ is determined by $C_{i-1\_b}$. When $G_i = 1$, $C_{i\_b} = 0$. Thereby, we have:

$$C_i = C_{i-1}P_i + G_i$$

b. TGs don't have weak 0 or weak 1 problem. We will also see later during the semester that while turned on, The TG's output resistance throughout the whole input voltage swing is relatively constant.

c. There are two transmission-gate based XOR gate. The first stage outputs $A_i \oplus B_i$. The Boolean equation for the circuit is $S_i = A_i \oplus B_i \oplus C_i$. Thereby, we can use it to generate sum bit.

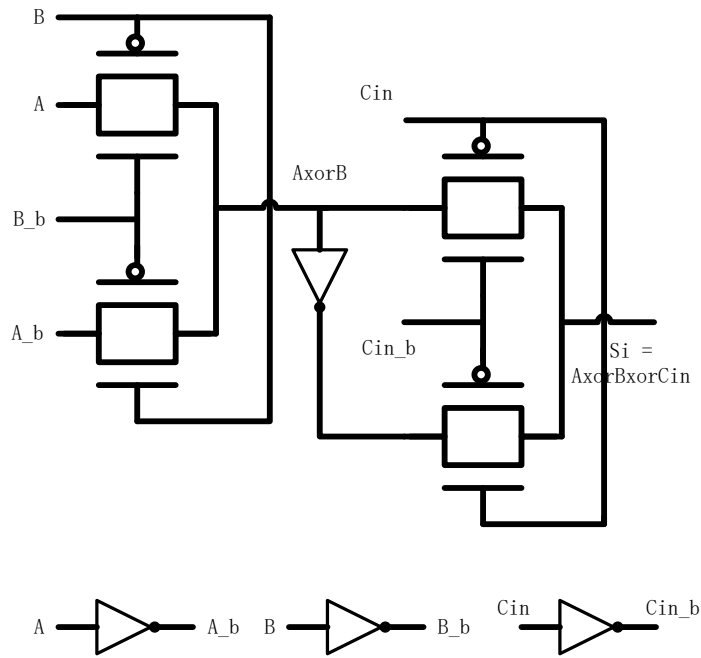

**37.**

a. $Out1 = \overline{In3 + In2 + In1 In0}$, $Out0 = \overline{In2 In1 + In2 In0 + In1 In0}$

b. The Euler path for Out1 and Out0 are listed as follows:

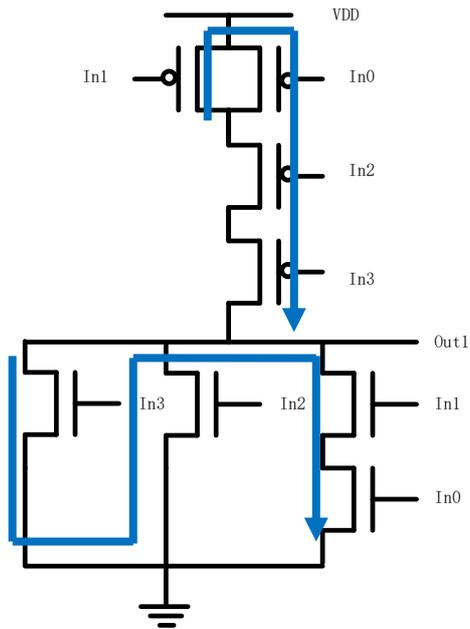

One possible common Euler path is In3->In2->In1>In0

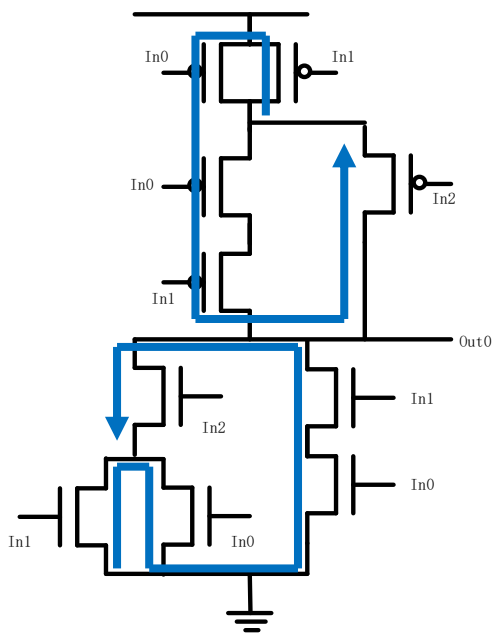

One possible common Euler path is In1->In0->In0->In1->In2

**38.**

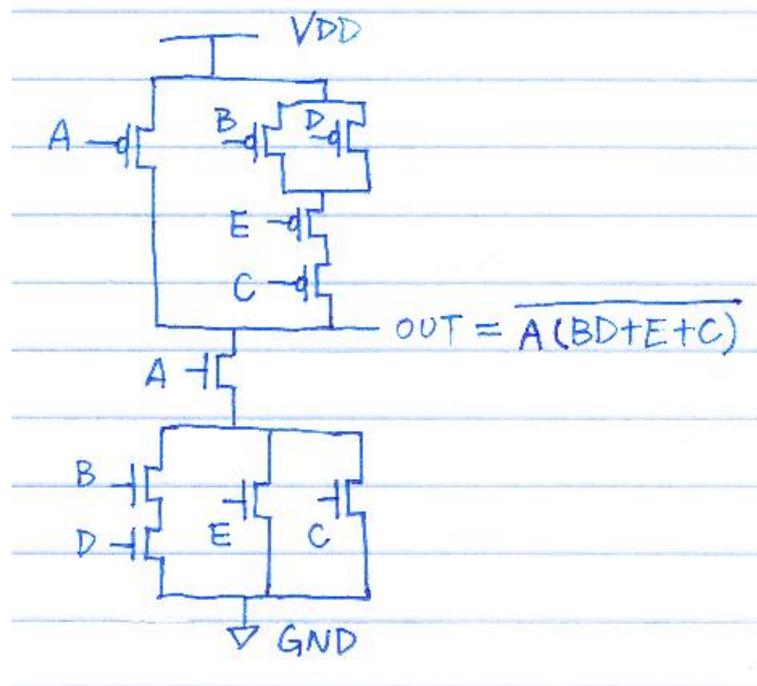

$OUT = \overline{A(BD+E+C)}$

**39.**

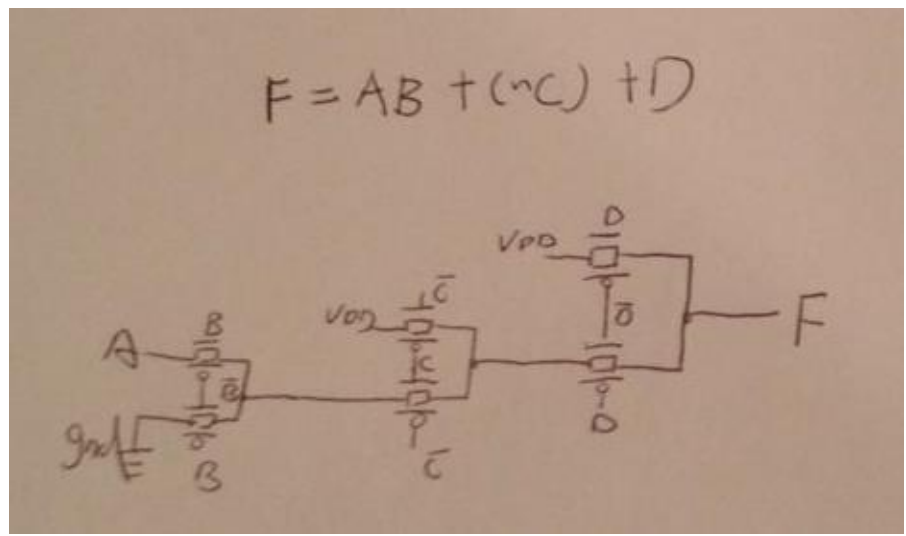

$F = AB + (\sim C) + D$

Shahin Nazarian USC 

# 20.3: Chapter 3 – Sequential Logic

**40.** For HOLD :

Delay in Data path = min(wire delay to the clock input of FF1) + min(Clk-to-Q delay of FF1) +min(cell delay of inverter) + min(2 wire delay- "Q of FF1-to-inverter" and "inverter-to-D of FF2") **=Td = 1+9+6+(1+1)=18ns**
Clock path Delay = max(wire delay from CLK to Buffer input) + max(cell delay of Buffer) + max(wire delay from Buffer output to FF2/CLK pin) + (hold time of FF2)**=Tclk = 3+9+3+2 = 17 ns**
**Hold Slack = Td - Tclk = 18ns -17ns = 1ns**
**Since Hold Slack is positive-> No hold Violation.**
For SETUP :
Delay in Data path = max(wire delay to the clock input of FF1) + max(Clk-to-Q delay of FF1) +max(cell delay of inverter) + max(2 wire delay- "Q of FF1-to-inverter" and "inverter-to-D of FF2")**=Td = 2+11+9+(2+2) = 26ns**
Clock path Delay = (Clock period) + min(wire delay from CLK to Buffer input) + min(cell delay of Buffer) + min(wire delay from Buffer output to FF2/CLK pin) - (Setup time of FF2)
**=Tclk = 15+2+5+2-4=20ns**
**Setup Slack = Tclk - Td = 20ns - 26ns = -6ns.**
**Since Setup Slack is negative -> Setup violation.**

**41.** Max Register to Register Delay

= (clk-to-Q delay of U2) + (cell delay of U3) + (all wire delay) + (setup time of U1)
= 5 + 8 + 3 = 16 ns.
Max freq = 1/16 GHZ

**42.** 1.3ns+5ns+1.1ns+1.15ns > 7.5ns  => setup violation

1 + 0.5 < 0.9ns + 1.15ns  => hold violation

**43.**

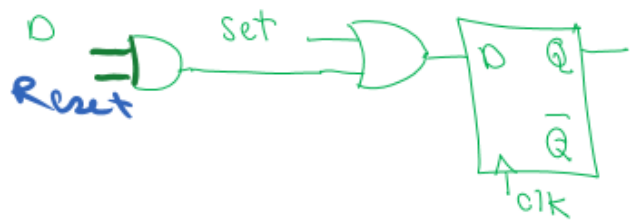

# 20.4: Chapter 4 – Delay

**44.**

a) Delay = RC = 1ns

b) If the wire is cut in half, both R and C are reduced to half. Hence delay of each part of the wire is 25k*10fF = 0.25ns. Buffer has delay of 0.25ns. The total delay is therefore 0.75ns; which is less than 1ns. Hence buffer insertion helped here.

c) Using buffer of 1ns will make the overall delay greater than what it was originally. Hence buffer insertion didn't help this time.

d) 0.33ns+2*.25=0.83ns, therefore it works better than no buffer. However since part (b) resulted a lower delay, one buffer is the optimal number of buffers in this case.

e) it does not help to add 3 buffers, therefore it seems the optimal number of buffers in this case is 2.

**45.** ACC: 24.31ns, DIFF: 21.82ns

**46.** $t_{d\text{-fallingout}} = C * \Delta V / I_{avg\text{-fallingout}} = C * VDD/2 / I_{avg\text{-fallingout}} = 50fF * 0.5V / 10uA = 2.5ns = 250ps$.

$t_{d\text{-risingout}} = C * VDD/2 / I_{avg\text{-risingout}} = 50fF * 0.5V / 8uA = 312.5ps$

**47.** a )

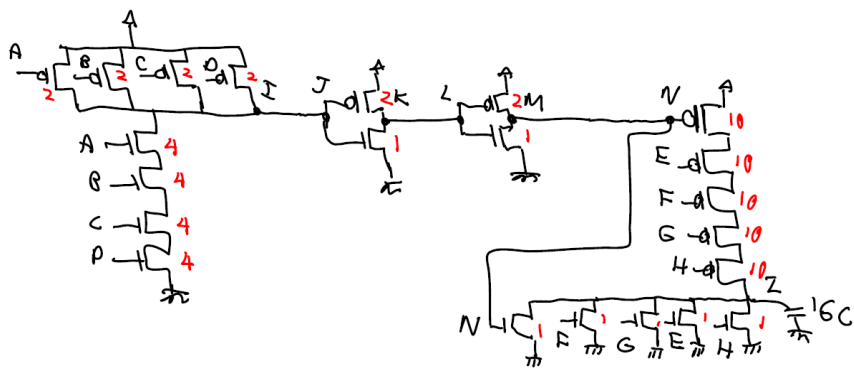

b)

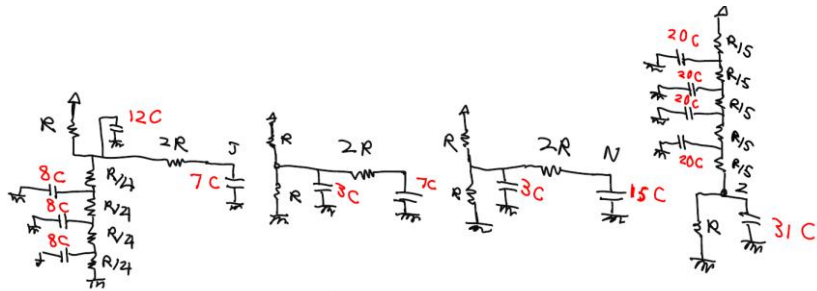

The above delays can be properly added to get the path delay.

C )
NAND:
Falling: ABCD: 1110->1111
Rising: ABCD: 1111->1110

NOR:
Falling: NEFGH: 00000->10000
Rising: NEFGH: 10000-> 00000

### 48.

B to A = 110RC
C to A = 75RC
B to C = 137RC
C to B = 118RC

### 49.

To be conservative, let's use the time constant:
a ) R1.(C1+C2+C3+C4+C5+C6) + R2.C2

b ) R1.(C1+C2+C3+C4+C5+C6) + R3.(C3+C4+C5+C6) + R4.C4

c ) R4(C3+C5+C6+C1+C0+C2) + R6.C6

d ) R5(C3+C4+C6+C1+C0+C2) + R3(C1+C2+C0) + R1(C0)

# 20.5: Chapter 5 – Ring Oscillator

**50.** For the one marked in red:

When OUTPUT is high, it takes 60 + 50 + 40 + 60 + 50 ns = $260ns$ high

When OUTPUT is low, it takes 40 + 50 + 60 + 40 + 50 ns = $240ns$ low.

$T = 240 + 260ns = 500\ ns$.

Also the duty cycle is 260 /(260+240) = 52%

**51.**

$THigh = NTPLH+(N+1)TPHL = N(5)+(N+1)(3) = 8N+3$
$TLow = NTPHL+(N+1)TPLH = N(3)+(N+1)(5) = 8N+5$
$Tout = (2N+1)(TPLH+TPHL) \rightarrow Freq=1/Tout=1/\{(2N+1)(3+5)\ ns\} = 1/(252N+1)$ MHz
$Duty\ Cycle = THigh\ /T = (8N+3)/8(2N+1)$

**52.**

$THigh = 0.45 \times 2ns = 0.9ns$
$TLow = 0.55 \times 2ns = 1.1ns\ (2N+1) = 5 \rightarrow N = 2 \rightarrow Calculate\ the\ equations\ in\ question\ \#2.$
$TLH = 0.3ns$
$THL = 0.1ns$

**53.**

$T = 5 \times (Tdf + Tdr) = 400ns$

$3Tdf + 2Tdr = 90 + 100 = 190ns$

## 20.6: Chapter 6 – Dynamic Logic

**54.**

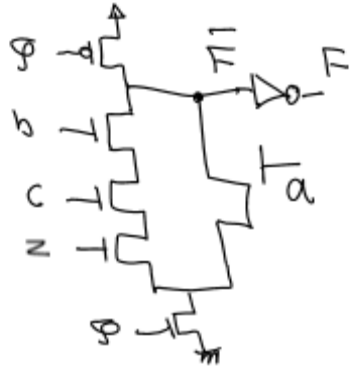

c) There is no unique solution. This is one of the solutions:

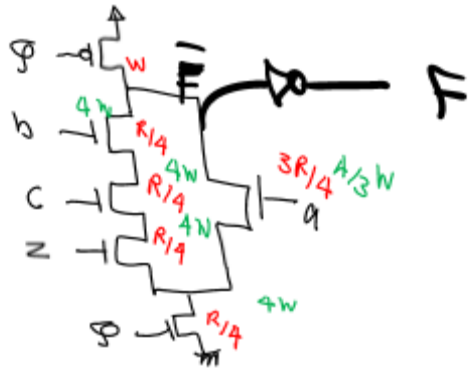

d) There is no unique solution. This is one of the possible solutions:

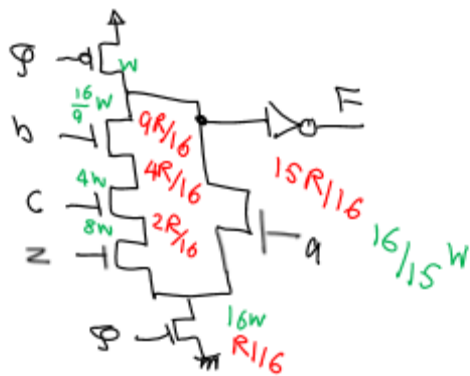

e) Based on the above circuit the input abcz appearing as 0110 results in the worst case charge sharing. It happens only during evaluation. During the precharge the pullup path would prevent any charge sharing issues and output will eventually rise to $V_{DD}$.

f) In the best case, we should not see a charge sharing problem. These are the possibilities: abcz = 1xxx or 00xx or 0111

g) Use a keeper pMOS transistor with size W/L (or even with higher L). Using one precharge transistor for every internal nodes 2 and 3 or using the zipper solution which use a different clock for precharge transistor which during evaluation is set to $V_{DD} - |V_{tpMOS}|$

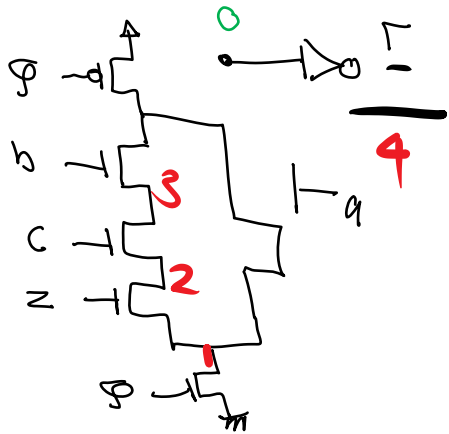

h) $C_1 = (16+8+16/15)C = 25.1$
$C_2 = (8+4)C = 12C$
$C_3 = (4+16/9)C = 5.78C$
$C_O = (1+16/9+16/15+3)C = 6.84C$  // here we assumed the sizing of the static gate is 2,2W and the input gate cap per transistor of size W is C, therefore 3C is the input cap of the static inverter.

i) The degraded voltage (at the output of dynamic) is $C_O / (C_2 + C_3 + C_O) V_{DD}$
$= 6.84/(12+5.78+6.84) V_{DD} = 0.28 V_{DD}$

j) If $0.28 V_{DD}$ is less than $V_{IH}$ (of the static inverter) then noise margin is violated which might result in Boolean error.

k) To simplify the calculation, the sizes are rounded up as show in the following:

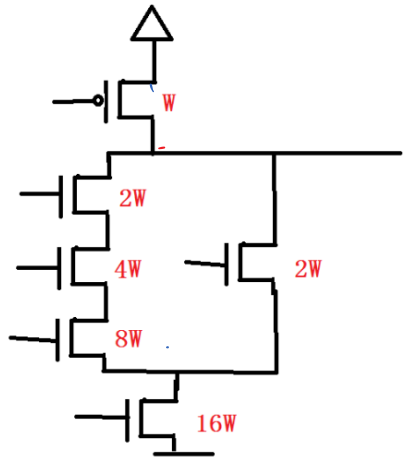

The following figure shows the equivalent RC model of the circuit.

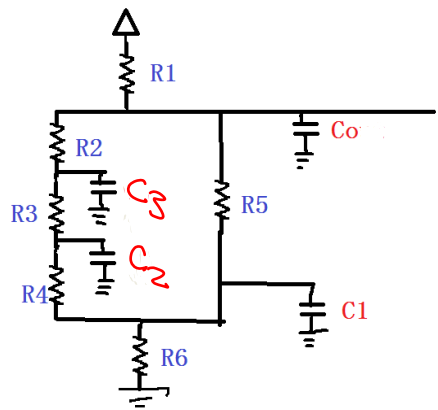

Assuming that every W in NMOS and PMOS results in cap C, then we have:
$C_O=8C$ [assumed 3C for the input of the static inverters] $C_3=6C$ $C_2=12C$
$C_1=26C$

Assuming that the resistance of PMOS with size W is $R1=2R$, resistance of an NMOS with size W is R we then have: $R2=0.5R=R5$ $R3=R/4$ $R4=R/8$ $R6=R/16$

Rising Elmore delay (worst case)
$\tau_r = R1*(C_O+C_1+C_2+C_3) = 2R*(8C+6C+12C+26C)$

Falling Elmore delay (worst case): From the time clock rises (to evaluation) to the time output falls:
$\tau_f = R6(C_1+C_2+C_3+C_O) + R4(C_2+C_3+C_o) + R3(C_3+C_O) + R2*C_O$

k)
For the rising output (during precharge,) the worst happens if abcz, goes from x111 or 111x to 1110.
For the worst case falling output (during evaluation) abcz: x111 or 111x to 1110.

**55.**

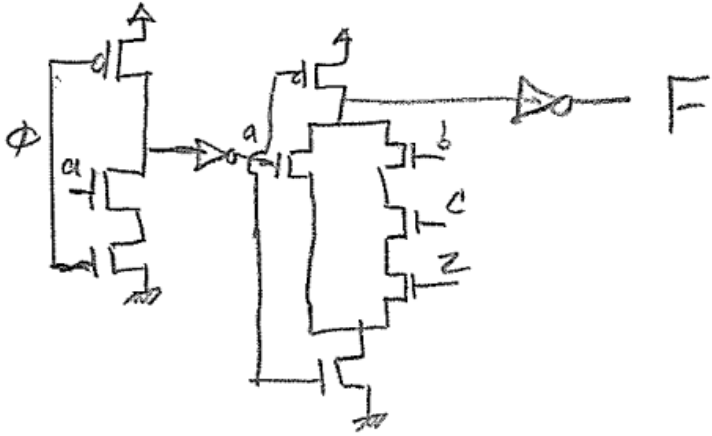

**56.**

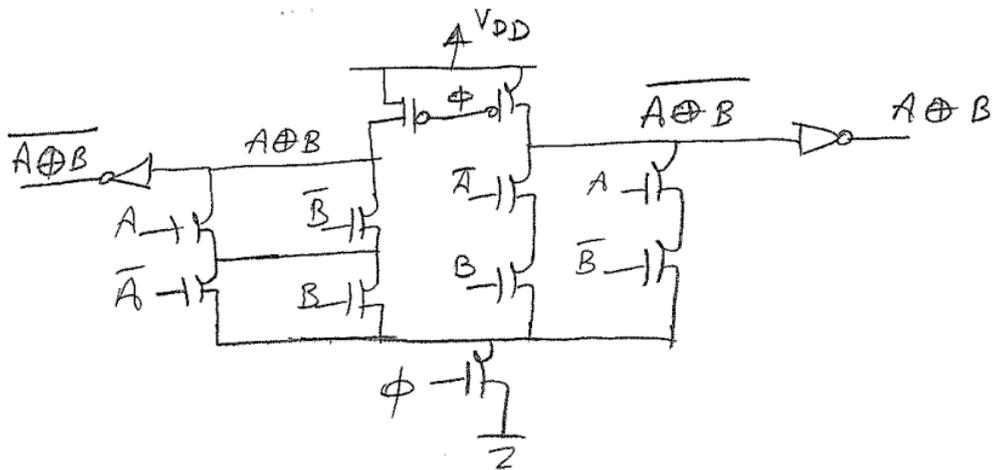

**57.**

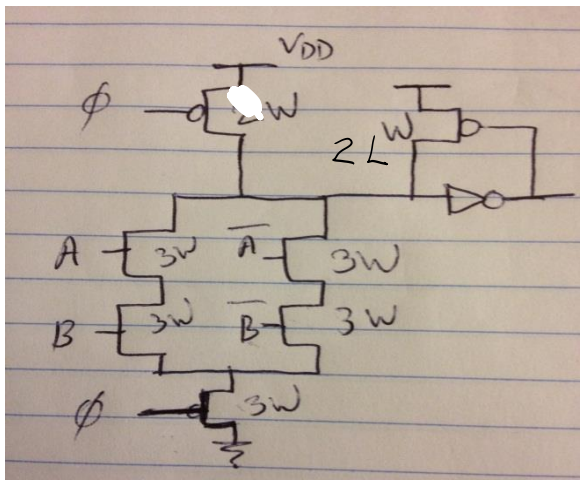

**58.** Design in one or higher number of stages.

Problems: 1) Charge sharing. 2) 1-to-0 transition at the output of the dynamic stage

Solutions:

1) Charge sharing: Multiple Output Domino, 2) using weak PMOS (keeper transistor) to feed the output and 3) Zipper circuit

2) Remove the possibility of 1-to-0 transition at the inputs of the next NMOS logic by adding a static inverter.

**59.**

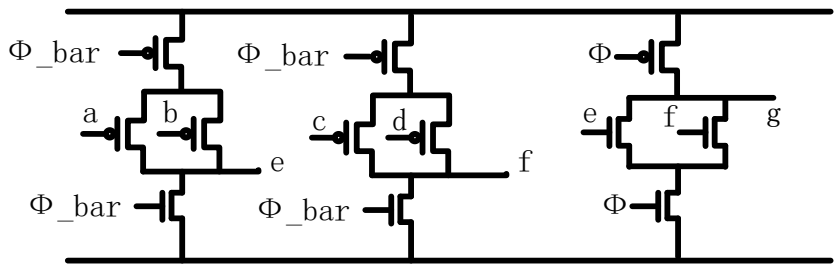

**60.**

The answer is not unique. Note that we accept if you decide to do all functionality in one stage and leave only an inverter for the second stage.

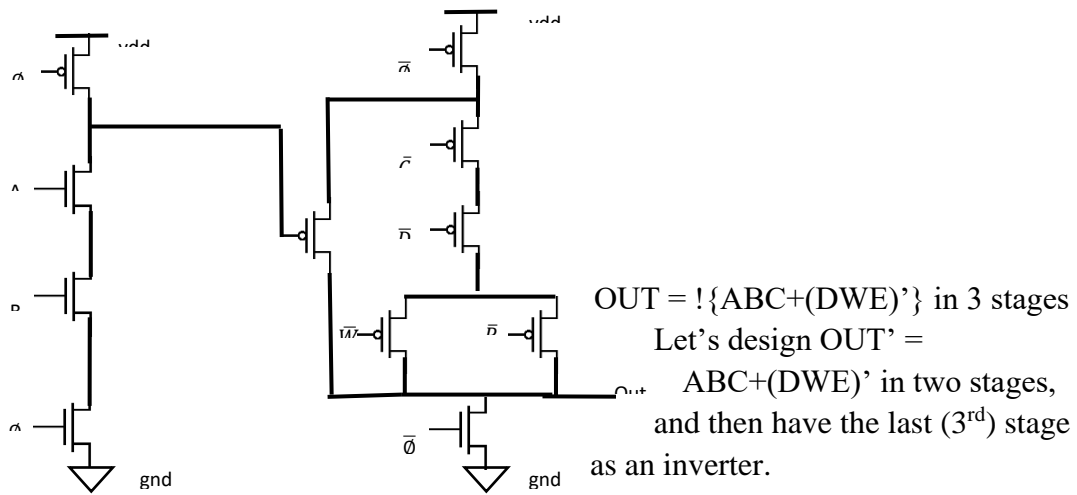

OUT = !{ABC+(DWE)'} in 3 stages
Let's design OUT' = ABC+(DWE)' in two stages, and then have the last (3$^{rd}$) stage as an inverter.

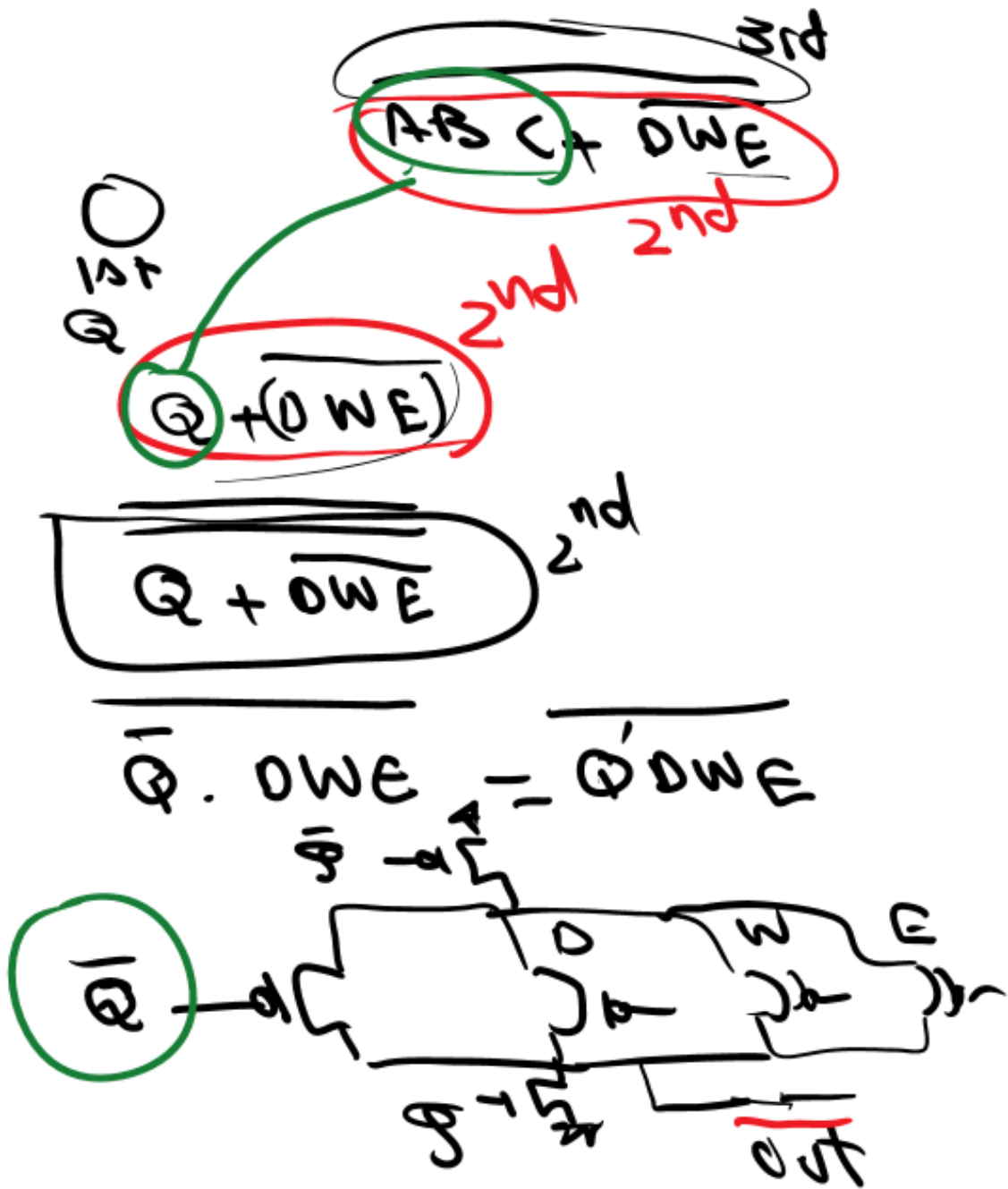

!{ABC+(DWE)'} in 3 stages

1st: $\overline{Q}$

$\overline{Q} = \overline{ABC}$ in 1st

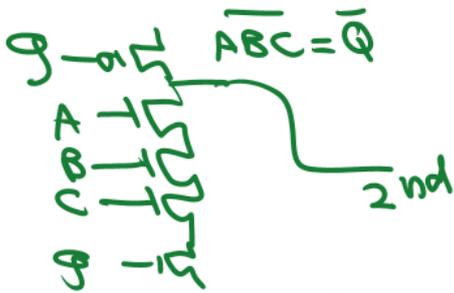

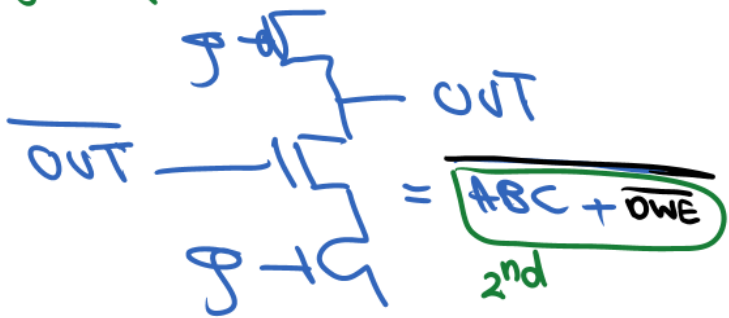

$= \overline{ABC + \overline{DWE}}$   2nd

= !{ABC+(DWE)'} in

$\overline{OUT} = \overline{ABC + \overline{DWE}}$
$= \overline{ABC} + \overline{D} + \overline{W} + \overline{E}$

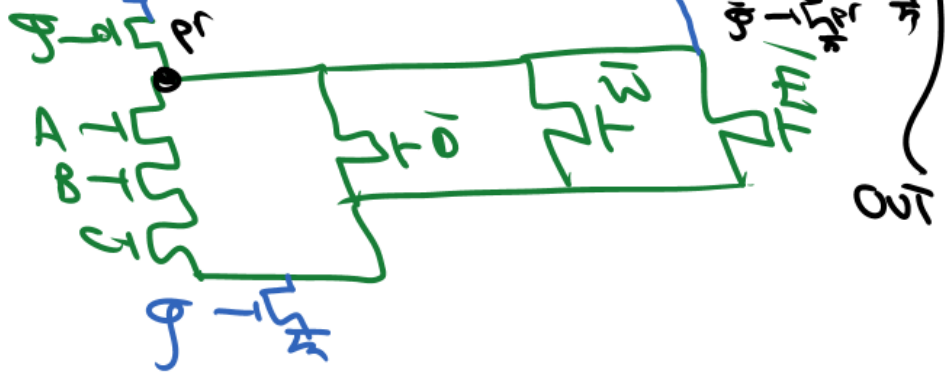

= !{ABC+(DWE)'} in

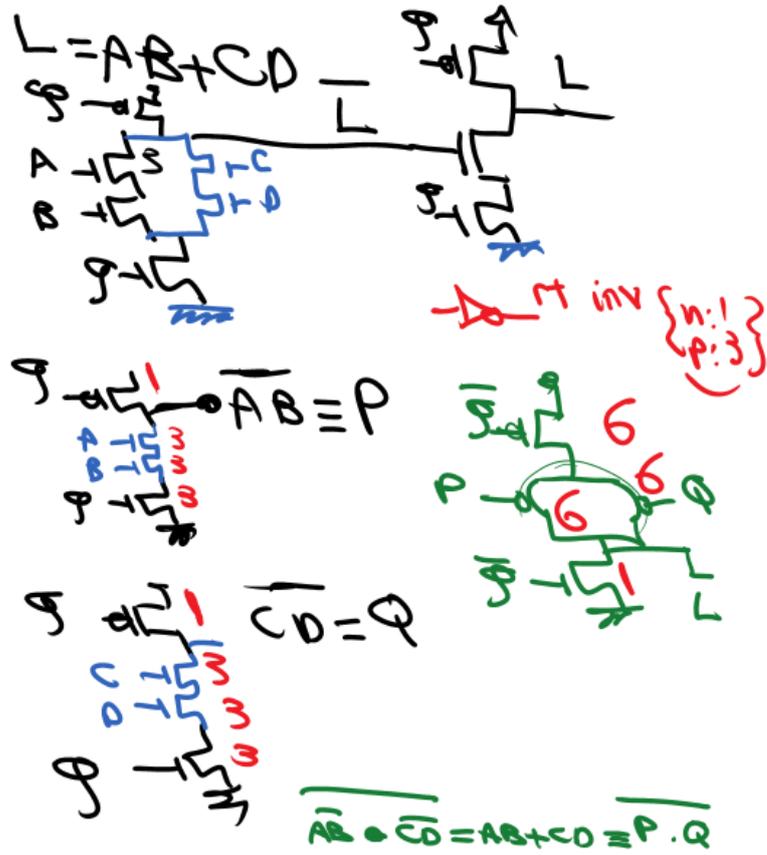

**61.**

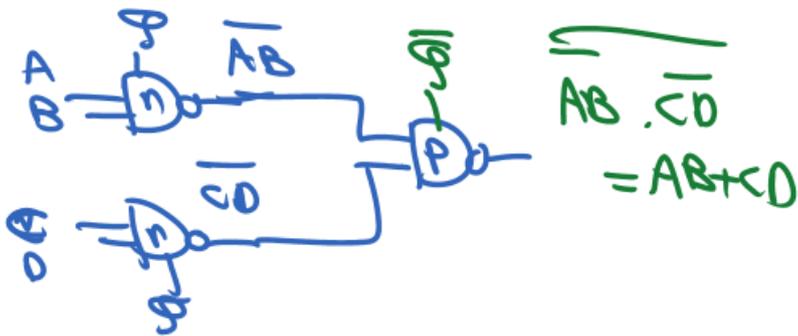

$$S_{477} = P_{477} \oplus G_{476:0}$$
$$G_{476:0} = G_{476:224} + P_{476:224} \cdot G_{223:0}$$

**62.**

**63.**

  1. **FF1 to FF2**

Setup time:
$$\Delta Comb < T \pm \Delta skew$$
$$\Delta Comb = \partial comb = 40 + 27 = 67ps$$
$$67 < T - (24 - 3)$$
$$T > 88ps$$

Hold time:
$$\partial Comb > \Delta Hold \pm \Delta skew$$
$$67 > \Delta Hold - 21$$
$$\Delta Hold < 88ps$$

  2. **FF2 to FF1**

Setup time:
$$\Delta Comb < T \pm \Delta skew$$
$$\Delta Comb = \partial comb = 32ps$$
$$32 < T - (3 - 24)$$
$$T > 11ps$$

Hold time:
$$\partial Comb > \Delta Hold \pm \Delta skew$$
$$32 > \Delta Hold + 21$$
$$\Delta Hold < 11ps$$

  3. **FF3 to FF2**

Setup time:

$$\Delta Comb < T \pm \Delta skew$$
$$\Delta Comb = 67ps$$
$$\partial comb = 27ps$$
$$67 < T - (13 - 3)$$
$$T > 77ps$$

Hold time:
$$\partial Comb > \Delta Hold \pm \Delta skew$$
$$27 > \Delta Hold - 10$$
$$\Delta Hold < 37ps$$

### 4. FF3 to FF4
Setup time:
$$\Delta Comb < T \pm \Delta skew$$
$$\Delta Comb = 12 + 36 = 48ps$$
$$\partial comb = 48ps$$
$$48 < T - (13 - 4)$$
$$T > 57ps$$

Hold time:
$$\partial Comb > \Delta Hold \pm \Delta skew$$
$$48 > \Delta Hold - 9$$
$$\Delta Hold < 57ps$$

### 5. FF4 to FF4
Setup time:
$$\Delta Comb < T \pm \Delta skew$$
$$\Delta Comb = 23 + 12 + 36 = 71ps$$
$$\partial comb = 71ps$$
$$71 < T$$
$$T > 71ps$$

Hold time:
$$\partial Comb > \Delta Hold \pm \Delta skew$$
$$71 > \Delta Hold$$
$$\Delta Hold < 71ps$$

### 6. FF4 to FF1
Setup time:
$$\Delta Comb < T \pm \Delta skew$$
$$\Delta Comb = 18 + 32 = 50ps$$
$$\partial comb = 50ps$$
$$50 < T - (4 - 24)$$
$$T > 30ps$$

Hold time:
$$\partial Comb > \Delta Hold \pm \Delta skew$$
$$50 > \Delta Hold + 20$$
$$\Delta Hold < 30ps$$

For each hold inequality note that it shows the max hold of the capturing FF associated with that inequality.
Also note that T overall has to be at least 88ps for work.
Finally note that this is similar to what you saw in MI, both focusing on skew calculation; however we will not have a complicated circuit like what you see here or what you had in MI, again in MII. Instead, it would be much smaller problem and also the focus will not be on skew.

**64.**

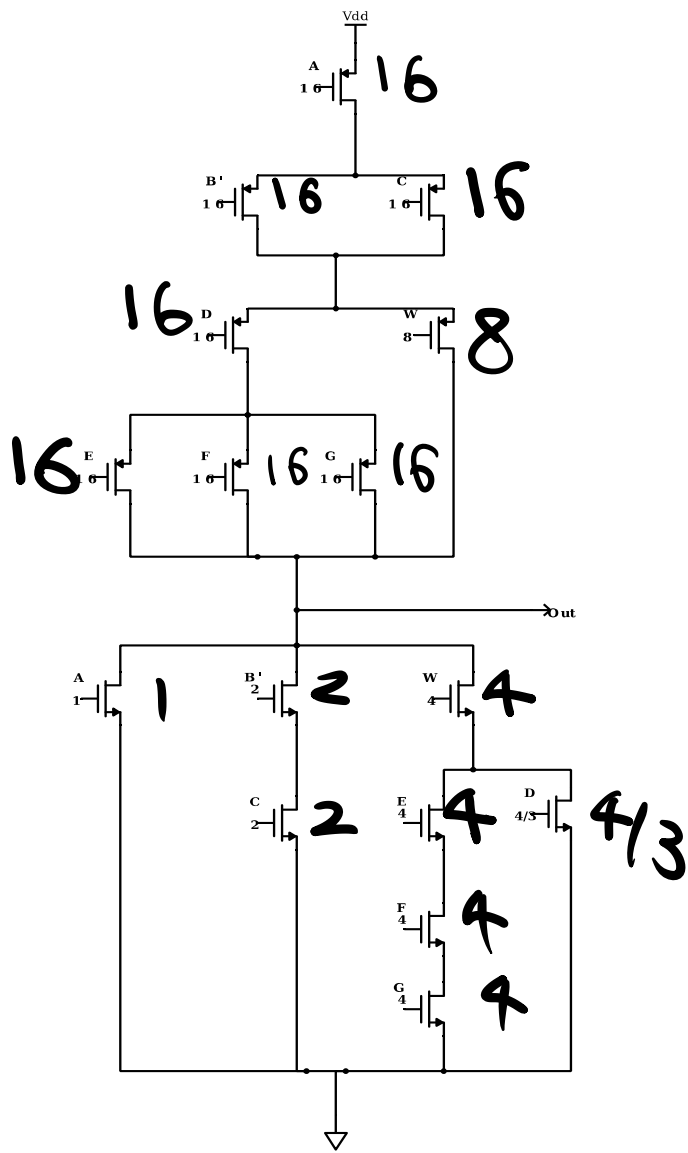

Worst Falling Output:

Shahin Nazarian

|     | A | B' | C | D | E | F | G | W |
|-----|---|----|---|---|---|---|---|---|
| 1st | 0 | 1  | 0 | 0 | 1 | 1 | 0 | 1 |
| 2nd | 1 | 1  | 0 | 0 | 1 | 1 | 0 | 1 |

t= R(7C+4C+ 8C+4/3C+8C+8C+ 56C + 64C + 56C +48C  ) = R (259C + 4/3C) = 781RC/3 ≈ 260RC

Worst Rising Output:

|     | A | B' | C | D | E | F | G | W |
|-----|---|----|---|---|---|---|---|---|
| 1st | 1 | 1  | 0 | 0 | 1 | 1 | 0 | 1 |
| 2nd | 0 | 1  | 0 | 0 | 1 | 1 | 0 | 1 |

t= $RA^p$(48C+56C+64C+63C+4C+28C/3+8C+8C)+ $RC^p$(56C+64C+63C+4C+28C/3+8C+8C)
 +$RD^p$(64C+63C+4C+28C/3+8C+8C) + $RG^p$(63C+4C+28C/3+8C+8C) = 541RC/12

**65.**

a) CMOS circuit diagram:

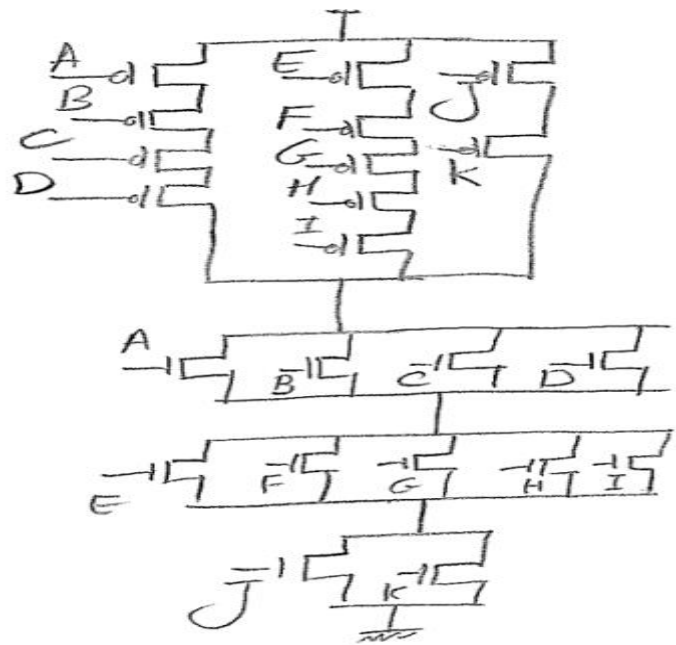

b) Domino CMOS logic circuit:

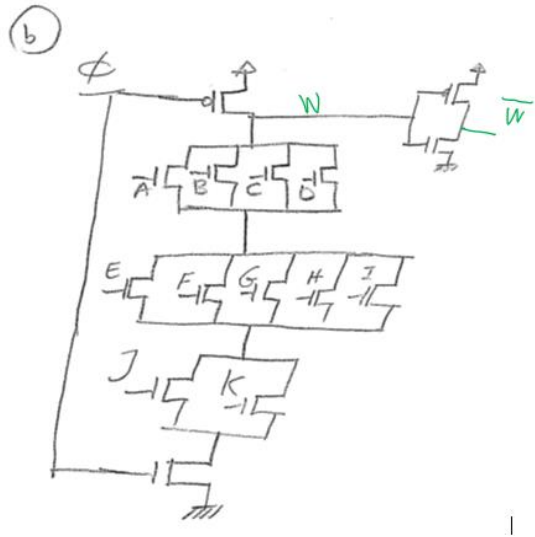

c) Equivalent circuit:

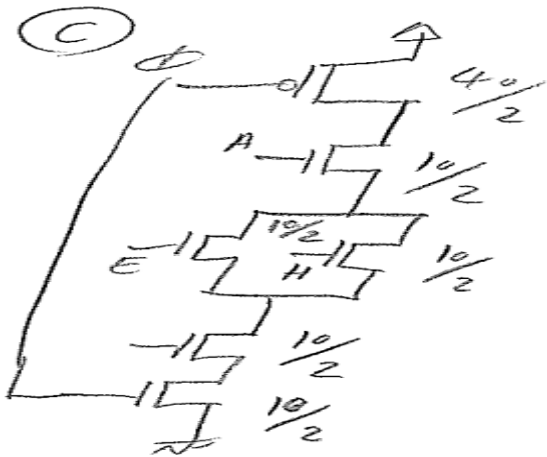

d) For dynamic stage, we first calculate its rising delay.
$I_{dynamic}(V_{out}=0) = 0.5*K'_p*(W/L)_p*(V_{sg}-V_t)^2$ (saturation)
$= 0.5*20uA/V^2*20*(2V-0.4V)^2$
$= 0.512mA$
$I_{dynamic}(V_{out}=0.5V_{dd}) = 0.5*K'_p*(W/L)_p*[2*V_{sd}*(V_{sg}-V_t)-V_{sd}^2] =$
$0.5*20uA/V^2*20*(2*1V*1.6V-1V^2) =$
$0.440mA$
So, the rising delay of dynamic stage is
$t_{dynamic}=C_{load}*0.5V_{dd}/I_{avg}=100fF*1V/0.476mA=210.1ps$
Also, we need to calculate the transition time of dynamic state.
$I_{dynamic}(V_{out}=0.2V_{dd}) = I_{dynamic}(V_{out}=0)$ (saturation) $= 0.512mA$
$I_{dynamic}(V_{out}=0.8V_{dd})=0.5*K'_p*(W/L)_p*[2*V_{sd}*(V_{sg}-V_t)-V_{sd}^2] =$
$0.5*20uA/V^2*20*(2*0.4V*1.6V-0.4V^2) = 0.224mA$

So the transition time of dynamic stage is
$t_{transition}=C_{load}*0.6Vdd/I_{avg}'=100fF*1.2V/0.368mA=326.1ps$
We then focus on the falling delay of the static stage(inverter) with a step input.
$I_{static}(Vout=Vdd)=0.5*k'_n*(W/L)_n*(Vgs-Vt)^2$ (Saturation) $= 0.5*40uA/V^2*5*(2V-0.4V)^2 = 0.256mA$
$I_{static}(Vout=0.5Vdd)=0.5*k'_n*(W/L)_n*[2*Vds*(Vgs-Vt)-Vds2]$
$=0.5*40uA/V^2*5*(2*1V*1.6V-1V2)= 0.22mA$
So the ideal falling delay of the static stage is
$t_{ideal,static}=0.5Vdd*C_{load'}/I_{avg''}=1V*100fF/0.238mA =420.2ps$
The non-ideal delay of static state is $t_{static}=\sqrt{t_{ideal,static}^2+(0.5* t_{transition})^2}=450.7ps$
At last, the total delay count as $t_{total}=t_{dynamic}+t_{static}=660.8ps$

e) Dual rail Domino logic: Note: A-H is the same as A. A-L is A' …

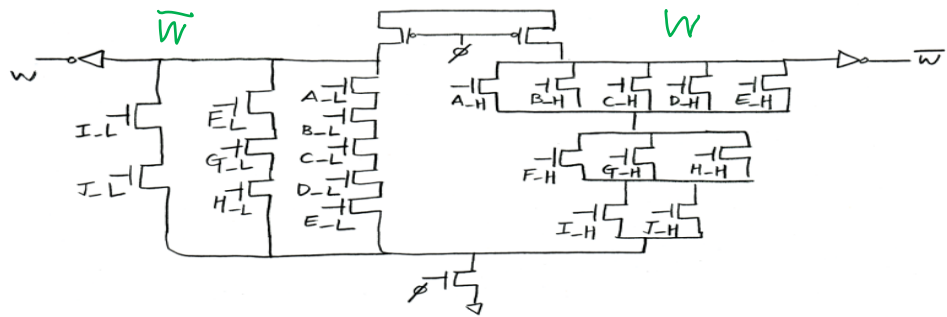

**66.**

Ignore $T_{dq}$
1.3ns+5ns+1.1ns+1.15ns > 7.5ns => setup violation
1 + 0.5 < 0.9ns + 1.15ns => hold violation

# 20.7: Chapter 7 – Super Buffer

**67.**

- $\alpha (\ln \alpha - 1) = 1 \rightarrow \alpha = 3.59 \ (3 \ or \ 4 \ are \ both \ OK)$

- $F = CLoad/Cg = 1000$

- $N + 1 = \ln F \ \ln \alpha = \ln 1000 \ \ln 3.59 = 5.40 \cong 6 \rightarrow N = 5$

*Hence 6 inverters in total, 5 stages.*

**68.**

$$R_{wire} = \frac{9mm}{3 \times 0.125 \mu m} \times 0.025 \Omega/sq = 600 \Omega$$

$$C_{wire} = 9mm \times 50 \ fF/mm = 450 \ fF$$

If no inverter is added:

$$delay = R_{Pull\_up}(2Cd + C_{wire} + 2Cg) + R_{wire}(C_{wire} + 2Cg)$$
$$= (2 \times 40 + 450 + 2 \times 100) \times 10^{-15} \times 1000 + (450 + 2 \times 100) \times 10^{-15} \times 600 = 1.12ns$$

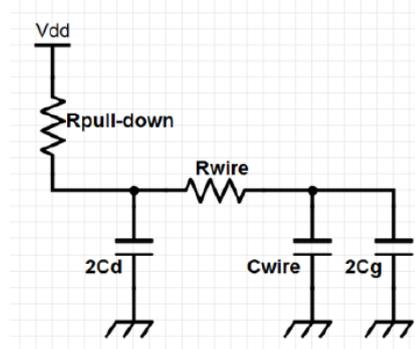

If inverters are added:

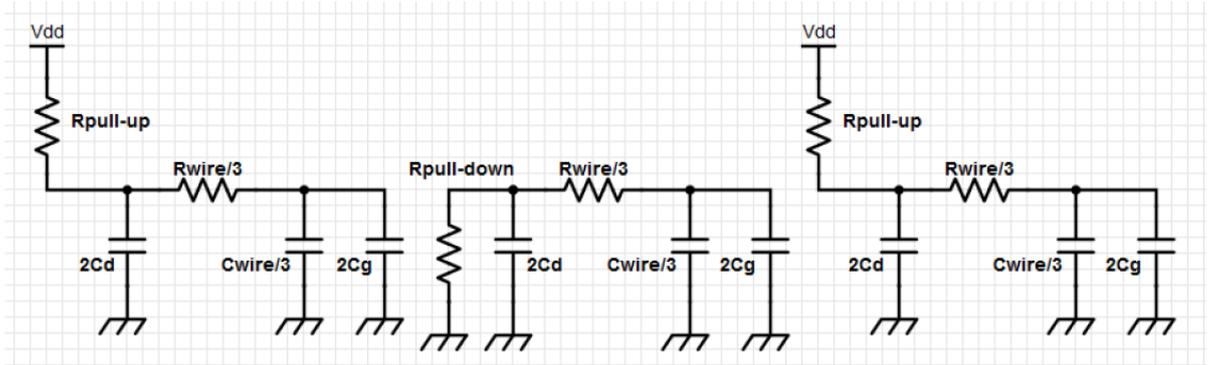

$$delay = \left[ R_{Pull\_up}\left(2Cd + \frac{C_{wire}}{3} + 2Cg\right) + \frac{R_{wire}}{3}\left(\frac{C_{wire}}{3} + 2Cg\right)\right] \times 3$$
$$= \left[(2\times 40 + 150 + 2\times 100)\times 10^{-15} \times 1000 + (150 + 2\times 100)\times 10^{-15} \times 200\right]\times 3 = 1.5ns$$

Adding 2 buffers in this case doesn't help. It worsened the delay by 1.5 ns – 1.12 ns = 0.38 ns

**69.** Fringing cap = .23pF, substrate cap=.032pF => total cap = .262pF.

Total R=2KOhms.

0.9RC=.471ns.

With 1 buffer (inverter) check that the delay is .286ns

With 2 buffers decreases to .257ns

With 3 buffers increases to .268ns. Therefore, we know that 2 buffers give the minimum delay.

## 20.8: Chapter 8 - SRAM

70. Assume during this operation, the storage bit changes from 1 to 0. Also assume the cell will change state for $V_b = V_{T,n} = 0.7V$ (b is the storage node). The initial condition is $V_b = 5V$, M1 is off. At the transition point, for M5:

$V_{GS} = 0 - 5 = -5V$

$V_{DS} = 0.7 - 5 = -4.3V$

Thus M5 is saturated.

For M3:

$V_{GS} = 5 - 0.5 = 4.5V$

$V_{DS} = 0.7 - 0.5 = 0.2V$

$V_{T,n} = 0.7 + 0.4(\sqrt{0.6 + 0.5} - \sqrt{0.6}) = 0.81V$

Thus, M3 is in the linear region.

$I_{D,M5} = I_{D,M3}$

The transistor W/L for M5 and M6 are thus:

$(\frac{10}{2})(W/L)_p(-5 + 0.7)^2 = 20 \left(\frac{2}{4}\right)\left(4.5 - 0.81 - \frac{0.2}{2}\right)0.2$

$(W/L)_p = 0.078$

71.

a)

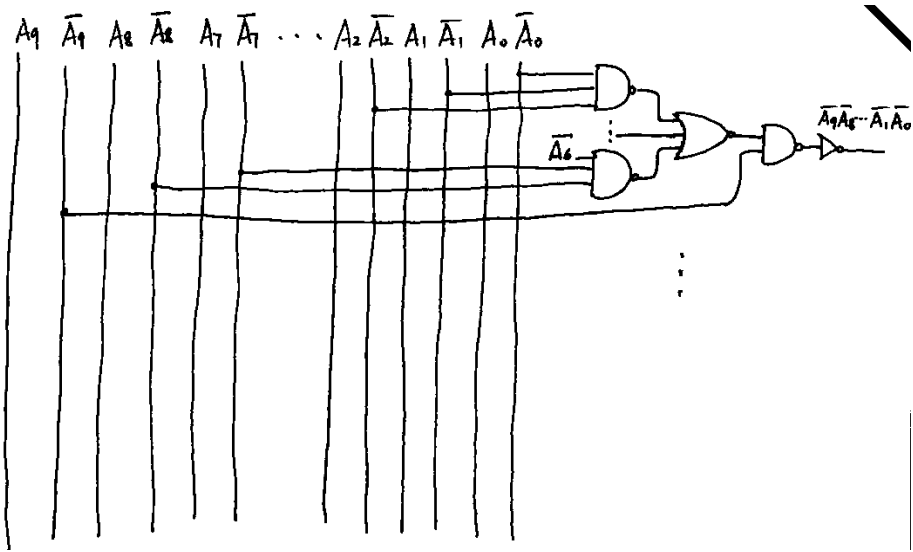

10 inverters + ( 3*3-input NAND + 3-input NOR + 2-input NAND + inverter ) * 2^10

Transistor number = 10 * 2 + ( 3*6 + 3*6 + 2*4 + 2) * 2^10 = 47124

b)

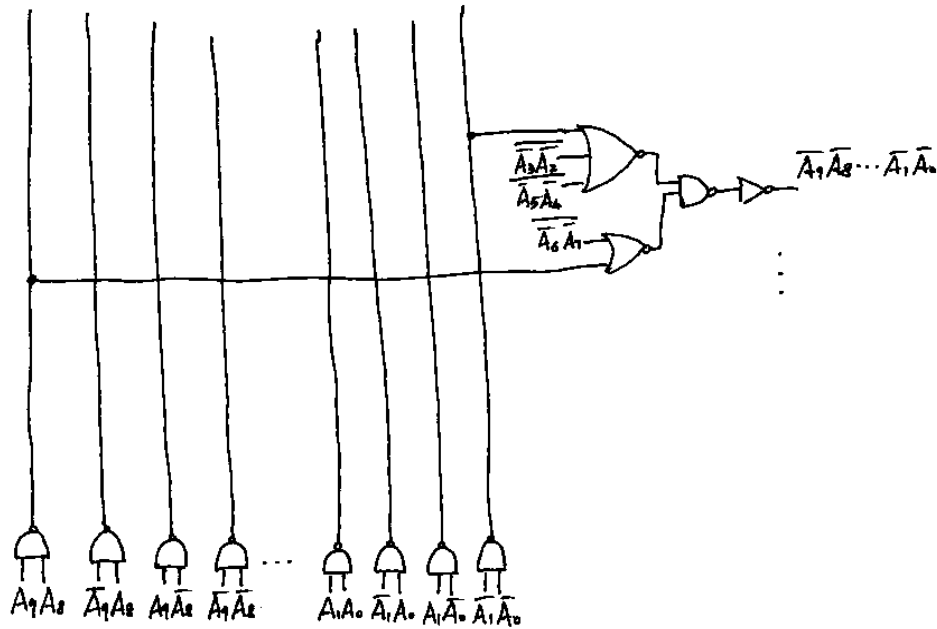

In Pre-Decoding circuitry, we need (4×4) × 5 = 80 transistors for Pre-Decoding part, and 1 3-input NOR, 1 2-input NOR, 1 2-input NAND, and 1 Inverter to decode each address from Pre-Decoding part. Hence total amount of transistors are 80 + {[(1×6) + (1×4) + (1×4) + (1×2)] × 1024} = 16464. Hence 14256 transistors have been saved with Pre-Decoding idea compared to Part (a).

**72. SRAM Memory**

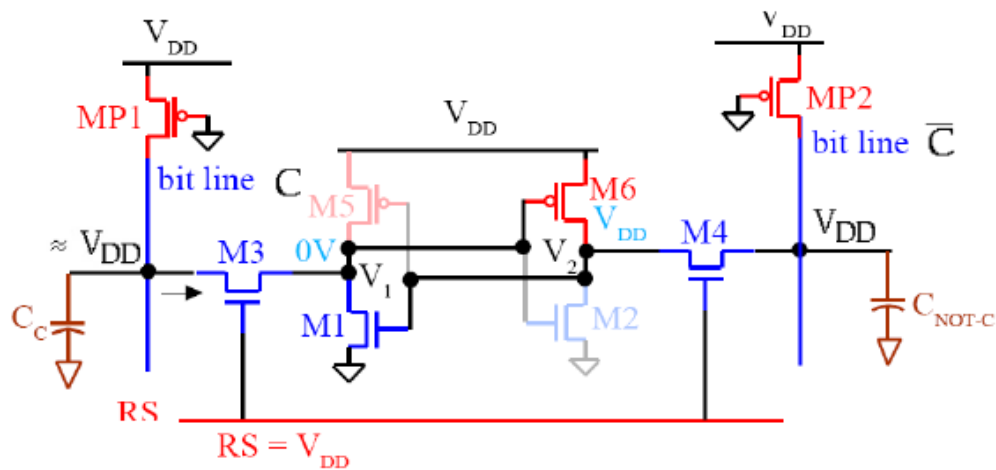

(a) M3 is in saturation mode, and M1 is in linear mode.

$k_{n,3} / 2 \times (V_{DD} - V_1 - V_{t,n})^2 = k_{n,1} / 2 \times [2(V_{DD} - V_{t,n})V_1 - V_1^2]$

Apply the parameter values, We have

$120\mu A/V^2 / 2 \times (2V - V_1 - 0.5V)^2 = 240\mu A/V^2 / 2 \times [2(2V-0.5V)V_1 - V_1^2]$

So $V_1$ = 2.72V (unreasonable answer) or 0.275V.

(b) M3 is in linear mode, and M5 is in saturation mode

$k_{n,3} / 2 \times [2(V_{DD} - V_{t,n})V_1 - V_1^2] = k_{p,5} / 2 \times (V_{DD} - |V_{t,p}|)^2$

Apply the parameter values, We have

$120\mu A/V^2 / 2\times[2(2V-0.5V)V_1 - V_1^2] = 45\mu A/V^2 / 2\times(2V - 0.5V)^2$

So $V_1$ = 2.69V (unreasonable answer) or 0.314V.

### 73. SRAM Memory Solution:

(a) The diffusion cap: $256\times(0.25\mu m\times 1fF/\mu m)$ = 64fF

The wire cap: $256\times 1.5\mu m\times 0.2\mu m\times 0.1fF/\mu m^2 + 2\times 256\times 1.5\mu m\times 0.05fF/\mu m/edge$ = 46.08fF

So the total capacitance loading each bitline in this memory is 64fF + 46.08fF = 110.08fF.

(b) The resistance of the bitline: $0.1\Omega/square\times(256\times 1.5\mu m) / 0.2\mu m = 192\Omega$

$\tau$ = RC / 2 = 110.08fF$\times$192$\Omega$ / 2 = 10.5

**74. SRAM Memory Solution:**

To make the drivability of M3 strong, M6 medium, M4 weak, the size of the three transistors should be: 6w, 4w and 5w for M3, M6 and M4. Why not make M4 4w? Then the ratio of M6 and M3 is 6/5, even weaker than the original 5/4.

**75. SRAM Memory Solution:**

See professor's slide 3-16 in unit 3 (SRAM)

**76. SRAM Memory Solution:**

Pmos, as Pmos passes strong one while Nmos passes weak one.

**77. SRAM Memory Solution:**

For 64×16 structure, decoding circuitry has longest part as NAND-INV-NAND-NOR, and there is no mux part. Hence worst case delay is

4Dgate + 0.69× [(16×17/2)Rword Cword + (64×65/2)Rbit Cbit] = 4Dgate + 93.84Rword Cword + 1435.2Rbit Cbit;

For 32×32 structure, decoding circuitry has longest part as NAND-INV-NAND-NOR, and there is 1 stage of mux. Hence worst case delay is

4Dgate + 0.69× [(32×33/2)Rword Cword + (32×33/2)Rbit Cbit] + Dmux = 4Dgate + 364.32Rword Cword + 364.32Rbit Cbit + Dmux;

For 16×64 structure, decoding circuitry has NAND-NOR, and there is 2 stages of mux. Hence worst case delay is

2Dgate + 0.69× [(64×65/2)Rword Cword + (16×17/2)Rbit Cbit] + 2Dmux = 2Dgate + 1435.2Rword Cword + 93.84Rbit Cbit + 2Dmux.

**78.**

$V_1$ will finally be low $\Rightarrow$ M3 in lin (+1)

M5 in sat (+1)

$\frac{15}{2} \times \frac{4}{3}(-1.8+0.4)^2 = \frac{30}{2} \times \frac{4}{2}\left[2(1.8-0.4)V_1 - V_1^2\right]$

(+1) for writing the correct eqn & KCL as above.

$\Rightarrow 3V_1^2 - 8.4V_1 + 1.96 = 0 \quad (1.4)^2 = 8.4V_1 - 3V_1^2 = 1.96$

$\begin{cases} V_1 = 0.24V \\ V_1 = 2.54 \text{ impossible} \end{cases}$ (+0.5) for a=3, b=−8.4, c=1.96

(−2) for choosing high $V_1$

b) this is the same as (a), therefore there is no need to repeat! All we wanted from students to mention that (b) is the same as (a)

**79.**

N3 is in saturation region and N2 in linear region. KCL at node x:

$$\frac{K_n}{2} \cdot (V_{dd} - V_x - V_{tn})^2 = \frac{K_n}{2} \cdot [2(V_{dd} - V_{tn}) \cdot V_x - V_x^2]$$

$$(0.7 - V_x)^2 = (1.4V_x - V_x^2)$$

$$2V_x^2 - 2.8V_x + 0.49 = 0$$

$$V_x = \frac{1.4 \pm \sqrt{1.4^2 - 2 \times 0.49}}{2} = \frac{1.4 \pm 0.99}{2} = 1.195 \ (unreasonable) \ or \ 0.205$$

The acceptable answer is $V_x$ = 0.205V

**80.**

$M_1$ is Saturation, $M_2$ is Linear

$(k'_n/2)(W/L)_1(V_{DD}-V_Q-V_{TN})^2 + (V_{DD}-V_Q)/R_L \leq (k'_n/2)(W/L)_2[2(V_{DD}-V_{TN})V_Q-V_Q^2]$

$50\mu/2 \times 1.5 \times (2.5-0.5-0.5)^2 + (2.5-0.5)/R_L \leq 50\mu/2 \times 3 \times [2 \times (2.5-0.5) \times 0.5-0.5^2]$

$R_L \geq 42.67 k\Omega$.

**81.**

**Solution:** The important aspects of the specifications of the decoder are as follows.
  i) Number of inputs and outputs: Number of outputs= number of SRAM rows   Number of inputs = $\log_2$ (number of SRAM rows)   **(1 point)**
  ii) Logic function: The decoder must implement a logic function where if the address is the binary encoding of decimal value $i$, then when that address is applied at the decoder inputs then the $i^{th}$ output of the decoder is low and every other of its outputs is high.   **(2 points)**
  iii) Transition from one address to another: Whenever the address value applied at the inputs is changed from $u$ to $v$, then the output that was low when address $u$ was applied must rise before the output that becomes low for address $v$ falls. In general, at any time no more than one decoder output must be low.
  This is required to avoid any possibility of corruption of stored values by ensuring that at no time, even for a short duration, the values stored within SRAM cells in different rows interact with each other via *BIT* and $\overline{BIT}$ lines.   **(2 points)**
  iv) The load capacitance on each decoder output is equal to the gate capacitance of (2×number of bits in each row) *p*MOS transistors driven by the word-line plus the routing capacitance of the word line.   **(1 point)**

**82. (this is just an example):**

| Unused | Row | Bank | Col. | Chip (byte) |
|---|---|---|---|---|
| A31…A29 | A28…A15 | A14, A13 | A12…A3 | A2…A0 |
| 000 | 00 0000 1001 1110 | 00 | 11 0101 1011 | 000 |

Note: A2 to A0 can be used to address a certain byte (chip) among the 8 bytes coming from the 8 chips.

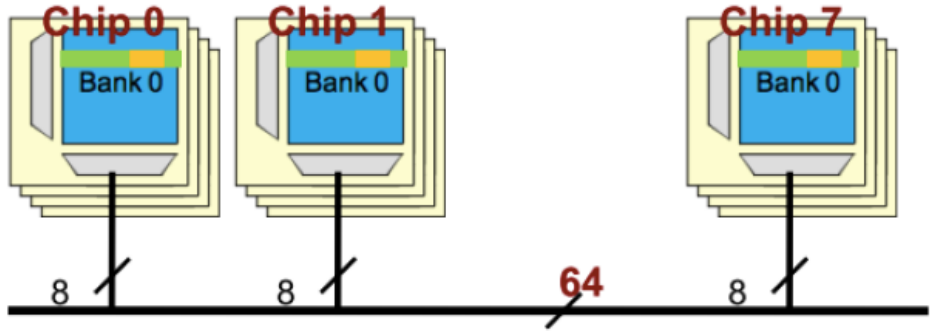

## 20.9: Chapter 9 – Logic Effort

**83.**

(1 point) $R_{AB} = \dfrac{1}{3} + \dfrac{1}{5} = \dfrac{8}{15}$

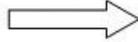

Path AB

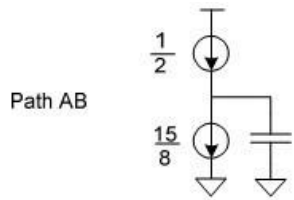

(2 points) $EF_{AB} = \dfrac{1}{\dfrac{15}{8} - \dfrac{1}{2}} = \dfrac{8}{11}$

(2 points) $EF_C = \dfrac{1}{5 - \dfrac{1}{2}} = \dfrac{2}{9}$

Effective Resistances

$g_{A,falling} = \dfrac{EF_{AB} \times 3}{1 \times 3} = \dfrac{8}{11}$ (1 point)

$g_{B,falling} = \dfrac{EF_{AB} \times 5}{1 \times 3} = \dfrac{40}{33}$ (1 point)

$g_{C,falling} = \dfrac{EF_C \times 5}{1 \times 3} = \dfrac{10}{27}$ (2 point)

$P_{AB,falling} = \dfrac{\dfrac{8}{11} \times (1+3+5)}{1 \times 3} = \dfrac{24}{11}$ (2 points)

$P_{C,falling} = \dfrac{\dfrac{2}{9} \times (1+3+5)}{1 \times 3} = \dfrac{2}{3}$ (2 points)

$P_{rising} = \dfrac{2 \times (1+3+5)}{1 \times 3} = 6$ (0.5 point)

$g_{A,rising} = \dfrac{2 \times 3}{1 \times 3} = 2$ (0.5 point)

$g_{B,rising} = \dfrac{2 \times 5}{1 \times 3} = \dfrac{10}{3}$ (0.5 point)

$g_{C,rising} = \dfrac{2 \times 5}{1 \times 3} = \dfrac{10}{3}$ (0.5 point)

**84.**
G = (5/3) * (4/3) * (4/3) = 2.96, F = GBH=2.96*512*20=30310.4
P = 3 + 1 + 2 + 1 + 2 + 1 + 1 + 1 = 12
D = NF^(1/N) + P = 41.1

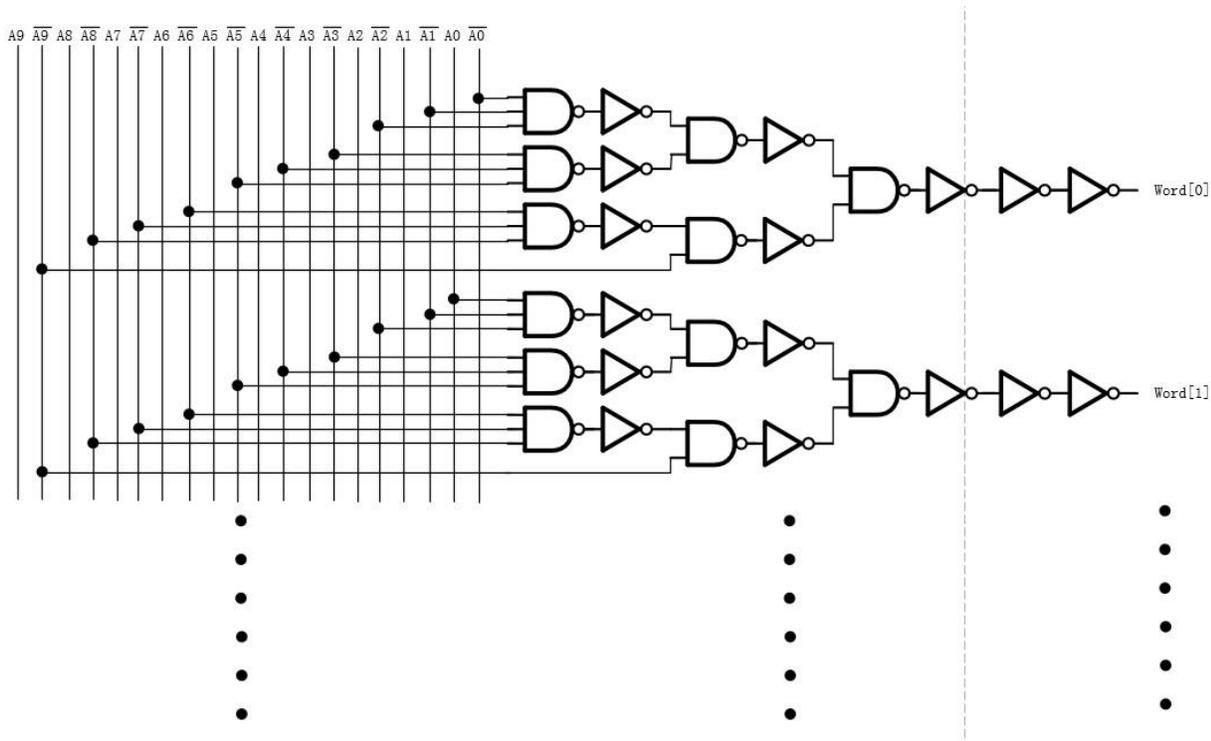

**85.**
We only need to consider 2 successive inverters, one for rising transition and the other for falling transition. Use design 1) as the template For design 1) g_rise=g_fall=1, p_rise=p_fall=1, h=1,total delay = 4. For design 2) R_rise=2R, R_fall=R, h=1, g_rise=p_rise=2*2/3=4/3, g_fall=p_fall=2/3, h=1, total delay = 4. So the 2 designs have the same delay

**86.**
For the total 3 stages, we have H=64C/C=64. Each stage is an inverter, so p1=p2=p3=1, g1=g2=g3=1. So we need to minimize h1+h2+h3. H=h1*h2*h3=64. When h1=h2=h3=4, the total delay can be minimized, hence the 3 inverters are sized as 1, 4 and 16.

**87.**
(1) 8-NAND: 10/3;  INV: 1;  4-NAND: 2;  2-NOR: 5/3;  2-NAND: 4/3;  2-NAND: 4/3;
(2) (a) $2(3.33H)^{1/2} +9$ ; (b)  $2(3.33H)^{1/2} +6$  ; (c)  $4(2.96H)^{1/4} +7$  ;
(3) For H =1, (b) is optimal; for  H=12, (c) is optimal

**88.**
(a) $\frac{n(n+\mu)}{1+\mu}$ (b) $\frac{n+\mu}{1+\mu}$ (c) $\frac{n(1+n\mu)}{1+\mu}$ (d) $\frac{1+n\mu}{1+\mu}$

**89.**
$$g_{rA} = \frac{2R \times 3C_g}{R \times 3C_g} = 2, \quad p_{rA} = \frac{2R \times 6C_d}{R \times 3C_g} = 4$$
$$g_{rC} = \frac{2R \times 3C_g}{R \times 3C_g} = 2, \quad p_{rC} = \frac{2R \times 6C_d}{R \times 3C_g} = 4$$
$$g_{fA} = \frac{3R \times 3C_g}{R \times 3C_g} = 3, \quad p_{fA} = \frac{3R \times 6C_d}{R \times 3C_g} = 6$$
$$g_{fC} = \frac{2R \times 3C_g}{R \times 3C_g} = 2, \quad p_{fC} = \frac{2R \times 6C_d}{R \times 3C_g} = 4$$
So, $d_f = 2 \times \frac{3C_g}{3C_g} + 4 + 2 \times \frac{20C_g}{3C_g} + 4 = 23.33$
$d_r = 3 \times \frac{3C_g}{3C_g} + 6 + 2 \times \frac{20C_g}{3C_g} + 4 = 26.33$

**90.**

H = 4.5. Thus F = GBH = 64, and D^ = 3(64)1/3 + 3(2pinv) = 18.0 delay units
**91.**

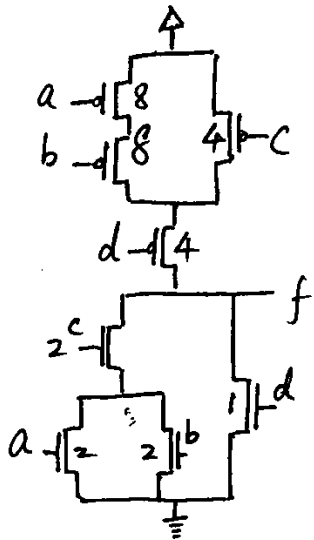

| Input | $C_{in}$ | g | $C_p$ | p |
|---|---|---|---|---|
| a | $10C_g$ | 10/3 | | |
| b | $10C_g$ | 10/3 | $7C_d$ | $7C_d/3C_g$ |
| c | $6C_g$ | 2 | | |
| d | $5C_g$ | 5/3 | | |

**92.**
a)
g1=2, p1=6
H=1650/20
g2=1, p2=1
D=32.7
b)
4 stages
D=23.34
c)

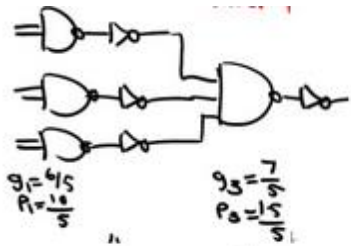

$$\hat{D} = 4\sqrt[4]{6/5 \times \frac{7}{5} \times \frac{1650}{20}} + 2+1+3+1 = 13.72 + 7 = 20.72$$

As far as I checked this is the best. Give 4 extra if all the G, H, and P values are calculated correctly. Give +1 more if the delay value approximated to be close to 21.

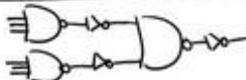

This gives the same delay, however it has more transistor (area) ⇒ not the best design.
⇒ Give 2 out of 5 extra credit points

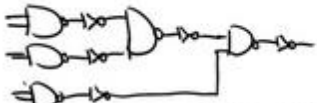

6/5
10/5
higher delay ⇒ give 0.5 out of 5 for creativity!

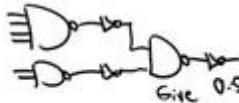

higher delay ⇒
Give 0.5 extra credit for creativity

d)

$C_{in} = 2C_g$
$R_{fall} = 6R \Rightarrow g_f = \frac{6 \times 2}{5}$
$R_{rise} = 4R \Rightarrow g_r = \frac{4 \times 2}{5}$
rising out = $g_1 f \cdot g_2 r = 12/5 \times 8/5$ ⇒ rising out is the worst
falling out = $g_1 r \cdot g_2 f = 8/5 \cdot 12/5$
$p_1 f = 6 \cdot \frac{7}{5} = 42/5$
$p_1 r = 4 \cdot 7/5 = 28/5$

$C_{in} = 2C_g$
$R_{fall} = R$  $g_f = \frac{2}{5} = 0.4$
$R_{rise} = 4R$  $g_r = \frac{4 \times 2}{5} = \frac{8}{5} = 1.6$

$g_1 f h_1 + p_1 + g_2 r h_2 + p_2$
12/5 * 2.5/2a + 42/5 + 8/5 * 1650.5/2 + 8/5 = 1333.4

(+2) for $C_{out} = 1650$   (+2) for $C_{in} = 2$

**93.**
1) G= 2(4/3)(4/3)(5/3)=160/27

B=4*2=8
H=192/10
P=4+2+2+3=11
Delay= $N(GBH)^{1/N} + P = 4[(160/27) \times 8 \times (192/10)]^{1/4} + 11 = 32.97$
$$g_1h_1 = g_2h_2 = g_3h_3 = g_4h_4 = (GBH)^{1/N} = 5.49$$
2)
Begin from load and go backward and calculate the input capacitance for each gate.
$C_{in4} = 58.29C$
$C_{in3} = 28.31C$
$C_{in2} = 6.87C$
$N_{opt} = \log(GBH)/\log 3.59 = 5.33$
For N=6, delay=31.46
We have to add two inverters (one buffer)

**94.**
Design 1: F=GBH = [(6/3)*1*(4/3)*1]*(16*8)*9 = 3072
Design 2: F = GBH = [(4/3)*1*(6/3)*1]*(2*64)*9 = 3072
For an FO4 inverter, d = gh + p = 1*4 + 1 = 5
The logical effort delay for both designs are:
d = $4*(F)^{1/4}$ + 4 + 1 + 2 + 1 = 37.78
So the propagation delay is approximately:
100ps / 5 * 37.78 = 756ps
According to the condition, the gates of Design 2 can be sized as follows:
NAND2: $W_p = 2$, $W_n = 4$
INV: $W_p = 4$, $W_n = 1$
NAND4: $W_p = 2$, $W_n = 8$
So F=GBH = [1*(5/6)*(10/6)*(5/6)]*(64*2)*9 = 1333.33
P = 4/3 + 5/6 + 8/3 + 5/6 = 17/3 = 5.67
d = $4*(F)^{1/4}$ + 5.67 = 29.84
So the propagation delay is approximately:
100ps / 5 * 29.84 = 596.6ps

**95.**
If the upper inverter has size x, and the lower 100-x and the second upper inverter has the same stage effort as the first, the least delays are: D = $2(300/x)^{1/2}$ + 2 = 300/(100-x) + 1, hence x = 49.4, D = 6.9. The sizes for the upper inverters are 49.4 and 121.7, and for the lower inverter it is 50.6

**96. Solution**
We only need to consider 2 successive inverters, one for rising transition and the other for falling transition.
Use design 1) as the template
For design 1)
g_rise=g_fall=1, p_rise=p_fall=1, h=1
total delay = 4
For design 2)
R_rise=2R, R_fall=R, h=1
g_rise=p_rise=2*2/3=4/3
g_fall=p_fall=2/3
h=1
total delay = 4
So the 2 designs have the same delay

**97.**
We are given the following configuration. We must determine m, the scaling factor of each inverter template, and, finally, the width of every transistor.

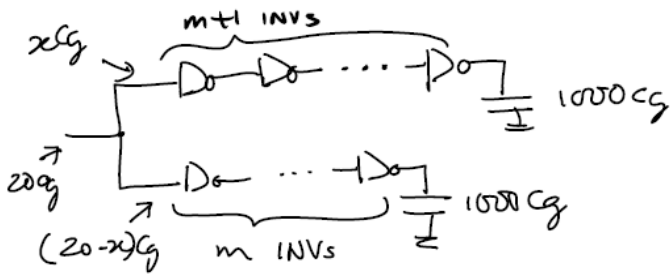

As described in the lectures, we will first use some approximation to estimate the value of m. Then we can designate the input capacitance of the first inverter in the long-branch-of-fork (lbf) as xCg. This gives the input capacitance of the first inverter in the short-branch-of-fork (sbf) as (20 – x)Cg. Using these input capacitance values, we can view each branch of the fork nearly independently, since they do not interact with each other except that their input capacitances must be constrained to a total of 20Cg. The optimal worst-case delay of the long-branch-of-fork for a given value of x can be written as:

$$\widehat{D}_{\text{lbf}}(x) = (m+1)\left[\frac{1000 C_g}{x C_g}\right]^{\frac{1}{m+1}} + (m+1).$$

Similarly, the delay of short-branch-of-fork, can be written as:

$$\widehat{D}_{\text{sbf}}(x) = m\left[\frac{1000 C_g}{(20-x) C_g}\right]^{\frac{1}{m}} + m.$$

The worst case delay of the fork is: $D_{\text{fork}} = maximum[\widehat{D}_{\text{lbf}}(x), \widehat{D}_{\text{sbf}}(x)]$.

Hence, for the selected value of m, we need to iterate over various values of x, to minimize Dfork.

First, I start with the following approximation: Ignore the fact that the two branches of the fork have different numbers of inverters, and assume that each branch has m inverters. This approximation gives us the following approximate view of the circuit.

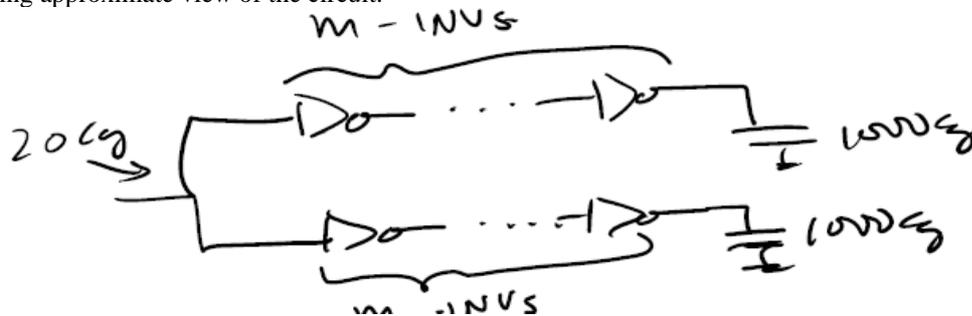

For this approximate view, we get G = 1, B = 2, H = 1000Cg/20Cg which gives GBH =100. We can now estimate the number of stages for this approximate view, m, by using the result derived (and the text) that shows that, ideally, the number of stages should be such that the effort per stage should be ≈ 3.59 (or in the range from 2-to-5, if practical
constraints make it necessary to deviate from 3.59). For the above approximate view, we get m = log3.59(GBH) = log3.59(100) = 3.6.

Since m must be an integer, we can try m = 3 and m = 4. (This is one common type of the practical constraints mentioned above for why we may need to use 2-to-5 instead of 3.59.)

In the second step, we iteratively explore different values of x to minimize Dfork as detailed above. For m = 3, we have (4, 3) fork. For this fork, the iterations to obtain optimal value of x is summarized below.

| x | $D_{lbf}(x)$ | $D_{sbf}(x)$ | $D_{fork}(x)$ | Next iteration? |
|---|---|---|---|---|
| 10 | 16.6 | 16.9 | 16.9 | Yes. Try lower x, since Dlbf(10) < Dsbf(10). |
| 9 | 17.0 | 16.5 | 17.0 | Yes. Try higher value of x, since we have gone too far. |
| 9.5 | 16.8 | 16.7 | 16.8 | Stop, since Dlbf(9.5) and Dsbf(9.5) are close. |

A word on approximations. Since we started with an approximate view of the circuit, when we arrive at a solution, like the one mentioned above, namely $m = 3$ and $x = 9.5$, we can check whether it satisfies our basic criterion. To this end, in the worksheet, we can compute the stage effort for each branch of the fork (shown as $\hat{f}_{lbf}$ and $\hat{f}_{sbf}$ in the worksheet). Since these values are within the 2-to-5 range for our final solution, we can stop. However, had that not been the case, we would return to the first step, adjust the choice made using the approximation (the value of $m$ in our case), and repeat the entire process.

In the third step, having obtained the value of $m$, $x$, $\hat{f}_{lbf}$, and $\hat{f}_{sbf}$ in previous two steps, we can now compute the input capacitance of each inverter in the design, as shown in the following figure.

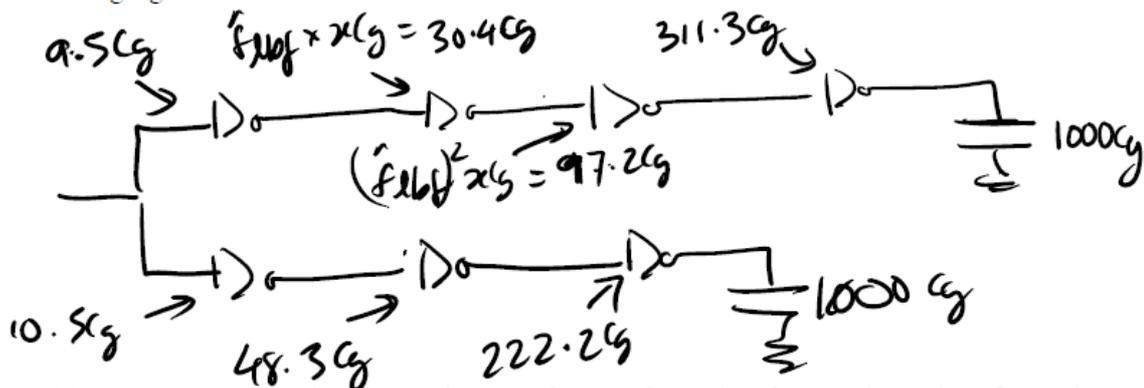

Now we can use the input capacitance values to determine the transistor sizes for each inverter. In particular, for our template inverter, the nMOS and pMOS transistors have widths 1 and 4, respectively, for total $C_{in}$ of $5C_g$. Since the first inverter in the long-branch-of-fork has input capacitance of $xC_g = 9.5$, we get a scaling factor of $9.5/5 = 1.9$. This in turn leads to the transistor sizes of $1.9\times1$ and $1.9\times4$ for the nMOS and pMOS, respectively. The computation of the width of every other transistor in the following design is left as an exercise. (Any difference between the capacitance values in the figure above and figure below is due to rounding.)

Finally, the reader is invited to use the input capacitance values to compute to size the transistors.

# 20.10: Chapter 10 – Power Optimization

**94. Dynamic Power**
1) p(F) = p(AB=1) + p(AB!=1 && C=1) = 0.2*0.2 + (1-0.2*0.2) * 0.667 = 0.68.
Therefore, β=2 * p(F) * (1-p(F)) = 0.4352
P = 0.5 $V_{DD}^2$ f $C_L$ β = 15.67 μW.
2) When A=B: F = B*B+C = B+C; When A=C': F = BC' +C = B+C. So whatever A is, F=B+C.
p(F) = 1 - (1-p(B))(1-p(C)) = 0.733.
β=2 * p(F) * (1-p(F)) = 0.391
P = 0.5 $V_{DD}^2$ f $C_L$ β = 14.08 μW.

**95. Power Supply Scaling**
   a) Due to the equation in slide page 7 and page 17, the switching power reduction factor is $(1.0/0.5)^2$=4, the short circuit power reduction factor is $(1.0*(1.0-2*0.2)^2)/(0.5*(0.5-2*0.2)^2)$=72.
   b) The delay will increase a lot.

**96. Leakage Power**
Compared with dynamic power, leakage power does not decrease so much with the reduction of supply voltage, so it plays a more important role in when the supply voltage is low.

**97. Dynamic Power**

   a) Make product terms mutually non-intersecting. $F = AB + AB'C + A'BCD$

   So, $S_F = S_A S_B + S_A(1-S_B)S_C + (1-S_A)S_B S_C S_D$ = 0.197 and β = $2S_F(1-S_F)$ = 0.3163
   $P_{dynamic}$ = 0.5 $V_{DD}^2$ f $C_{load}$ β = 11.38 μW

   b) $P_{sc}$ = 32.5nW

**98. Adiabatic Logic**
a)

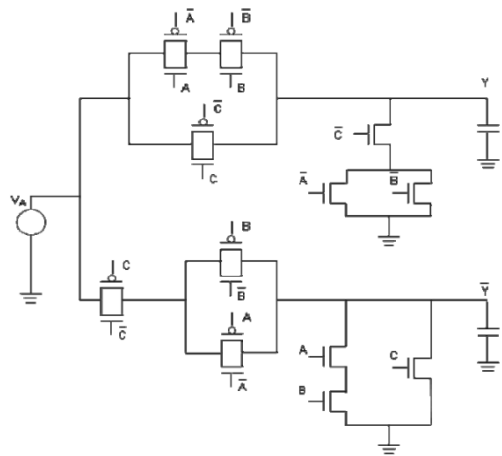

b) Only one output (Y) will change from 0 to 1 because when the clock is low both output is 0.
   R = (5K+5K)||5K = 3.3K

$$E = \frac{RC}{T} CV_{c,max}^2$$

$E = 3.3 \times 10^3 \times 200 \times 10^{-15} \times 200 \times 10^{-15} \times 4 / (100 \times 10^{-9})$ = 5.28fJ

**99. Bus Splitting**
The power saving is: (eqn from page 40)

$$PS\% = \left(1 - \frac{\frac{C_{BUS}}{m} \times 0.8 \times V_{DD}^2 + (2\frac{C_{BUS}}{m} + \frac{m.C_{BUS}}{N}) \times 0.2 \times V_{DD}^2}{C_{BUS} \times V_{DD}^2}\right) \times 100 = \left(1 - \left(\frac{1.2}{m} + \frac{0.2m}{N}\right)\right) \times 100$$

Next we take its derivate with respect to $m$, and find its maximum value:

$$\frac{dPS}{dm} = 0 \Rightarrow \frac{1.2}{m^2} = \frac{0.2}{N} \Rightarrow m = \sqrt{6N}$$

### 100. Switching Power

Make product terms mutually non-intersecting for the function: $F = A'BD + AC'D$.
$S_F = (1-S_A)S_BS_D + S_A(1-S_C)S_D = 0.258$, $\beta_g = 2S_F(1-S_F) = 0.382872$, $P_g = \frac{1}{2} C_{g,load} Vdd^2 f \beta_g = 0.1531488$ mW.

### 101. Glitch Reduction

Based on the circuit and K-Map, there might be glitch while B & D keep as 1, but A & C change from 0 to 1, **considering the situation when transition on A comes earlier than C, and path between NAND #2 and NAND #4 is much longer than the other two paths**.
Such glitch is shown in graph below.

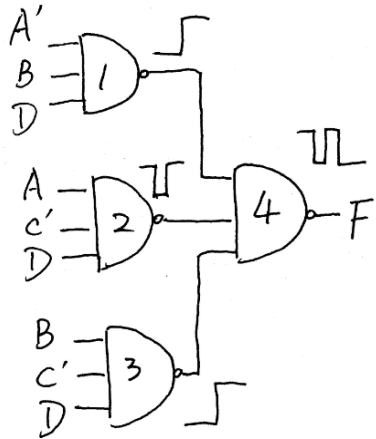

To avoid such glitch, we can change the circuit as below.

[K-map with AB across top (00, 01, 11, 10) and CD down the side (00, 01, 11, 10); 1's at AB=01,CD=01; AB=11,CD=01; AB=10,CD=01; AB=11,CD=11; groupings shown]

BD = 11
AC: 00 → 11

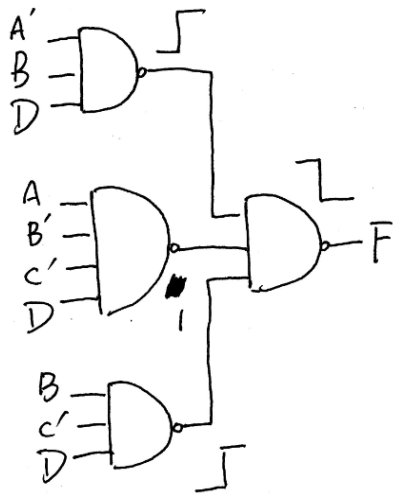

### 102. Bus Encoding

A simple algorithm to encode Binary Code to Gray Code is as below (Take 3-bit for example), denoting $B_n$ as the $n^{th}$ bit of Binary, and $G_n$ as the $n^{th}$ bit of Gray.

$G_0 = B_1$ XOR $B_0$
$G_1 = B_2$ XOR $B_1$
$G_2 = B_2$

Hence for 3-bit continuous Binary Code (000→111), corresponding Gray Code is 000, 001, 011, 010, 110, 111, 101, 100.
Then for a complete counting-up sequence, number of original transitions is 1+2+1+3+1+2+1 = 11, and number of new transitions is 1+1+1+1+1+1+1 = 7. Totally we save 4 transitions. Since Gray Code just changes one bit every time in representing continuous binary value, it saves transitions compared to original Binary Code, especially if number of bits is large.

### 103. Switching Power

Activity factor of NAND3 or NOR3 = $2 \times (7/8) \times (1/8) = 7/32$  **(4 points)**
NAND3 sizing: $(W/L)_n = 3$, $(W/L)_p = 2$   **(2 points)**
NOR3 sizing: $(W/L)_n = 1$, $(W/L)_p = 6$   **(2 points)**

$$P_{avg,nand3} = \sum \frac{1}{2} C_{load} V_{dd}^2 \cdot f \cdot \beta = \frac{1}{2} V_{dd}^2 f [3 \times \frac{5C_g}{3} \times \frac{1}{2} + 8C_g \times \frac{7}{32}] = \frac{17}{8} C_g V_{dd}^2 f$$

**(3 points)**

$$P_{avg,nor3} = \sum \frac{1}{2} C_{load} V_{dd}^2 \cdot f \cdot \beta = \frac{1}{2} V_{dd}^2 f [3 \times \frac{7C_g}{3} \times \frac{1}{2} + 8C_g \times \frac{7}{32}] = \frac{21}{8} C_g V_{dd}^2 f$$

**(3 points)**

$$\frac{P_{avg,nand3}}{P_{avg,nor3}} = \frac{\frac{17}{8} C_g V_{dd}^2 f}{\frac{21}{8} C_g V_{dd}^2 f} = \frac{17}{21}$$  **(1 point)**

Note: $C_g$ is the total input capacitance of a minimum-sized inverter, and $C_g$ is actually the gate cap of a NMOS(1/1) plus the gate cap of a PMOS(2/1). Hence a one-unit MOS gate cap is $C_g/3$.

### 104. Switching Power

$P = 0.5\beta \cdot C \cdot V^2 \cdot f = 0.1 * (150e{-}12 * 70) * (0.9)^2 * 450e6 = 0.38$ W.

Note: In literature you may find a different formula, where activity factor is calculated based on a pair of rising and falling transitions, e.g., activity of 1 means there is a pair of transitions per clock, which really means two transitions (one rising and one falling). In that case, you will see the formula written as activity_factor*C*V²*f, where the activity_factor refers to the number of pairs of transitions per cycle. Please also note that both formulae are essentially the same. We will however, use the definition presented , which is the number of transitions.

**105.** Switching Power

Pavg= $\frac{1}{2}$(Vdd − 2Vt)Vdd$f$clk$\beta$. Vdd is the power supply voltage and (Vdd-2Vt) is the output swing.

**106.** Leakage Power

|  | Sub-threshold | Gate | Reverse-biased Junction |
|---|---|---|---|
| Voltage decrement | decrease | decrease | decrease |
| Width decrement | decrease | decrease | decrease |
| Length decrement | increase | decrease | - |

However, the speed of the circuit would be limited when we scale down the voltage and dimension.

**107. Dynamic Power**

a) The gate or diffusion capacitance of minimum width is $C_g$ = 1.25fF/μm * 80nm = 0.1fF. The delay of option (A) is:

$g_1h_1 + p_1 + g_2h_2 + p_2 = (9/3) * (15 C_g / 9 C_g) + 4 + 1 * (10fF / 15C_g) + 1$

The delay of option (B) is: (assuming the NAND2 gate is scaled x times)

$g_1h_1 + p_1 + g_2h_2 + p_2 = (5/3) * (4xC_g / 10C_g) + 2 + (4/3) * (10fF / 4xC_g) + 2$

X = 3.2 or 15.8(unreasonable answer), so we can round up x to 4, and the size of NAND2 in option B is 4X.

b) Assume the inputs are toggled with the same probability. The activity factor and output capacitance load for both designs are the same. The activity factor at inputs of both designs are the same and the input gate capacitance of option (B) is larger than option (A) as a 2X NOR2's input gate capacitance is 10Cg = 1fF, while a 1X NOR4's input gate capacitance is 9Cg = 0.9fF.

For option (A), the activity factor $\alpha_{0->1}$ = 1/16 * 15/16 = 15/256 at the output of first stage NOR4 gate, and the load capacitance at that immediate node is 15$C_g$ = 1.5fF.

For option (B), the activity factor $\alpha_{0->1}$ = ¼ * ¾ = 3/16 at the output of first stage NOR2 gate, and there are two such immediate nodes, the load capacitance at each immediate node is 16$C_g$ = 1.6fF.

So apparently, option (A) saves more dynamic power than option (B).

**108. Dynamic Power**

a) According to the truth table, only when (A,B,C)=(0,0,1) and (0,1,0), the output is 1, so the activity factor is $p_0p_1$ = (6/8) * (2/8) = 3/16

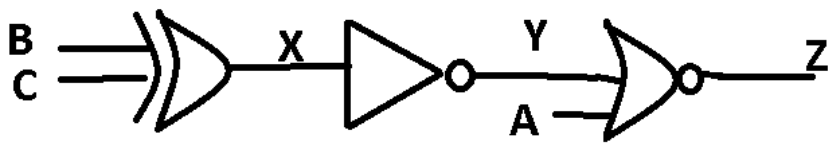

b)
X = B xor C, so $\alpha_{0->1}$ = 0.5 * 0.5 = 0.25
Y = (B xor C)', so $\alpha_{0->1}$ = 0.5 * 0.5 = 0.25

So P = ½ * $C_{load}$ * $V^2_{dd}$ * (2Σ $α_{0->1}$) * f = 0.5 * 50fF * (1V)² * 2 * (0.25 + 0.25 + 3/16) * 1GHz = 34.375μW

### 109. Leakage Power

a) $I_{leak,A} = I_0 10^{(0-|V_{th}|+\lambda_d(V_{dd}-V_x))/S}$

$I_{leak,B} = I_0 10^{((V_x-V_{dd})-|V_{th}|+\lambda_d V_x)/S}$

$I_{leak,A} = I_{leak,B}$ , so $V_x = \frac{1+\lambda_d}{1+2\lambda_d} V_{dd} = \frac{1.1}{1.2} = 0.91V$

b) $I_{leak,INV} = I_0 10^{(0-|V_{th}|+\lambda_d V_{dd})/S}$

$\frac{I_{leak,NOR2}}{I_{leak,INV}} = 10^{-\frac{1.1}{1.2}} < 1$, so $I_{leak,NOR2} < I_{leak,INV}$

# 20.11: Chapter 11 – Multi-Gate

**110.** This is because of their superior performance, better scalability lower leakage, and stronger
   immunization to process variations.

**111.** (i) the double gate (DG) mode, where the front gate and 47. the back gate of the fin are tied
   together to the input signal
      (ii) the independent gate (IG) mode, where one of the gate is driven by the input signal and the
   other is connected to a pre-defined biasing voltage or to the ground.

**112.** In (a), we use the double gate mode for both the N-type and P-type fins. In (b), we use the independent gate mode for both N-type and the P-type fin, where each back gate is tied to a certain biasing voltage to control the threshold voltage of the front gate. In (c), we use double gate mode for N-type fin while the two gates of the P-type fin are driven by different input signals.

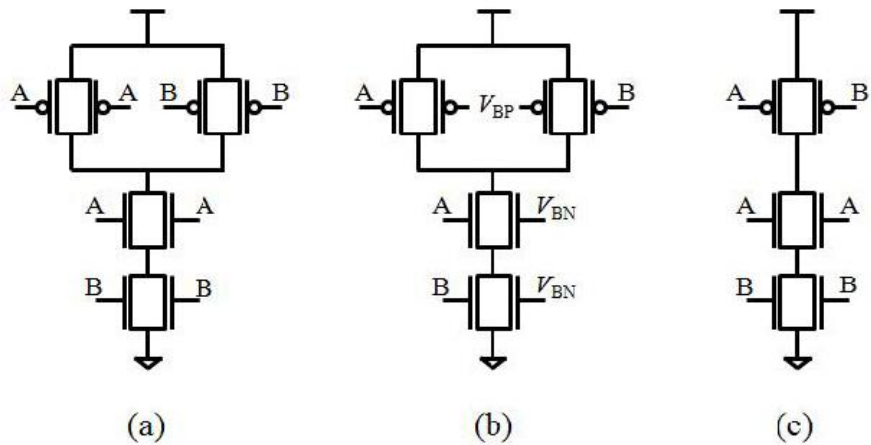

(a)   (b)   (c)

**113.** This is a FinFET NAND gate. The number of p-type Fin is 1 and the number of n-type Fin is 3

## 20.12: Chapter 12 – Logic Effort and Data Path

**114.**

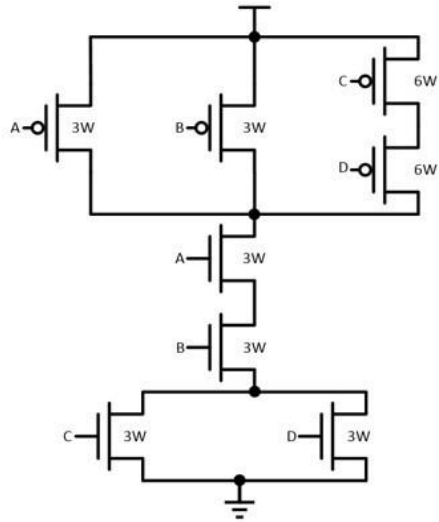

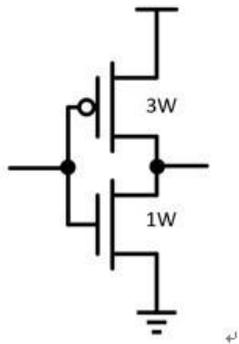

Reference template:

F=[AB(C+D)]'

$$p = \frac{RCp}{R^{rt}Cin^{rt}} = \frac{15}{4}$$

Falling worst case: A, B, C is on or A, B, D is on.

$$g_{A,fall} = g_{B,fall} = \frac{RCinA}{R^{rt}Cin^{rt}} = \frac{3}{2}$$

$$g_{C,fall} = g_{D,fall} = \frac{RCinA}{R^{rt}Cin^{rt}} = \frac{9}{4}$$

Rising worst case: A is on or B is on or C, D is on.

$$g_{A,rise} = g_{B,rise} = \frac{RCinA}{R^{rt}Cin^{rt}} = \frac{3}{2}$$

$$g_{C,rise} = g_{D,rise} = \frac{RCinA}{R^{rt}Cin^{rt}} = \frac{9}{4}$$

**115.** The critical path is from Cin to Cout, therefore all the transistors which are not on that
critical path can be minimum sized (to 1). This is shown in the following figure.

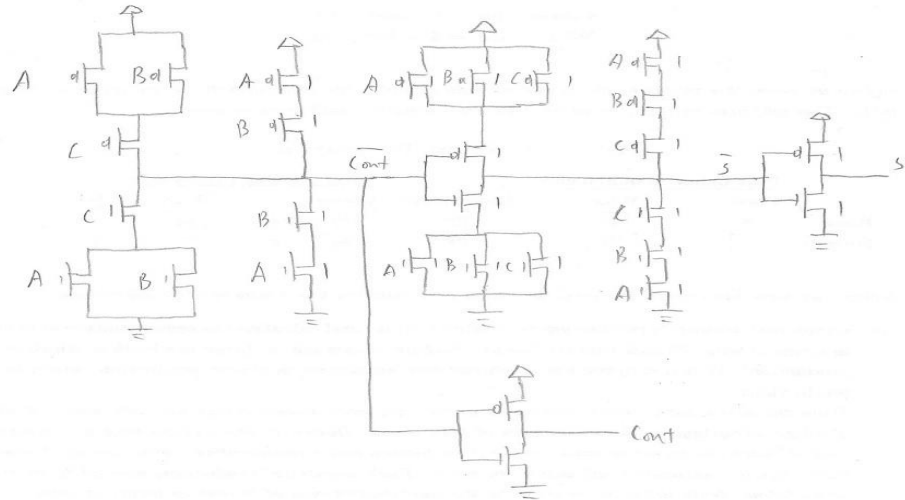

**116.**

The critical path is from C to Cout
so all the transistors outside the critical path should be minimal sized to 1

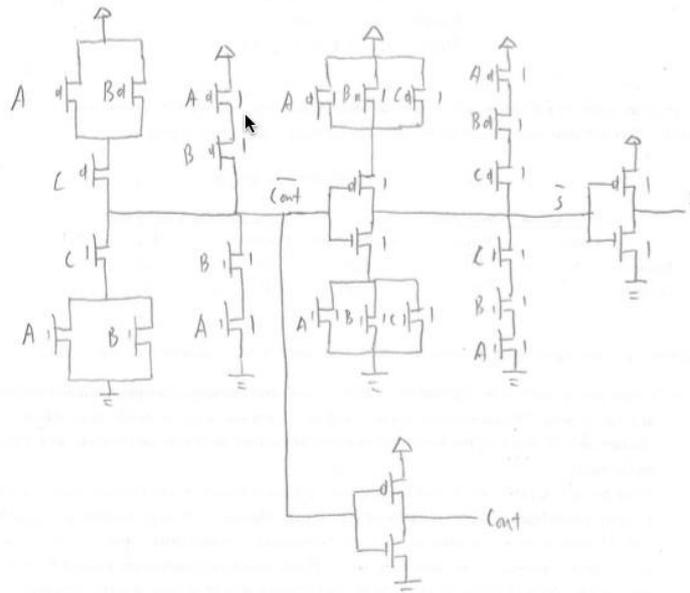

**117.**

For odd stages: 
$$\begin{cases} \overline{C_{out}} = \overline{AB + BC + AC} \\ \overline{S} = \overline{A \oplus B \oplus C} \end{cases} \quad \text{input: } A, B, C \quad \text{output: } \overline{C_{out}}, \overline{S}$$

For even stages:
$$\begin{cases} C_{out} = AB + BC + AC = \overline{\overline{A}+\overline{B}} + \overline{\overline{B}+\overline{C}} + \overline{\overline{A}+\overline{C}} = \overline{\overline{A}\overline{B} + \overline{B}\overline{C} + \overline{A}\overline{C}} \\ S = A \oplus B \oplus C = \overline{\overline{A} \oplus \overline{B} \oplus \overline{C}} \end{cases}$$
input: $\overline{A}, \overline{B}, \overline{C}$   output: $C_{out}, S$

The same logic can be used, so the adders are logically identical.

**118.**

*G=2.37, B=1, H=1, F = GBH = 2.37*

f = 2.371/3 = 4/3, z =C ×(4/3)/(4/3)=C, y=C: the transistor sizes in the three gates will be the same

**119.**

D = 3(18.96)1/3 + 3(2pinv) = 14.0 delay units

Each successive logic gate has twice the input capacitance of its predecessor.

**120.**

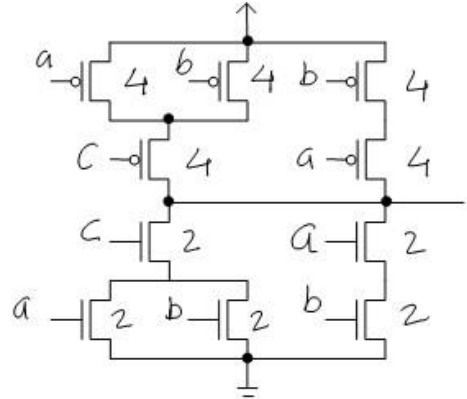

The worst-case resistance of the pull-up and pull-down are $R_u = R_d = R$. Also, for the reference template $R^{rt}C_{in}^{rt} = 3RC_g$ (from page-1). We use these along with other values shown in the following table to compute following g and p values.

| Input | $C_{in}$ | g | $C_p$ | p |
|---|---|---|---|---|
| a | $12C_g$ | 4 | | |
| b | $12C_g$ | 4 | $12C_d$ | $4C_d/C_g \approx 4$ |
| c | $6C_g$ | 2 | | |

**Assumptions, notations, etc.:**
1) Unless otherwise specified, assume that in your fabrication process $\mu = \mu_n/\mu_p = 2$.
2) Unless otherwise specified, an inverter with a minimum-size nMOS transistor and a pMOS transistor with minimum-length and twice the minimum-width is considered as the **reference template** with a logical effort ($g_{inv}$) of 1 and parasitic delay ($p_{inv}$) of 1.
3) Let $C_g$ denote the gate capacitance of a minimum-size transistor. Hence, the total capacitance at the input of the above reference template inverter is $3C_g$.
4) Let $C_d$ denote the diffusion capacitance of the drain of a minimum-size transistor. Hence, the total parasitic capacitance at the output of the above reference template inverter is $3C_d$.
5) Let $R$ denote the effective channel resistance of a minimum-size nMOS transistor.

## 121. :

*G11:0 = G11:8 + P11:8 * G7:0*

For each components of G11:8, P11:8 and G7:0, we should further expand to bit-wise:

*G11:8=G11+P11*G10:8*
*G10:8=G10+P10*G9:8*
*G9:8=G9+P9*G8*

P11:8 =
P11*P10:8
P10:8=P10*P9:
8 P9:8=P9*P8

G7:0 = G7:4 + P7:4*G4:0 We should similarly expand G7:4, P7:4, and G4:0 down to bitwise. I leave this part to our students.

The main reason CIA is faster than RCA is the parallelism used to calculate the group P and G values in CIA, while there is no parallelism in RCA. E.g., while G3:0 is being calculated, G/P7:4, also G/P11:8, and also G/P15:12 are calculated.

**122.** When we consider the optimal architecture, we should first try NAND2 and INV because their logical effort is small compare to others. Since the load cap is large, we need about 6 stages to achieve the optimal design. With 6 or more stages, it is possible to design this logic using only NAND2 and INV.

**123.** For equal rise and fall delays, NAND2 has two NMOS and two PMOS transistors each with size 2. For equal rise and fall delays, Inverter has one NMOS transistor with size 1 and one PMOS transistor with size 2. For equal rise and fall delays, NOR2 has two NMOS transistors with size 1 and two PMOS transistors with size 4.

$$C_{in\ inv} = 3C_g \qquad C_{in\ nand2} = 4C_g \qquad C_{in\ nor2} = 5C_g$$
$$R_{i\ inv} = R \qquad R_{i\ nand2} = R \qquad R_{i\ nor2} = R$$
$$C_{p\ inv} = 3C_d \qquad C_{p\ nand2} = 6C_d \qquad C_{p\ nor2} = 6C_d$$

$$C_d = C_g$$

$$g_{inverter} = \frac{3C_gR}{4C_gR} = 3/4 \qquad g_{nand2} = \frac{4C_gR}{4C_gR} = 1 \qquad g_{nor2} = \frac{5C_gR}{4C_gR} = 5/4$$

$$p_{inverter} = \frac{3C_dR}{4C_gR} = 3/4 \qquad p_{nand2} = \frac{6C_dR}{4C_gR} = 3/2 \qquad p_{nor2} = \frac{6C_dR}{4C_gR} = 3/2$$

**124.** Inverter has g = 1, p = 1.

Assume input cap of one branch (B1) with **y** Inverters is **x** $C_g$, then input of the other branch (B2) with **(y+1)** Inverters is **(5-x)** $C_g$. Then $D_{B1} = y(800/x)^{1/y} + y$, $D_{B2} = (y+1)[800/(5-x)]^{1/(y+1)} + (y+1)$. To find the most equal $D_{B1} = D_{B2}$, then we can get the minimum delay.

| | y = 3 | | | y = 4 | | | y = 5 | |
|---|---|---|---|---|---|---|---|---|
| x | $D_{B1}$ | $D_{B2}$ | x | $D_{B1}$ | $D_{B2}$ | x | $D_{B1}$ | $D_{B2}$ |
| 3 | 22.30979 | 21.88854 | 2.4 | 21.09148 | 20.7251 | 2 | 21.57227 | 21.22227 |
| 3.1 | 22.09988 | 22.11941 | 2.5 | 20.91794 | 20.84893 | 2.1 | 21.41134 | 21.30852 |
| 3.2 | 21.89882 | 22.36599 | 2.6 | 20.75287 | 20.97886 | 2.2 | 21.25936 | 21.39832 |

Hence if **x = 2.5** and **y = 4**, we have minimum delay.

For Branch B1, input cap of 1st Inv is $2.5C_g$, input cap of 2nd Inv is $10.57C_g$, input cap of 3rd Inv is $44.72C_g$, input cap of 4th Inv is $189.15C_g$;

For Branch B2, input cap of 1st Inv is $2.5C_g$, input cap of 2nd Inv is $7.92C_g$, input cap of 3rd

Inv is $25.12C_g$, input cap of 4th Inv is $79.62C_g$, input cap of 5th Inv is $252.38C_g$..

**125.**

(1) We can obtain such a template by using a NMOS transistor of minimum width ($W_{min}$) and minimum length ($L_{min}$) and a PMOS with width that is 4 times the minimum width ($4*W_{min}$) and length equal to $L_{min}$. Since in most designs, the length of transistors are minimum length, typically only the transistor width is specified, as shown in following figure.

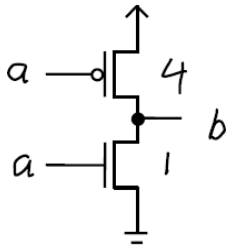

(2) The designs are shown in the figure below. Please note the transistor sizes in the template that implements ~(ab+c). In the pulldown, one NMOS transistor (with input c) is in parallel with a series connection of two NMOS transistors (with input a and b, respectively). We have selected the transistor size in such a manner that the worst-case falling delay is equal, independent of which of the above mentioned two parallel path conducts. In other words, the falling delay invoked is nearly identical when we apply 110 or 0x1 or x01 to inputs a, b and c, respectively. Increasing the width of the NMOS transistor with input c will not help improve the worst-case delay. In fact, it will increase the worst case delay by increasing the parasitic capacitance on the output. It will also increase the total input capacitance of input c and the area of this cell.

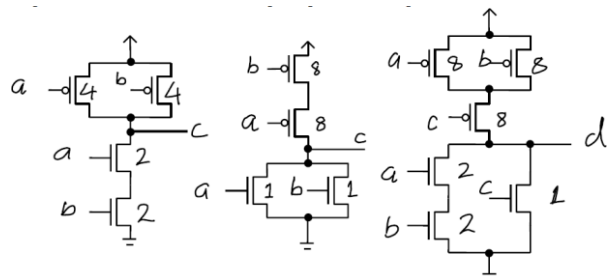

(3) For each of these four templates, we first compute the values of $C_{in}$, $C_p$, $R_u$ and $R_d$, as shown in the following table. Note that for the template AOI21, i.e., the template which implements ~(ab+c), the value of $C_{in}$ for three inputs are not identical and hence shown separately. As we see ahead, for this template we compute a different g value for each input.

|  | Inverter | NAND2 | NOR2 | AOI21 |
|---|---|---|---|---|
| $R_u = R_d$ | R | R | R | R |
| $C_{in}$ | $5C_g$ | $6C_g$ | $9C_g$ | Input $a$: $10C_g$<br>Input $b$: $10C_g$<br>Input $c$: $9C_g$ |
| $C_p$ | $5C_d$ | $10C_d$ | $10C_d$ | $11C_d$ |

Since we have selected the inverter (INV) as the reference template, we use $C_{in\_INV}R_{d\_INV}$ as the normalized quantity and compute the p and g values for each template as:

$$g_{INV} = \frac{C_{in\_INV} \cdot R_{d\_INV}}{C_{in\_INV} \cdot R_{d\_INV}} = 1$$

The above algebraic expression clearly shows why g is always 1 for the reference template.

$$p_{INV} = \frac{C_{p\_INV} \cdot R_{d\_INV}}{C_{in\_INV} \cdot R_{d\_INV}} = \frac{C_d}{C_g}$$

Note that the above p value is 1 only when we assume that $C_d = C_g$. Hence, since this assumption is not true for fabrication processes, the value of p for the reference template is not always guaranteed to be equal to 1.

$$g_{NAND2} = \frac{C_{in\_NAND2} \cdot R_{d\_NAND2}}{C_{in\_INV} \cdot R_{d\_INV}} = \frac{6}{5}$$

$$p_{NAND2} = \frac{C_{p\_NAND2} \cdot R_{d\_NAND2}}{C_{in\_INV} \cdot R_{d\_INV}} = 2 \cdot \frac{C_d}{C_g}$$

$$g_{NOR2} = \frac{C_{in\_NOR2} \cdot R_{d\_NOR2}}{C_{in\_INV} \cdot R_{d\_INV}} = \frac{9}{5}$$

$$p_{NOR2} = \frac{C_{p\_NOR2} \cdot R_{d\_NOR2}}{C_{in\_INV} \cdot R_{d\_INV}} = 2 \cdot \frac{C_d}{C_g}$$

$$g_{AOI21}^a = \frac{C_{in\_AOI21}^a \cdot R_{d\_AOI21}^a}{C_{in\_INV} \cdot R_{d\_INV}} = \frac{10}{5} = 2$$

$$g_{AOI21}^b = \frac{C_{in\_AOI21}^b \cdot R_{d\_AOI21}^b}{C_{in\_INV} \cdot R_{d\_INV}} = \frac{10}{5} = 2$$

$$g_{AOI21}^c = \frac{C_{in\_AOI21}^c \cdot R_{d\_AOI21}^c}{C_{in\_INV} \cdot R_{d\_INV}} = \frac{9}{5}$$

$$p_{AOI21} = \frac{C_{p\_AOI21} \cdot R_{d\_AOI21}}{C_{in\_INV} \cdot R_{d\_INV}} = \frac{11}{5} \cdot \frac{C_d}{C_g}$$

**126.**

NAND4 has g = 2, p = 4, NOR4 has g = 3, p = 4.
1) $D = 2(2\times3\times3)^{1/2} + 4+4 = 16.49$;

2) Since $F = 2\times3\times3 = 18$, $\log_4 F = 2.08$, calculate delay for N = 4, which is $4F^{1/4} + 4+4+1+1 = 18.24$, which is less than D. Hence we use N = 2, namely **no** Inverter is added. And input cap of NAND4 is $C_g$, input cap of NOR4 is $2.12C_g$;

3) Since $F = 2\times3\times300 = 1800$, $\log_4 F = 5.41$, calculate delay for N = 4, which is $4F^{1/4} + 4+4+1+1 = 36.05$, and for N = 6, which is $6F^{1/6} + 4+4+1+1+1+1 = 32.93$. Hence we use N = 6, namely **4** Inverters are added. And input cap of NAND4 is $C_g$, input cap of NOR4 is $1.74C_g$, input cap of 1st Inv is $2.03C_g$, input cap of 2nd Inv is $7.07C_g$, input cap of 3rd Inv is $24.66C_g$, input cap of 4th Inv is $86.02C_g$.

**127.**

A (m+1, m) fork in shown below

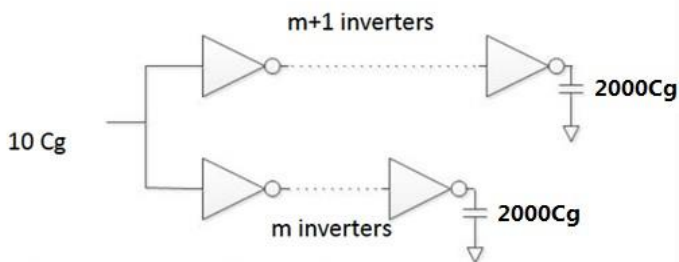

We can have an approximate circuit:

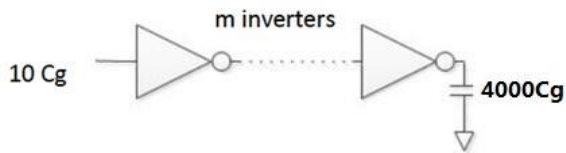

**m** can be estimated such that $(GBH)^{1/m} = (1\times1\times\frac{4000}{10})^{1/m} \approx 4$, $m \approx 4.3$

Try (5,4) fork first, say the longer path has delay D5 and input cap of $xC_g$, the shorter path has delay D4 and input cap of $(10-x)C_g$. Delay of the fork $D_F$ = max (D5, D4).

$$D5 = 5\times(1\times1\times\frac{2000}{x})^{1/5} + 5$$

$$D4 = 4\times(1\times1\times\frac{2000}{10-x})^{1/4} + 4$$

Solve above equations iteratively:

| x | D5 | D4 | $D_F$ |
|---|---|---|---|
| 5 | 21.6 | 21.9 | 21.9 |
| 4.5 | 21.9 | 21.5 | 21.9 |
| 4.8 | 21.7 | 21.7 | 21.7 |

Choosing x = 4.8 because of smallest $D_F$.

For the longer path, $f = (1 \times 1 \times \frac{1000000}{4.8})^{1/5} = 3.34$, the input cap of the inverters on the path is following (reverse order): $598.5C_g$, $179.1C_g$, $53.6C_g$, $16C_g$, $4.8C_g$.

For the shorter path, $f = (1 \times 1 \times \frac{1000000}{4.8})^{1/4} = 4.42$, the input cap of the inverters on the path is following (reverse order): $451.6C_g$, $102C_g$, $23C_g$, $5.2C_g$.

## 20.13: Chapter 13 – Flip-Flop

**128.**

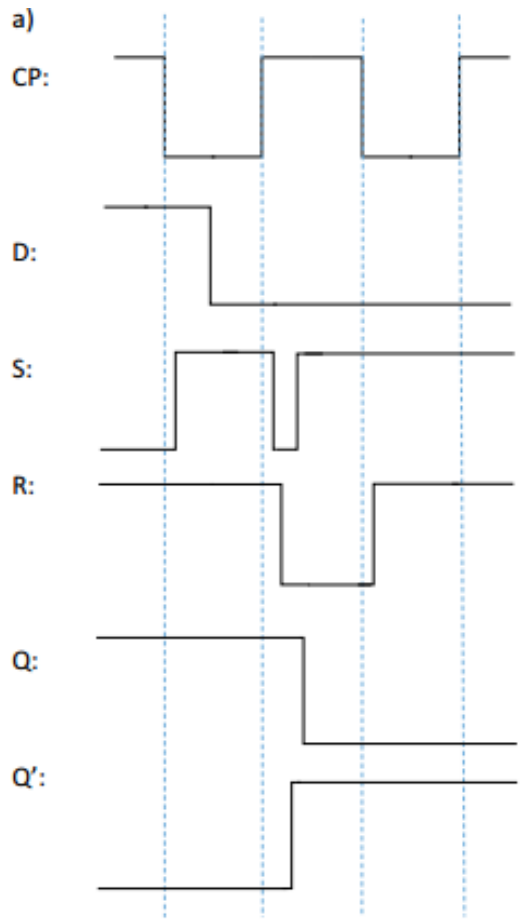

The functionality of this structure can be found at http://en.wikipedia.org/wiki/Flip-flop_(electronics).

b) NOTE: Please keep in mind that setup/hold calculations should be done very conservatively.

Let us denote the outputs of gate 1 and 2 as X and Y, respectively.

## SETUP:

We would like check on a case where changing D before the clock triggering edge, i.e., 0->1, can change the final result (i.e., Q) which would mean a setup violation:

$1^{st}$ scenario: D: 0->1: If this change comes early enough, such that gate 4 gets the chance to process it, then the Y becomes 0. That's enough to guarantee that the rest of the value settings would be done correctly, even if clock is already 1. This means any change within a gate 4 delay would be a setup violation, which means $t_{setup} = t_4$

$2^{nd}$ scenario: D: 1->0:

We will first calculate a less conservative setup value: If D: 1->0 comes early enough, such that gate 4 gets the chance to process it, then Y becomes 1. That's enough to guarantee that the rest of the value settings would be done correctly, even if clock is already 1. This means any change within a gate 4 delay would be a setup time violation, i.e., $t_{setup}$ can be considered as $t_4$. However we can have a more conservative definition using the following argument:  Initially while D is 1, S=R=1, also Y 0, and X is 1. When D changes to 0, after $t_4$ delay, Y goes to 1. Then to guarantee no issues such as racing, Y should be processed by gate1 to make X as 0, otherwise when clock goes to 1, and if X is still 1, then S goes to 0, and then there is a chance SR gets stuck at 01, which means the wrong output. Therefore a conservative setup time is $t_4+t_1$.

Therefore overall the setup can be conservatively defined as $t_4+t_1$.

## HOLD:

We want to check on a case where changing D after the clock 0->1 transition, can change the final result (i.e., eventually change Q) which means a hold violation:

First scenario: D: 0->1: If we give clock enough time to go through gate 3, it can change R to 0, in that case there is no chance for D:0->1 to make any impact on the rest of the circuit, so the key is to wait gate 3 delay to guarantee R changing to 0. Therefore $t_{hold}$ is $t_3$.

$2^{nd}$ scenario: D: 1->0: If we give clock enough time to go through gate 2, it can change S to 0, in that case there is no chance for D:1->0 to impact the values of R and S, so the key is to wait gate 2 delay to guarantee S changing to 0. Therefore $t_{hold}$ is $t_2$. Overall $t_{hold}$ is then the maximum of $\{t_2, t_3\}$.

c)
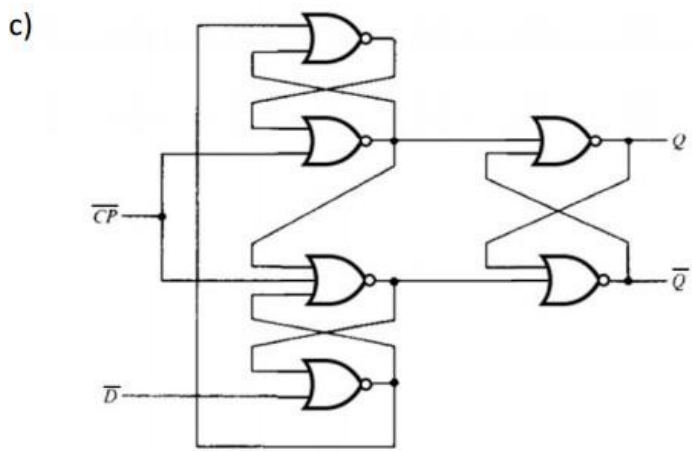

**129.**

a) I. the pulse generator in implicit P-FF is part of the Flip-Flop, it can't move out. While pulse generator in explicit P-FF can move out from Flip-Flop and share with neighbors.

   II. In EPFF, the precharge is controlled by clock, while in SEPFF, the precharge is controlled by D, the latter one consumes less power.

b) Three inverters are used to generate a delayed inverted clock. Before rising edge of clock: N1 N4 off, N3 N6 P1 on, X precharged to Vdd and Q holds previous value. When rising edge of clock arrives, N1 N4 on, N3 N6 is on for a while (because clock_bar is delayed). During this period, both '1' and '0' writing paths are transparent and the data D can be transmitted to node Q.

**130.**

a) If using HLFF, whenever the clock ticks, internal node X will be charged, which consumes a lot of power. While using CCFF, the unnecessary internal switching activities are reduced with the control from Q and Q', hence, it uses less power.

b) During the transparency window, if X has been precharged to Vdd, D =1 and Q = 0, then Q' = 1, N1 is on, thus X is discharged, P1 is on and Q = 1. If D =1 and Q = 1, then Q' = 0, N1 is off, no internal activity exists in this case and this is the idea of conditional discharge.

**131.**

a) Reduce the frequency to save switching power. E.g. Double Edge Triggered Flip-Flop (DET-FF). A single edge triggered flip-flop (SET-FF) changes its output on one of the two clock edges. A DET-FF changes its output on both of clock edges. SET has an idle clock edge with internal nodes switching which should be added to the total power consumption. DET keeps both clock edges busy, and if we are given the same amount of data and same frequency, the SET needs twice the total time therefore more energy.
Low power FF: Conditional Precharged FF and Static Explicit-Pulsed FF.

b) If there's setup time violation, we can lower the clock frequency to fix it. If there's hold time violation, there's not much more we can do. Maybe we can heat the chip, make the chip aging faster so the chip can be slower.

c) The process variation makes the paths slower than the expected design, so there may be setup time violation and the soft-edge flip flop can help resolve this problem.

d) Because if the fabricated chip has hold time violation, there's not much that can help to solve the problem and the chip is failed.

e) The claim is not valid. Because there may be a stage that borrows more than the W of transparency, however the overall delay is smaller, which means the N-stage constraint would not fail, but the 1-stage does fail. This happens because it is possible that not all stages are excited to be the slowest and it's quite possible some are very fast while others are not, so the N-stage seems not having setup time violation. Also for diagnosis purposes, even if N-stage constraint fails, one cannot tell which stage was the main reason, therefore writing 1-stage constraints would help debug the setup problem.

f) This claim is not valid. Because maybe all one-stage constraints pass, but when all stages are transparent and borrowing, overall, they borrow more than W, which results in N-stage constraint failure.

## 20.14: Chapter 14 – Time Borrowing

**132.**

$$\Delta_1 + \Delta_{CQ} \leq T - \Delta_{DC} - t_{skew}$$

$$\Delta_2 + \Delta_{CQ} \leq T - \Delta_{DC} - t_{skew}$$

$$\Delta_3 + \Delta_{CQ} \leq T - \Delta_{DC} - t_{skew}$$

$$\Delta_4 + \Delta_{CQ} \leq T - \Delta_{DC} - t_{skew}$$

$$\Delta_1 + \Delta_2 + \Delta_{CQ} + \Delta_{DQ} \leq \mathbf{1.4}T - \Delta_{DC} - t_{skew}$$

$$\Delta_2 + \Delta_3 + \Delta_{CQ} + \Delta_{DQ} \leq \mathbf{1.6}T - \Delta_{DC} - t_{skew}$$

$$\Delta_3 + \Delta_4 + \Delta_{CQ} + \Delta_{DQ} \leq \mathbf{1.4}T - \Delta_{DC} - t_{skew}$$

$$\Delta_1 + \Delta_2 + \Delta_3 + \Delta_{CQ} + 2\Delta_{DQ} \leq 2T - \Delta_{DC} - t_{skew}$$

$$\Delta_2 + \Delta_3 + \Delta_4 + \Delta_{CQ} + 2\Delta_{DQ} \leq 2T - \Delta_{DC} - t_{skew}$$

$$\Delta_1 + \Delta_2 + \Delta_3 + \Delta_4 + \Delta_{CQ} + 3\Delta_{DQ} \leq \mathbf{2.4}T - \Delta_{DC} - t_{skew}$$

**133.** Based on the maximal delay inequality, we should have

$$D \leq T$$

$$2D \leq 1.5T$$

$$3D \leq 2T$$

$$\ldots\ldots$$

$$nD \leq (0.5 + 0.5n)T, \quad n = 1, 2, \ldots,$$

In order to satisfy these inequalities for any n, consider n->∞, we should have T≥2D.

## 20.15: Chapter 15 – Domino and Dual Rail Dynamic Logic

**134.** F can be implemented using one stage and also multiple stages.

    Problems: 1) Charge sharing. 2) 1-to-0 transition for next NMOS logic

    Solutions: 1) Charge sharing: Multiple Output Domino, 2) using weak PMOS (keepertransistor) to feed the output and 3) Zipper circuit,

    Problem 2  Remove the possibility of 1-to-0 transition at the inputs of the nextNMOS logic by adding inv.

**135.**

a) (i)

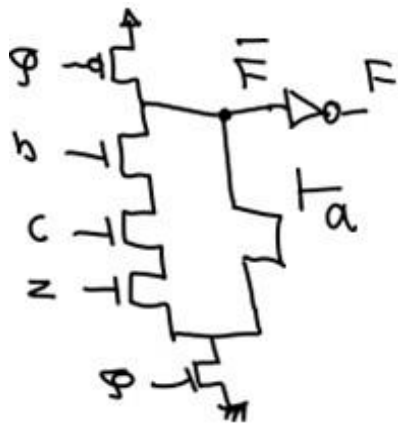

a )(ii)

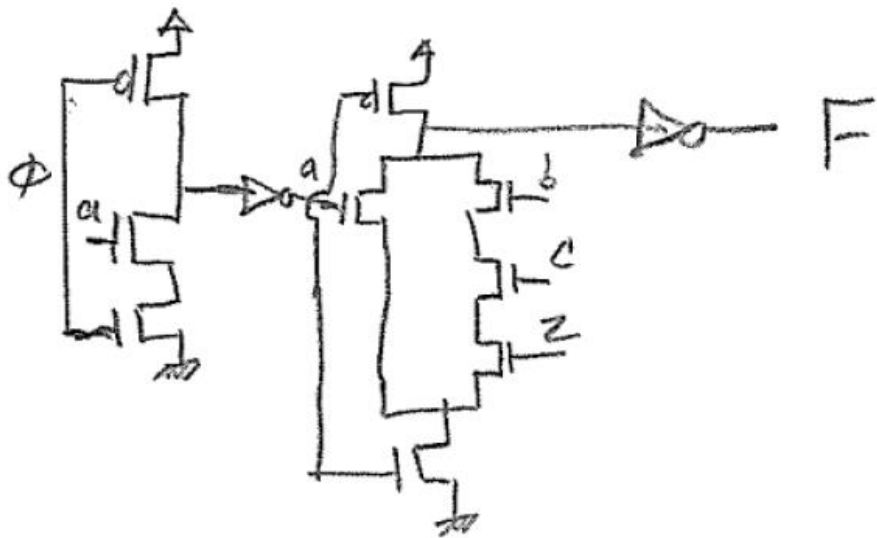

b) Note: There is no unique solution to this problem. We could implement F using one dynamic stage following by a static stage, but here we show a 2-stage domino and the first stage has two dynamic stages that work in parallel.

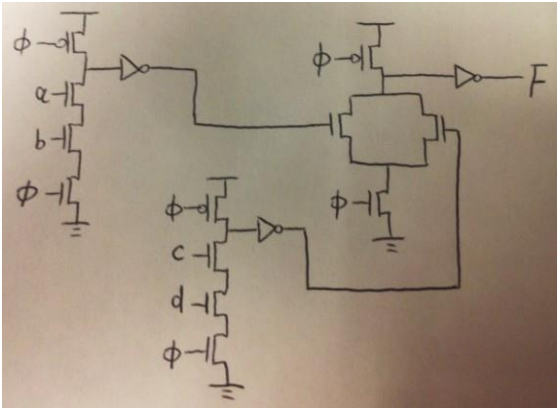

c) Two-stage domino logic in Figure 1.

Input combinations: ABC=000->100.

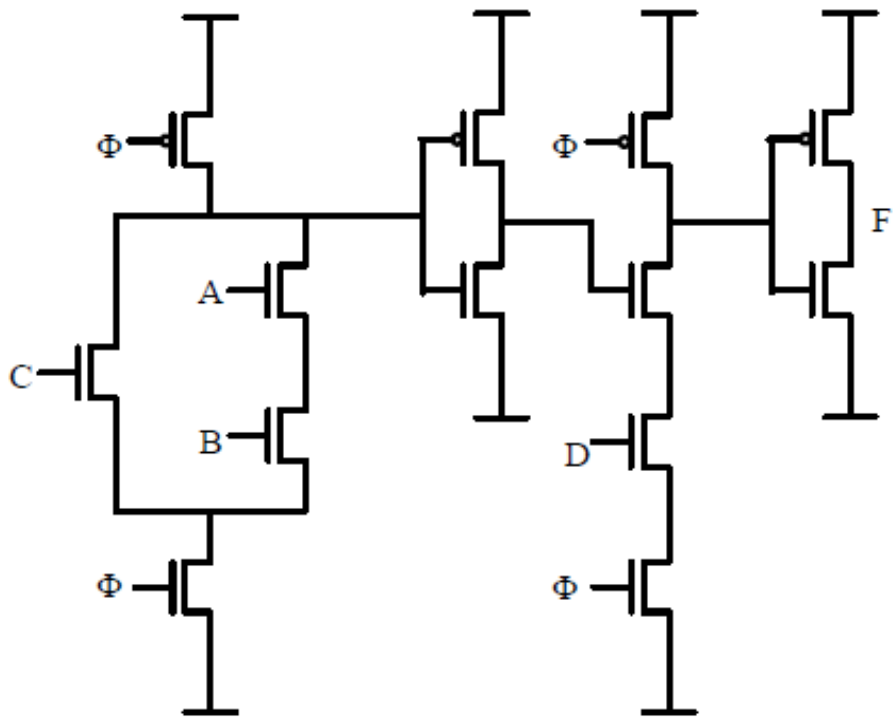

Figure 1

Solution to charge sharing:

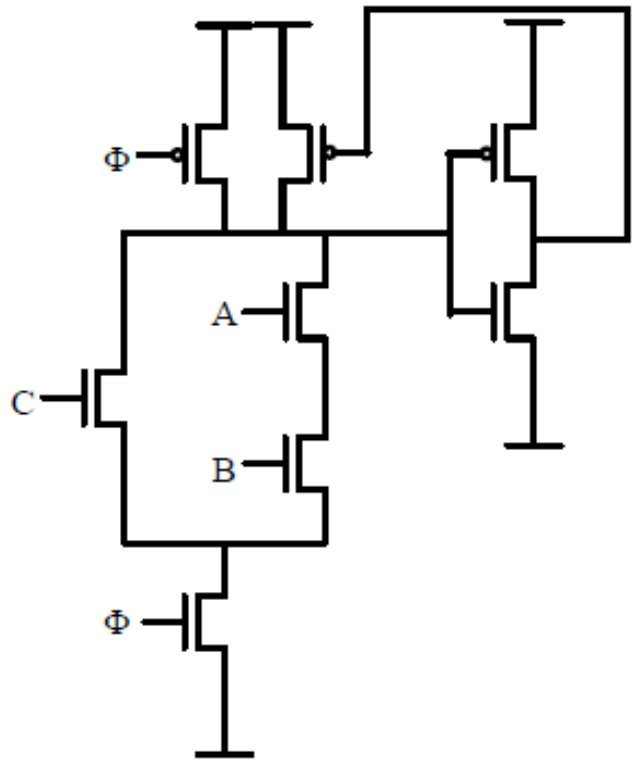

Figure 2

d)

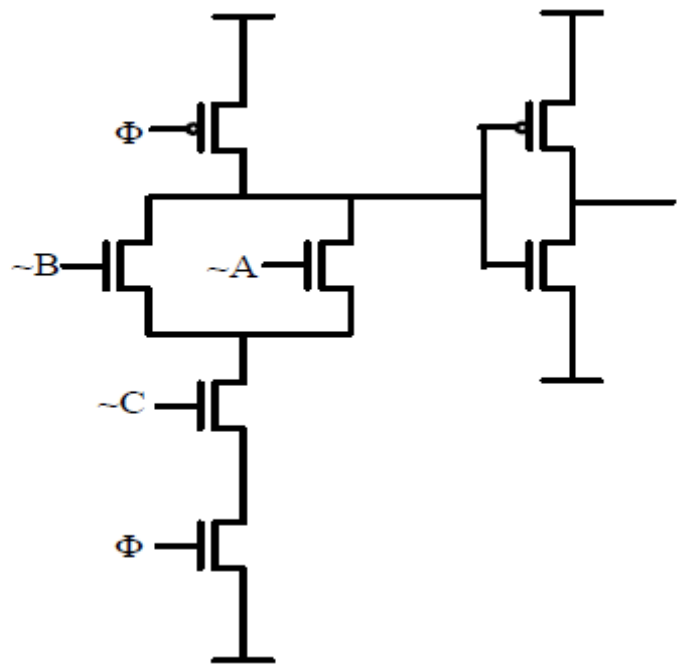

Note: The dual rail dynamic stages can share the footer (evaluation) transistor or can have separatefooter transistors as shown.

e ) Note: The dual rail dynamic stages can share the footer (evaluation) transistor or can have separate footer transistors as shown.

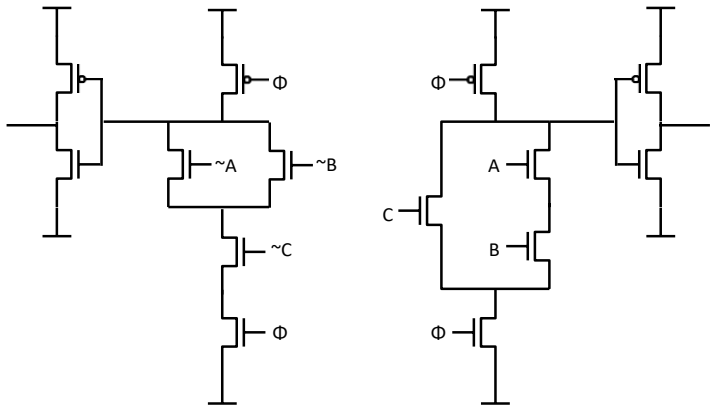

Figure 4

f )

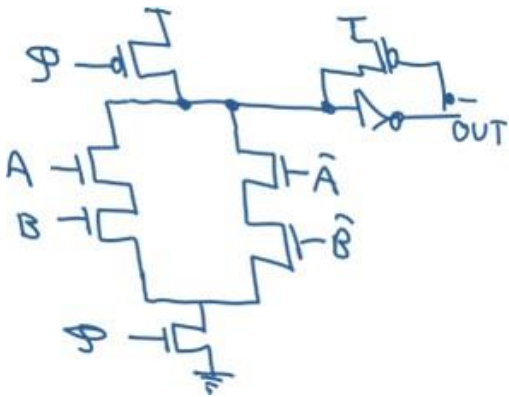

g)

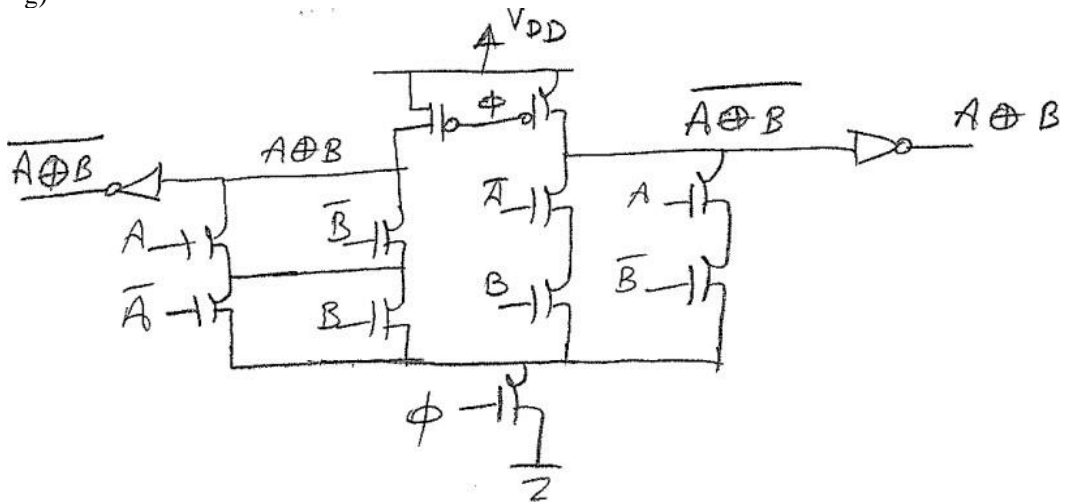

h )

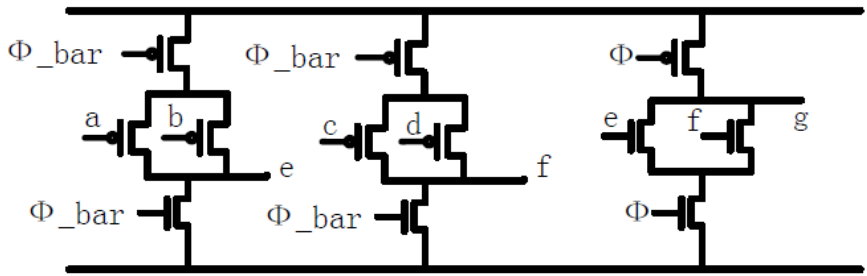

**136.** a) CMOS circuit diagram

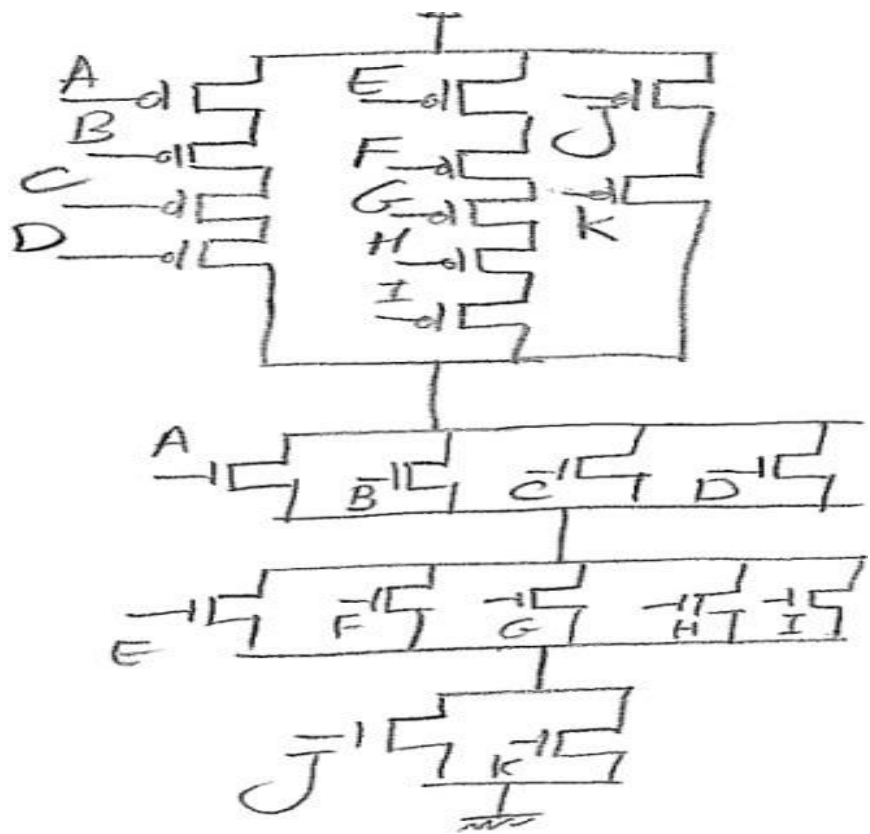

b) Domino CMOS logic circuit:

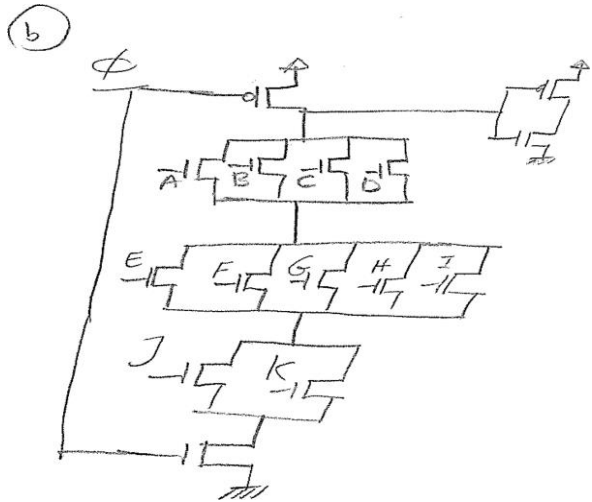

c) Dual rail Domino logic: Note: A-H is the same as A. A-L is A' ...

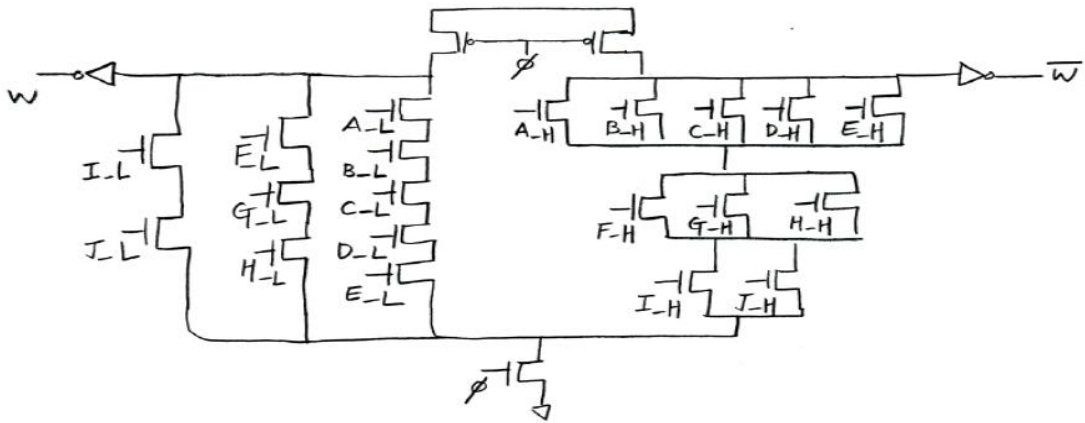

**137.** 1) Yes
2)   Yes, they precharge properly but sequentially. The second dynamic gate precharges when the first domino gate precharge is over and y=0.
3)   Second one since there is no footer transistor and less resistance and capacitance in the pull-down network

**138.**

$x = \overline{AB}$

$y = \overline{x + C} = \overline{\overline{AB} + C} = AB\overline{C}$

1) When the circuit comes out of the precharge phase, x will be high, which will cause y to discharge. Later x evaluates to low, but then it cannot change the value of y, which has erroneously been discharged. There is a race condition between the transition of x from 1 to 0 and the evaluation of y.

2) The new design:

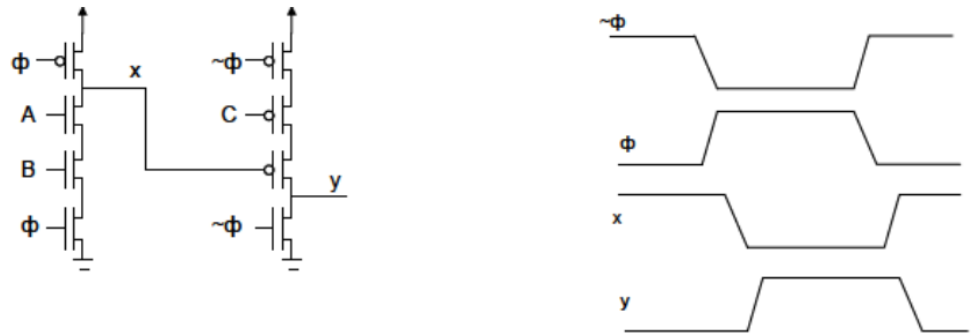

**139.**

(Note, this solution based on footless design at every stage. However, in general, when a stage is connected to primary input, it must be footed.)

$H = 500 / 30 = 16.7$ First assume $G = 1$, $F = GBH = 16.7$, If we assume $\rho = 2$, then $N = \log_2 F = 4$. So we design a four-stage Domino circuit.

$G = g_1 * g_2 * g_3 * g_4 = (2/3) * (5/6) * (1/3) * (5/6) = 25/162$

$F = GBH = 16.7 * 25 / 162 = 2.58$. $\log_2 F = 1.37$.

Compare the delay of two stages or four stages circuit:

**Two stages:**

$G = g_1 * g_2 = (2/3) * (5/6) = 5/9$

$F = GBH = 16.7 * 5 / 9 = 9.28$

$D = 2(F)^{1/2} + 14/9 + 5/6 = 8.5$

**Four stages:**

$D = 4(F)^{1/4} + 8/9 + 5/6 + 7/9 + 5/6 = 8.4$ → So the four-stage design is the optimum.

$f = (F)^{1/4} = 1.27$

C4 = 500 * (5/6) / 1.27 = 328. C3 = 328 * (1/3) / 1.27 = 86. C2 = 86 * (5/6) / 1.27 = 56

**The sizing of each transistor is as follows: (the unit is λ)**

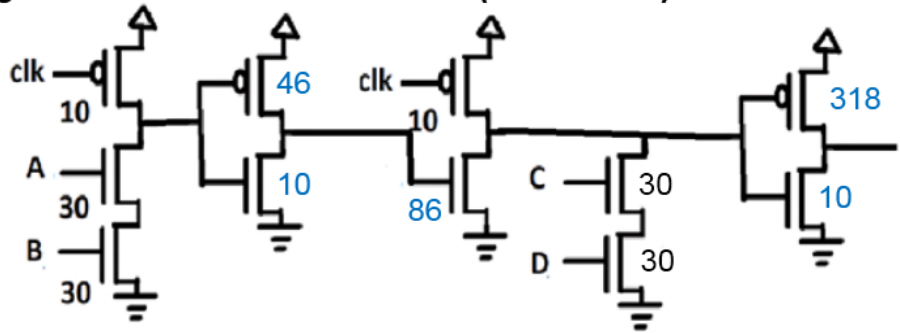

**140.**

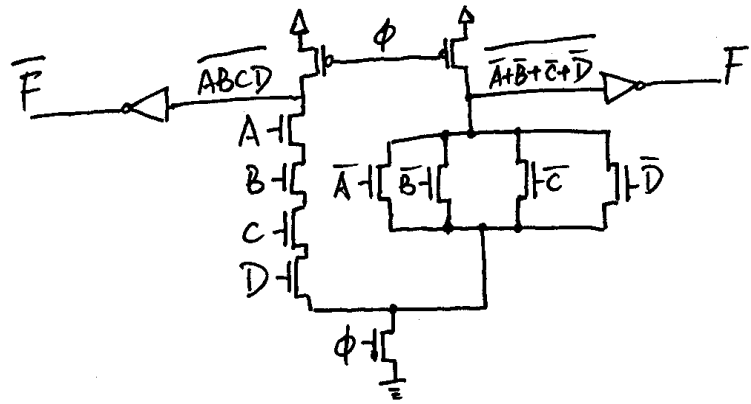

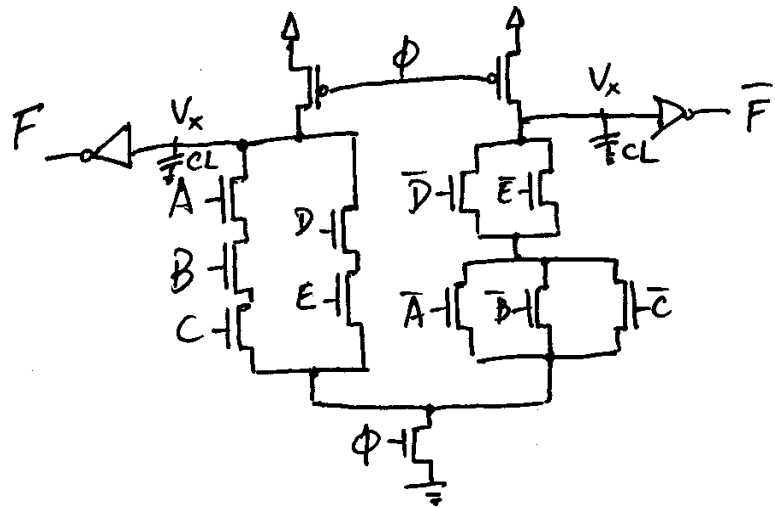

**141.**

For output F

If (A = 1, B = 0, D = 0), or (A = 0, D = 1, E = 0), $V_{DD} C_L = V_X (C_L+2C_P)$, $V_X = V_{DD} C_L / (C_L+2C_P)$;

If (A = 1, B = 0, D = 1, E = 0) or (A = 1, B = 1, C = 0, D = 0), $V_{DD} C_L = V_X (C_L+4C_P)$, $V_X =$

$V_{DD} C_L / (C_L+4C_P)$;

If (A = 1, B = 1, C = 0, D = 1, E = 0), $V_{DD} C_L = V_X (C_L+6C_P)$, $V_X = V_{DD} C_L / (C_L+6C_P)$.

For output F'

If (D = 0 or E = 0, A = 1, B = 1, C = 1), $V_{DD} C_L = V_X (C_L+5C_P)$, $V_X = V_{DD} C_L / (C_L+5C_P)$.

# 20.16: Chapter 16 – Scripting

**142.**

For perl:
```perl
#!/usr/bin/perl

@pizzas = qw(pepperoni cheese veggie Works Anchovie sausage);
@sorted_pizzas = reverse sort { "\L$a" cmp "\L$b" } @pizzas;
print "@sorted_pizzas\n";
```

For python:
```python
lst = ['pepperoni', 'cheese', 'veggie', 'Works', 'Anchovie', 'sausage']
print lst
# lst = sorted(lst, key=lambda s: s.lower())
lst.sort(key=lambda y: y.lower(), reverse=True)
print lst
```

**143.** For perl, use ".".    For python, use "append"

**144.**

For perl,
```
$out = $ARGV[0];
if ($ARGV[0]<$ARGV[1]){
    $out = $ARGV[1];
}
if ($out <$ARGV[2]){
    $out = $ARGV[2];
}
print $out;
```
For python, similar…

**145.** a) Warnings are one of the most basic ways in which you can get Perl/Python to check the quality of the code that you have produced. Mandatory warnings highlight problems in the lexical analysis stage. Optional warnings highlight cases of possible anomaly. To turn warnings on: use warnings; (note that for python, you may import the lib: import warnings)
b)
For perl:
Scalars : store single values and are preceded by $ sign
Arrays: store a list of scalar values and are preceded by @ sign

Hashes: store associative arrays which use a key value as index instead of numerical indexes.
Use % as prefix.
For Python:
Numeric types: int, long, float
Sequences: str, list, tuple
Sets: set
Mappings: dict

c)

```
for Perl:
open(FILE," input.txt ");
@array= ;
$wor="memory";
$count=0;
foreach $line (@array)
{
@arr=split (/s+/,$line);
```

d)
```
foreach $word (@arr)
{
if ($word =~ /s*$wors*/i)
$count=$count+1;
}
}
print "The word memory occurs $count times";
```
For python:
Similar approach

similar for python

# 20.17: Chapter 17 – Miscellaneous

**146)** a) SRAM.v: interface to read/write the memory.list; SRouter.v: main router; tb.v: testbench to initialize both modules
b) address for target/source can be put in the memory at a certain location
c) hint: try via as another neighbor grid
d) hint: Most likely the Verilog code has to redone, with more balancing and parallelism to shorten the critical path. It may also be possible buffer insertion or refining tool constraints could work out.
e) same as d)

**147)** a) more parallelism
b) hint: add some stage reg variables
c) hint: lookup table method

**148)** a) Please refer to the tutorial
b) Placement: the portion of the physical design flow that assigns exact locations for various circuit components within the chip's core area.
 Routing: connecting each block using physical wires based on the logic connection
c) hint: FPGA tool has its own router
d) routing and parasitic cap added. Critical path may get increased

**149) a)** Synthesis is the process of converting a high-level description of the design into an optimized gate-level representation given a standard-cell library and certain design constraints. Inputs: HDL specification, constraints (area, power, timing), library. Output: structural netlist of standard cells (often in a structural Verilog format).
 The basic flow:
 1. Configuration variables 2. Specify Libraries 3. Read design 4.Set design Constraints 5. Compile 6. Reports 7. Write design
 **b)** Target Library: A technology library that Design Compiler maps to during optimization. Along with the link_library and search_path variables, you need to specify the logical library that will be used for mapping/optimization. It is the library you synthesis to.
 Link Library: The technology library that contains the definition of the cells used in the mapped design. In principle should be the same as target_library unless a technology translation is being performed. It is the library you use components from.
 **c)** There are three basic constraints specified in sdc. Clock, input delay and output delays.
 DC makes a best effort attempt to synthesize your design while still meeting the two types of constraints: user specified constraints and design rule constraints. User specified constraints can be used to constrain the clock period (as was done with the create clock command) as well as the arrival of certain input signals, the drive strength of the input signals, and the capacitive load on the output signals. Design rule constraints are fixed constraints which are specified by the standard cell library.
 **d)** Applying timing constraints, flat the design, logic cloning, rebalance registers/flops, usage of LVT & HVT libs...

**150)** **a)** Using STA to find the hold violation path and add delays, such as adding buffer/ inverter pairs/ delay cells to the data path to fix the hold violation. Decreasing the size of certain cells in the data path. It is better to reduce the cells closer to the capture FF because there is less likelihood of affecting other paths and causing new errors.

**b)** STA is a method of validating the timing performance of a design by checking all possible paths for timing violations. Compared to dynamical timing analysis, the advantages are: 1. All timing paths are considered for the timing analysis. This is not the case in simulation. 2. Analysis times are relatively short when compared with event and circuit simulation.3. Timing can be analyzed for worst case, best case simultaneously. This type of analysis is not possible in dynamic timing analysis. 4. Static Timing Analysis (STA) works with timing models. STA has more pessimism and thus gives maximum delay of the design. DTA performs full timing simulation. The problem associated with DTA is the computational complexity involved in finding the input patterns (vectors) that produce maximum delay at the output and hence it is slow.

Disadvantages of STA: 1. All paths in the design may not run always in worst case delay. Hence the analysis is pessimistic. 2. Clock related all information has to be fed to the design in the form of constraints. 3. Inconsistency or incorrectness or under constraining of these constraints may lead to disastrous timing analysis. 4. STA does not check for logical correctness of the design. 5. STA is not suitable for asynchronous circuits.

C) A false path, as its name denotes is a timing path not required to meet its timing constraints for the design to function properly.

A Multicycle path in a sequential circuit is a combinational path which doesn't have to complete the propagation of the signals along the path within one clock cycle. For a Multicycle path of N, design should ensure the signal transition propagated from source to destination within N clock cycle.

A virtual clock is a clock that has been defined, but has not been associated with any pin/port. A virtual clock is used as a reference to constrain the interface pins by relating the arrivals at input/output ports with respect to it with the help of input and output delays.

A timing arc defines the propagation of signals through logic gates/nets. Timing arc is one of the components of a timing path. Static timing analysis works on the concept of timing paths. Each path starts from either primary input or a register and ends at a primary output or a register. In-between, the path traverses through what are known as timing arcs. We can define a timing arc as an indivisible path from one pin to another that tells EDA tool to consider the path between the pins.

**151)** **a)** CAM stands for Content Addressable Memory. Unlike standard computer memory (random access memory or RAM) in which the user supplies a memory address and the RAM returns the data word stored at that address, a CAM is designed such that the user supplies a data word and the CAM searches its entire memory to see if that data word is stored anywhere in it. If the data word is found, the CAM returns a list of one or more storage addresses where the word was found (and in some architectures, it also returns the contents of that storage address, or other associated pieces of data).

Binary CAM is the simplest type of CAM which uses data search words consisting entirely of 1s and 0s. Ternary CAM (TCAM) allows a third matching state of "X" or "don't care" for one or more bits in the stored dataword, thus adding flexibility to the search.

For example, a ternary CAM might have a stored word of "10XX0" which will match any of the four search words "10000", "10010", "10100", or "10110". The added search flexibility comes at an additional cost over binary CAM as the internal memory cell must now encode three possible states instead of the two of binary CAM. This additional state is typically implemented by adding a mask bit ("care" or "don't care" bit) to every memory cell.

**b)** Yes. Require additional logic or encoding to handle "don't care". The solution is not unique. One example is

```verilog
// Ternary match
wire [DATA_WIDTH-1:0] bit_match;
genvar i;
generate
   for (i=0; i<DATA_WIDTH; i=i+1)
   begin : bmt
      assign bit_match[i] = !care[i] | !(data[i] ^ lookup_data[i]);
   end
endgenerate

always @(posedge clk) begin
   if (rst) match <= 1'b0;
   else match <= (& bit_match) & cell_used;
end

endmodule
```

**c)** Whenever there are setup and hold time violations in any flip-flop, it enters a state where its output is unpredictable: this state is known as metastable state (quasi stable state); at the end of metastable state, the flip-flop settles down to either '1' or '0'. This whole process is known as metastability.

Clock Domain Crossing Issues:

   a.   **Metastability**

For single and multi-bit control signals and single bit data signals in the design, use two FF synchronizers. For multi-bit data signals, use MUX recirculation, handshake and FIFO.

   b.   **Data loss**

Whenever a new source data is generated, it may not be captured by the destination domain in the very first cycle of the destination clock because of metastability

In order to prevent data loss, the data should be held constant in the source domain long enough to be properly captured in the destination domain. In other words, after every transition on source data, at least one destination clock edge should arrive where there is no setup or hold violation so that the source data is captured properly in the destination domain. There are several techniques to ensure this. For example, a finite state machine (FSM) can be used to generate source data at a rate, such that it is stable for at least 1 complete cycle of the destination clock. Handshake and FIFO are also the solution.

   c.   **Data Incoherency**

Gray-encoded.

**152) a)** 3 input AND, the only syntax error is "1" in line 2
**b)** all correct
    **c)** Chaining is a simple solution and hierarchical hashing a more complicated but efficient one.

d) O(n) to find and O(1) to delete, so overall O(n)

e) This source lists some information:
http://www.thecoffeebrewers.com/article2.html

But let's use the opportunity and talk about the difference between Espresso and Quinne-McCluskey. Q-M is an exact algorithm, but with exponential complexity, however Espresso is a polynomial time heuristic with sufficiently good results.

f) Many steps of the flow (transformation, simulation, and verification) are proven to be hard problems, meaning polynomial time algorithms would not exist for them, therefore as a compromise we use time/space efficient heuristics that deliver reasonably good results.

## 153) Timing

**a)** at the beginning of initialization

**b)** Assume we need n stages, $3.74ns/n + 100ps < 1/(2GHz)$, find n=10

**c)** $x(x/2*0.5fF+15fF) = (40um-x)\{(40um-x)/2*0.5fF+20fF\}$

**d)** jitter/hold. Jitter is assumed to affect both capturing and launching clocks of a certain edge, as the noise would typically not last for more than a cycle. This means the hold time would not be violated as it depends to the current clock transitions. However there is a chance setup time is violated, as it depends to current and next clock transitions.

**e)** Clock skew is the difference in clock arrival times at two spatially distinct points. Clock jitter is the difference in clock period. Clock is generated by an electronic circuit inside or outside the chip which is sensitive to different types of noise. Although not agreed by all, skew is typically looked at as static and spatial and jitter as dynamic and temporal. Timing uncertainty that needs to be accounted during design (e.g., while running DC) and verification (e.g., while running PrimeTime STA) should account for both spatial skew and temporal jitter. Quantizing timing uncertainty as a function of skew and jitter is not straightforward. Assuming that skew remains fairly constant from cycle to cycle, one idea is to monitor the skew between each two clock pairs of interest for a long time. Any perturbation cause by cycle time of each one of the two clock signals (as compared to rather constant skew between them) could be presented as the jitter for the corresponding signal.

**f)** Clock skew along different paths is monitored and if the difference increases, the paths are dynamically modified, e.g., their frequency is adjusted. This is called Clock De-skewing. Clock de-skewing techniques compensate for device and interconnect within-die variation. Ex: An 8-level digitally adjustable delay line - in Pentium II a phase comparator checks the arrival time of clocks and adjust digitally controlled delay lines to have simultaneous clock arrival times.

## 154) DDR

a)   Check whether (ready =1).
b)

      a.     The size of blocks
      b.     The internal command FIFO is not full
      c.     Output valid: High when the address and data at the output of the controller is valid
      d.     Goes high after the DDR2 initialization is complete
      e.     This input tells the controller that the data and address at its output is being fetched this cycle
      f.     Data strobe signal.

c) NOP command

d) design

e) The interface. Because of SI issues, interface can be the bottleneck of performance.

f) Not correct, because the system frequency of DDR3 is 4 times of that of DDR. However, the total CAS latency time has not improved so much. As total CAS latencytime = clock period * CAS latency cycle, CAS latency cycle of DDR3 should be higher than that of DDR.

g) Using posted CAS, we can give read/write command right after active command, then AL compensates for the RAS-CAS delay. Total read latency will be determined by AL+CL, and similarly the write latency.

h) DDR2 memory controller is the main interface to external DDR2 memory. It performance all memory-related background tasks such as opening or closing banks, refreshing, and command arbitration. In nutshell, it is to operate DDR2 correctly.
DDR2 controller is used to assert commands, write in data and receive the read data at correct time.

i) Like DDR before it, DDR2 cells transfer data both on the rising and falling edge of the clock (a technique called double pumping). DDR2 is a multi-bank architecture which allows for concurrent operation. The key difference between DDR and DDR2 is that in DDR2 the bus is clocked at twice the speed of the memory cells, so four words of data can be transferred per memory cell cycle. Thus, without speeding up the memory cells themselves, DDR2 can effectively operate at twice the bus speed of DDR. (Refer to http://en.wikipedia.org/wiki/DDR2_SDRAM).

j) Project related DDRx technologies

a. Precharge can be initiated by either a manual PRECHARGE command or by an autoprecharge in conjunction with either a READ or WRITE command. During a manual PRECHARGE command, the A10 input determines whether one or all banks are to be precharged.
b. Only one row
c. We should set up M2-M0 of MR to 011(BL=8); E5-E3 of EMR to 010(AL=2); M6-M4 of MR to 101(CL=5).
d. ODT is to improve signal integrity of the memory channel by allowing the DDRx SDRAM controller to independently turn on/off ODT for any devices. OCD is for driver calibration. OCT and ODT can coexist, so having one does not cover for another.
e. SSTL_18 is used and its reference voltage Vref is MIN: 833mV, Nom: 900mV, Max: 969mV.
f. RAS# = L, CAS#=L, AL and BA are don't care,i.e. AL=X, BA=X.
g. With interleaved addressing mode, memory addresses are allocated to each memory bank in turn. The interleaved burst mode computes the address using an xor operation between the counter and the initial address.

h.  Find the commands in the following waveform.

[Waveform diagram showing DDR2 SDRAM initialization sequence with signals CK#/CK, CKE, ODT, Command, DM, Address, DQS, DQ, Rtt. Commands shown: NOP, PRE, LM, LM, LM, LM, PRE, REF, REF, LM, LM, LM, Valid. Timing markers: T=200μs (MIN), T=400ns (MIN), tRPA, tMRD, tMRD, tMRD, tMRD, tRPA, tRFC, tRFC, tMRD, tMRD, tMRD. Annotations: EMR(2), EMR(3), EMR, MR with DLL RESET, MR without DLL RESET, EMR with OCD default, EMR with OCD exit. Note: "200 cycles of CK are required before a READ command can be issued". Normal operation at end.]

i.  BL/2
j.  We could use data mask to mask the word that we do not write.
k.  FIFOs are designed to decouple flow between processor and memory controller. FIFOs are used to stores fetched commands, input data to memory and output data from memory.
l.  -The scalar read/write operations operated on a single word out of the accessed burst length number of words. For scalar read, the controller stores the very first value read to the output data FIFO and ignores others. For the scalar write command, data mask is applied to prevent subsequent words from being overwritten.
    -Burst mode is when you send one address to the memory, but rather than reading/write the data only for the specified address, you also read/write some number of consecutive locations (typically 4 or 8).

m.  M15(BA1) and M14(BA0)

n.  For return address FIFO and return data FIFO, they can share the same put/get signal so we can simply do a width expansion to combine them.
    For input data FIFO and input command/address FIFO, they cannot share the same put/get signals because read command has no data to write into the data FIFO.

o.
```
always@(CMD, READY, RESET, CMD_NOTFULL)
begin
    if(RESET)
    begin
            IN_put = 0;
            CMD_put = 0;
    end
    else if(READY && CMD_NOTFULL)
            case(CMD)
                    3'b001:
                    begin
                            CMD_put = 1;
                            IN_put = 0;
                    end
```

```verilog
                        3'b010:
                        begin
                                CMD_put = 1;
                                IN_put = 1;
                        end
                        default:
                        begin
                                CMD_put = 0;
                                IN_put = 0;
                        end
                    endcase
            else
            begin
                    CMD_put = 0;
                    IN_put = 0;
            end
    end
end
```

p ) The buffers will be removed during the synthesis step and we will not be able to delay the strobe.

q ) We can remove the "full" signal of the return FIFO. Because the testbench is trying to fetch data from return FIFO all the time, which means the "get" signal of return FIFO is active whenever the return FIFO is not empty. As a result, return FIFO will never get full.

## 155) Testing

a) A = 0, B = 1, C = 1, E =1
b) 0111 does not. E=1 excites a D, but k=0 would not let it be observed at the PO (i.e., L).
c) In part (a) test generation was done. Part (b) is like an example of fault simulation for a single fault: Given a vector we tried to see what faults it could test. If we tried part ( b) on all faults of our interest, then we could call that a complete fault simulation done, before moving to ATPG for the rest of the fault.
d) The voltage at the short could be any value, depending on the characteristics of the PMOS and NMOS that would ultimately define their strengths.
e) IDDQ testing would be able to detect that short.

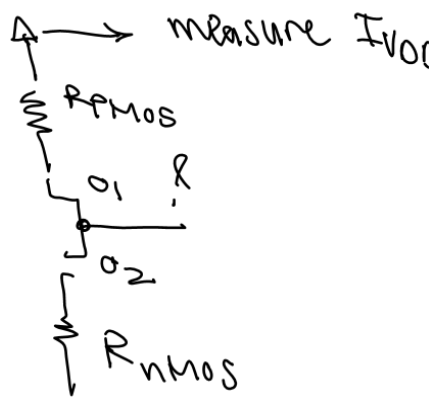

f)

    I ) SA0 at G and SA1 at M are detected.

    II ) Note of the 3 faults can be detected.

158
Shahin Nazarian    USC    All Rights Reserved.

g)

I ) e.g., 10X1X or 1X01X       II) 10000

III) e.g., 10X1X       IV ) X111X and then measuring I of VDD

## 156)   DLL & PLL

a)   DLLs use variable "phase" (delay) to achieve lock, i.e. they lock onto a fixed phase difference whereas PLLs use variable frequency block, i.e. they adjust their "frequency" until there is a lock. PLLs use a VCO which DLLs don't need. DLLs are more recently developed than PLLs and are used more in digital applications.
A DLL compares the phase of its last output with the input clock to generate an error signal which is then integrated and fed back as the control to all of the delay elements. The integration allows the error to go to zero while keeping the control signal, and thus the delays, where they need to be for phase lock. Since the control signal directly impacts the phase this is all that is required.
A PLL compares the phase of its oscillator with the incoming signal to generate an error signal which is then integrated to create a control signal for the VCO. The control signal impacts the "frequency" of the oscillator, and phase is the integral of frequency, so a second integration is unavoidably performed by the oscillator itself.
The simple answer is that DLL that uses phase instead of frequency. However as explained earlier, even PLL ultimately needs to deal with phase as well.

b)   DLL block diagram:

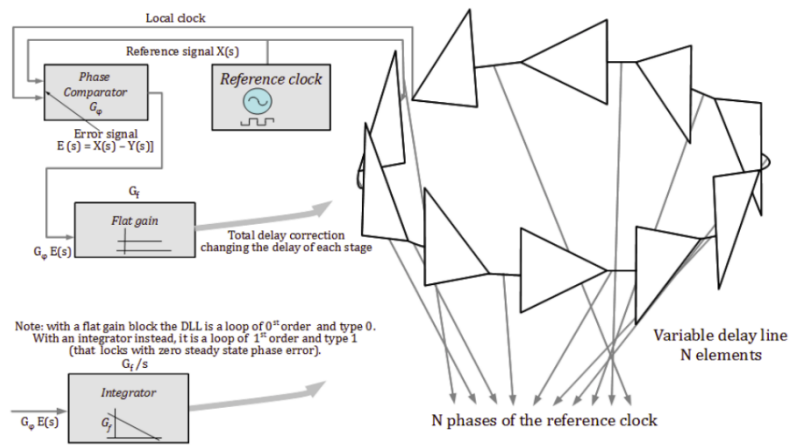

c)   Digital PLL:
   TDC: Time-to-Digital Converter
   DCO: Digitally-Controlled Oscillator

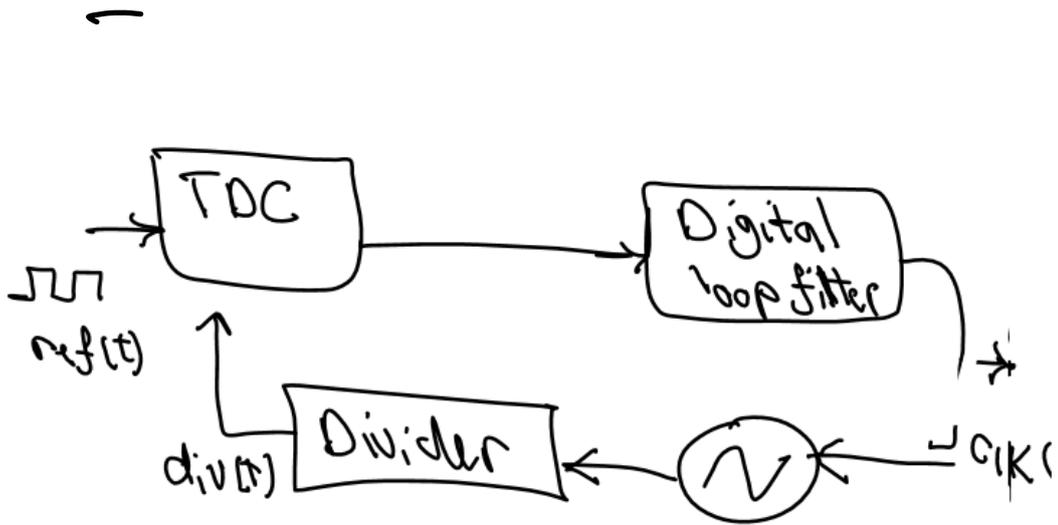

### 157) MISR & BIST

There are many ways to MISR the BIST out of something :P This is just one of them:

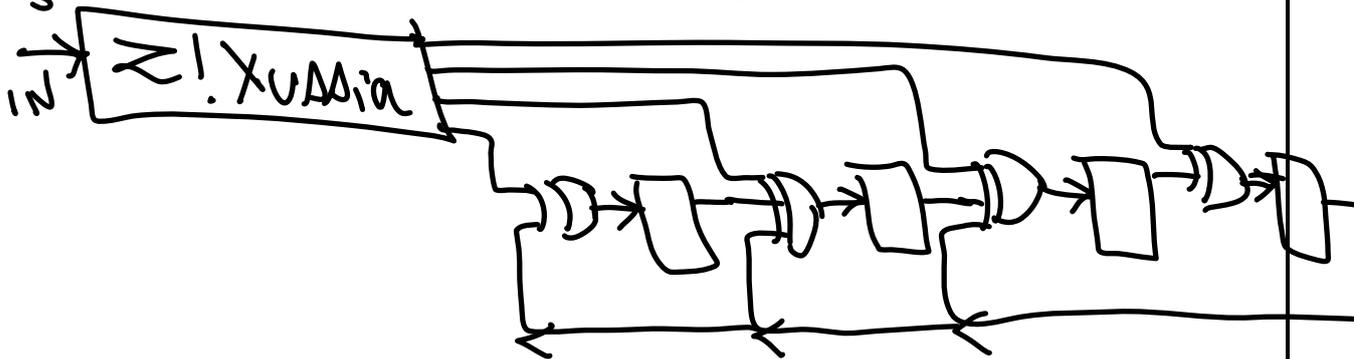

### 158) FIFO
False

### 159) Twisted Bitline
I

### 160)
The idea is to send not only the data, but also the clock of the sending domain to the receiving domain. The receiving domain would then use the sending clock to sample the data. This increases the signal integrity. Having sampled the data with the sending clock, the receiving domain would later on sample it using its own clock, but until then, the data could remain in an elasticity buffer or FIFO.

### 161)
always @ (posedge clk)
    if (reset) ck =0;
    else ck=~ck;

### 162)

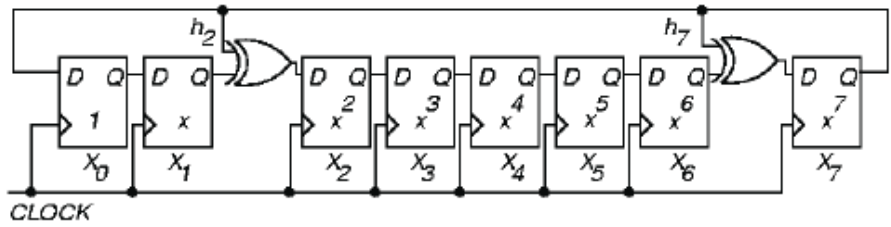

$$M = \begin{bmatrix} 1 & 0 & 0 & 0 & 0 & 0 & 0 \\ 0 & 1 & 0 & 0 & 0 & 0 & 0 \\ 0 & 0 & 1 & 0 & 0 & 0 & 0 \\ 0 & 0 & 0 & 1 & 0 & 0 & 0 \\ 0 & 0 & 0 & 0 & 1 & 0 & 0 \\ 0 & 0 & 0 & 0 & 0 & 1 & 0 \\ 0 & 0 & 0 & 0 & 0 & 1 & 1 \end{bmatrix}$$

$$\begin{bmatrix} X_0(t+1) \\ X_1(t+1) \\ \vdots \\ X_\tau(t+1) \end{bmatrix} = M * \begin{bmatrix} n_0(t) \\ n_1(t) \\ \vdots \\ n_\tau(t) \end{bmatrix}$$

**163)** This must be a NAND array. The operation is erase of array based on FN. By the way, sources and drains are open through the BLs and source lines and select lines that are open. Also the gates are the wordlines that are feeding the control gates.

Trench capacitance will get taller, ECC will be more complicated, etc.

**164) FSM Solution:**

1)

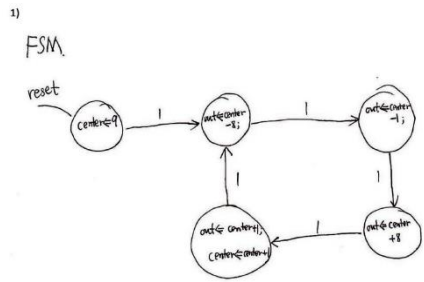

2)

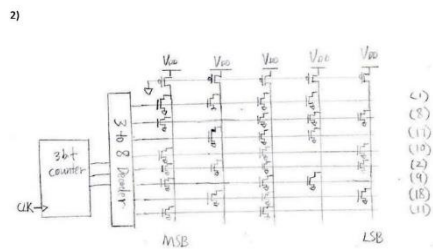

3) The size of decoder will be a big overhead.

## 20.18: Chapter 18 – Verilog Basics

**165)**
- **a)** value is 4. missing endcase, four undefined, missing a variable definition and missing a bus width,
- **b)** 36,10
- **c)** 4'b0000    1'b0   4'b0110    8'b110000xx    9'b111100100
- **d)** The statement b = #2 a+c evaluates the value of a+c at time t =0 and then wait for 2 time units and then assigns the value to b.
- **e)** The equality operators ( = = , ! = ) will yield an x if either operand has x or z in its bits. Where as the case equality operators ( = = = , ! = = ) compare both operands bit by bit and compare all bits, including x and z.
- **f)** Example code:

always @(s1 or s0 or i0 or i1 or i2 or i3)

   case ({s1, s0})

     2'd0 : out = i0;

     2'd1 : out = i1;

     2'd2 : out = i2;

   endcase

How to avoid latches: To avoid inferring latches make sure that all the cases are mentioned or add a default condition.

# 20.19: Chapter 19 – Verilog Coding

## 166)
### a) Behavioral RTL Verilog – Transparent Latch

```
module TR_LATCH( in, clk, out, reset)
    input in, clk;
    output out
       always @ (clk or reset or in)
      begin
       If (reset==1)  out <= 0;
       else if (clk==1) out <= in ;
      end
    //your code here
endmodule
```

### b) Behavioral Dataflow Verilog – 8-to-1 Multiplexer

```
module MUX_8TO1( d, select, q );
input[7:0] d;
input[2:0] select;
output    q;
        assign out = d[select];  //your code here
endmodule
```

### c) Behavioral RTL Verilog – 4-bit Shift Register

```
module SR4(clk, rst, dir, D, SER_IN, Q);
  input clk, rst, dir, SER_IN;
  input  [3:0] D;
  output [3:0] Q;
  reg [3:0] temp;
    always @(posedge clk)
     begin
       if(!rst)        temp[3:0] = D[3:0];
        else
             if (dir)    temp[3:0] = {SER_IN, temp[3:1]};
             else    temp[3:0] = {temp[2:0], SER_IN};
     end
      assign Q[3:0] = temp[3:0];
//your code
endmodule
```

### d) Behavioral and Structural RTL Verilog Combined – 16-bit Shift Register

```
module SR16(clk, rst, dir, D, SER_IN, Q);
  input        clk, rst, dir, SER_IN;
  input  [15:0] D;
  output [15:0] Q;
```

Version I                                                              Version II

163
Shahin Nazarian                              USC                           All Rights Reserved.

```verilog
    reg   C0, C1, C2, C3;
    SR4 s0(rst, clk, dir, D[3:0],   C0, Q[3:0]);
    SR4 s1(rst, clk, dir, D[7:4],   C1, Q[7:4]);
    SR4 s2(rst, clk, dir, D[11:8],  C2, Q[11:8]);
    SR4 s3(rst, clk, dir, D[15:12], C3, Q[15:12]);
  always @ (*)
    begin
        if(dir) begin
            C0 = Q[4];
            C1 = Q[8];
            C2 = Q[12];
            C3 = SER_IN;
        end else
        if(!dir) begin
            C3 = Q[11];
            C2 = Q[7];
            C1 = Q[3];
            C0 = SER_IN;
        end
    end
  //your code here
  endmodule
```

```verilog
    wire C0, C1, C2, C3
    SR4 s0(rst, clk, dir, D[3:0],   C0, Q[3:0]);
    SR4 s1(rst, clk, dir, D[7:4],   C1, Q[7:4]);
    SR4 s2(rst, clk, dir, D[11:8],  C2, Q[11:8]);
    SR4 s3(rst, clk, dir, D[15:12], C3, Q[15:12]);

    assign C0 = (dir==1'b1)?Q[4]:SER_IN;
    assign C1 = (dir==1'b1)?Q[8]:Q[3];
    assign C2 = (dir==1'b1)?Q[12]:Q[7];
    assign C3 = (dir==1'b1)?SER_IN:Q[11];
    //your code here
  endmodule
```

e) **Behavioral RTL Verilog– FSM Design**

```verilog
module FSM2 (input wire Reset, input wire In, output wire Out);
    localparam Jaleh = 2'b00, Neda = 2'b01, Mo = 2'b10;
    reg [1:0] CurrentState, NextState;
  always @ (posedge Clock) begin
      if (Reset) begin
          CurrentState < = Jaleh;                         // missing
      end
      else begin
          CurrentState <= NextState;                      // missing
      end
    End
  always @ ( * ) begin
      NextState = CurrentState;
      case (CurrentState)     // case statements are missing
      Sue: begin NextState = (In) ? Neda : Jaleh; end
Ken: begin NextState = (In) ? Mo : Jaleh; end
Bob: begin NextState = (In) ? Mo : Jaleh; end
Default: begin NextState = Jaleh;

      Endcase
```

```
   End
 assign Out =    (CurrentState == Neda);                              // missing
 Endmodule
```

### f) FSM – Verilog Design of a Crosswalk Controller
Answer not unique. It is recommended to simulate your version to realize your possible design issues more effectively.

### g) Hierarchical Parametric FIFO Design and Verification
(a) Answer is not unique. It would be a good idea to do a width expansion first and create a width expanded FIFO and then go for depth expansion. It is recommended to simulate your version to realize your possible design issues more effectively. You need to consider synthesizability, and simplicity and efficiency of design into account.
(b) It is impossible to test all the possible cases, therefore you should use the ideas to increase coverage.
Note: A non-synthesizable design can in the worse-case receives 0 points, however we would follow a more generous grading policy and give partial credit to non-synthesizable work depending how the rest of the students perform. Regardless of the grading policy, you should follow our recommendations and write synthesizable design. Of course this is just for the design (and not the testbench).

### h) 33-Dimensional Maze Router
Although other faster algorithms could be used, especially for multi-dimensional routers where the time and space complexity would be exponential; the idea is to modify Lee's Maze Router algorithm of Lab1 (just a two node (single source, single target) version). It would be very slow, but doable. The core of the algorithm would not change, in the sense that starting from source, one could propagate the cost in all the multiple 33 dimensions, and since Zaydaf comes with negative cost, most likely that would contribute significantly to the path from source to target. The only reason Zaydaf dimension could not be used is any blockage in that dimension. In any case, the wave propagation in multiple dimensions and tracing back from target back to source (which would be a path with multiple dimensions) would be the core of the algorithm. You may assume a very small grid size for each dimension in case you plan to simulate your Verilog code. You do not need to synthesize your code. The memory management should be like what you did in Lab1, for for 33 dimensions. The answer is not unique.

### i) Write a Verilog code of the following design.
Solution not unique. Hint: You may use structural design (i..e, include tri-state primary gate)

### j) Verilog Design – Gray Counter:
Design a 6bit gray counter. This counter has a synchronous reset signal.

```
module gray_counter (
   out , // counter out
   enable , // enable for counter
   clk , // clock
```

```verilog
        rst // active hight reset
        );
        input clk, rst, enable;
        output [ 5:0] out;
        wire [5:0] out;
        reg [5:0] count;
        always @ (posedge clk)
            if (rst)
                    count <= 0;
            else if (enable)
                    count <= count + 1;

   assign  out  =  {count[5],(count[4]  ^  count[3]),(count[3] ^ count[2]),  (count[2] ^ count[1]),(count[1] ^ count[0]) };
end
```